\documentclass[12pt,a4paper]{article}
\usepackage{epsfig}
\usepackage{axodraw}
\def\nl{\nonumber\\}
\def\beq{\begin{equation}}
\def\eeq{\end{equation}}
\def\beqar{\begin{eqnarray}}
\def\eeqar{\end{eqnarray}}
\def\bfi{\begin{figure}}
\def\efi{\end{figure}}
\def\btab{\begin{table}}
\def\etab{\end{table}}
\def\bce{\begin{center}}
\def\ece{\end{center}}
\def\bit{\begin{itemize}}
\def\eit{\end{itemize}}

\def\scrs{\scriptscriptstyle}
\def\text{\textstyle}

\def\al{\alpha}

\def\ga{\gamma}
\def\de{\delta}
\def\veps{\varepsilon}
\def\la{\lambda}
\def\si{\sigma}

\def\refeq#1{\mbox{(\ref{#1})}}
\def\reffi#1{\mbox{Fig.~\ref{#1}}}
\def\reffis#1{\mbox{Figs.~\ref{#1}}}

\def\refse#1{\mbox{Sect.~\ref{#1}}}
\def\refses#1{\mbox{Sects.~\ref{#1}}}

\def\refapp#1{\mbox{Appendix~\ref{#1}}}
\def\citere#1{\mbox{Ref.~\cite{#1}}}
\def\citeres#1{\mbox{Refs.~\cite{#1}}}

\newcommand{\GeV}{\unskip\,\mathrm{GeV}}

\newcommand{\TeV}{\unskip\,\mathrm{TeV}}

\def\mathswitchr#1{\relax\ifmmode{\mathrm{#1}}\else$\mathrm{#1}$\fi}

\newcommand{\PW}{\mathswitchr W}

\def\mathswitch#1{\relax\ifmmode#1\else$#1$\fi}

\newcommand{\sw}{\mathswitch {s_{\scrs\PW}}}
\newcommand{\cw}{\mathswitch {c_{\scrs\PW}}}
\newcommand{\rw}{{\mathrm{W}}}

\newcommand{\cew}{C^{\ew}}

\newcommand{\Mathematica}{{\sc Mathematica}}

\def\ie{i.e.\ }

\def\wrt{wrt.\ }

\newcommand{\ord}{{\cal O}}
\newcommand{\Oa}{\mathswitch{{\cal{O}}(\alpha)}}

\newcommand{\lsim}
{\;\raisebox{-.3em}{$\stackrel{\displaystyle <}{\sim}$}\;}

\newcommand{\ew}{\mathrm{ew}}

\newcommand{\ri}{\mathrm{i}}

\newcommand{\rr}{\mathrm{r}}

\newcommand{\rR}{\mathrm{R}}
\newcommand{\rL}{\mathrm{L}}
\newcommand{\rT}{{\mathrm{T}}}
\newcommand{\rS}{{\mathrm{S}}}
\newcommand{\rd}{{\mathrm{d}}}
\newcommand{\rc}{{\mathrm{c}}}

\newcommand{\rF}{{\mathrm{F}}}

\newcommand{\pT}{p_{\mathrm{T}}}
\newcommand{\pTcut}{p_{\mathrm{T}}^{\mathrm{cut}}}

\newcommand{\M}{{\cal {M}}}
\newcommand{\calL}{{\cal L}}
\newcommand{\cD}{{\cal D}}

\newcommand{\NNLLa}{\stackrel{\mathrm{NNLL}}{=}}
\newcommand{\NLLa}{\stackrel{\mathrm{NLL}}{=}}

\newcommand{\shat}{{\hat s}}
\newcommand{\that}{{\hat t}}
\newcommand{\uhat}{{\hat u}}
\newcommand{\rhat}{{\hat r}}
\newcommand{\sphat}{\hat{s}'}
\newcommand{\tphat}{\hat{t}'}
\newcommand{\uphat}{\hat{u}'}

\newcommand{\rar}{{\rightarrow}}
\newcommand{\lrar}{{\leftrightarrow}}

\newcommand{\ps}{p\hspace{-0.42em}/\hspace{-0.08em}}

\newcommand{\MSBAR}{\overline{\mathrm{MS}}}
\newcommand{\msbar}{$\MSBAR$}
\newcommand{\qbar}{{\bar q}}
\newcommand{\A}{{\mathcal{A}}}
\newcommand{\F}{{\mathcal{F}}}
\newcommand{\smel}{\mathcal{S}}
\newcommand{\loops}{{J}}
\newcommand{\Mqq}{\M^{\qbar q'\to W^\si g}}

\def\draftdate{\relax}
\def\mpar#1{\relax}
\def\mua{\relax}
\def\mda{\relax}
\def\mla{\relax}
\def\draft{
\def\thtystars{******************************}
\def\sixtystars{\thtystars\thtystars}
\typeout{}
\typeout{\sixtystars**}
\typeout{* Draft mode!
         For final version remove \protect\draft\space in source file *}
\typeout{\sixtystars**}
\typeout{}
\def\draftdate{\today}
\def\mua{\marginpar[\boldmath\hfil$\uparrow$]%
                   {\boldmath$\uparrow$\hfil}%
                    \typeout{marginpar: $\uparrow$}\ignorespaces}
\def\mda{\marginpar[\boldmath\hfil$\downarrow$]%
                   {\boldmath$\downarrow$\hfil}%
                    \typeout{marginpar: $\downarrow$}\ignorespaces}
\def\mla{\marginpar[\boldmath\hfil$\rightarrow$]%
                   {\boldmath$\leftarrow $\hfil}%
                    \typeout{marginpar: $\leftrightarrow$}\ignorespaces}
\def\Mua{\marginpar[\boldmath\hfil$\Uparrow$]%
                   {\boldmath$\Uparrow$\hfil}%
                    \typeout{marginpar: $\Uparrow$}\ignorespaces}
\def\Mda{\marginpar[\boldmath\hfil$\Downarrow$]%
                   {\boldmath$\Downarrow$\hfil}%
                    \typeout{marginpar: $\Downarrow$}\ignorespaces}
\def\Mla{\marginpar[\boldmath\hfil$\Rightarrow$]%
                   {\boldmath$\Leftarrow $\hfil}%
                    \typeout{marginpar: $\Leftrightarrow$}\ignorespaces}
\def\mpar##1{\marginpar{\hbadness10000%
                      \sloppy\hfuzz10pt\boldmath\bf##1}%
                      \typeout{marginpar: ##1}\ignorespaces}

\overfullrule 5pt
\oddsidemargin -15mm
\marginparwidth 29mm
}

\begin{document}
\newcommand{\pantip}{\hspace{0.1em} p \hspace{-0.69em} ^{^{(-)}}}

\thispagestyle{empty}
\def\thefootnote{\fnsymbol{footnote}}
\setcounter{footnote}{1}
\null
\draftdate
\hfill   TTP07-19\\
\strut\hfill  SFB/CPP-07-43\\
\strut\hfill MPP-2007-102\\
\strut\hfill DESY 07-112\\
\vskip 0cm
\bigskip

\begin{center}
{\Large \bf
Electroweak corrections to hadronic production of $W$ bosons at large transverse momenta
\par}

\bigskip
\bigskip

{\large \sc
Johann~H.~K\"uhn$^a$,
A.~Kulesza$^b$,
S.~Pozzorini$^c$,
M.~Schulze$^a$}

\bigskip

\begin{it}

$^a$Institut f\"ur Theoretische Teilchenphysik, 
Universit\"at Karlsruhe,\\
D-76128 Karlsruhe, Germany\\
\bigskip
$^b$Deutsches Elektronen-Synchrotron DESY, Notkestrasse 85,\\
D--22607 Hamburg, Germany \\
\bigskip
$^c$Max-Planck-Institut f\"ur Physik, F\"ohringer Ring 6,\\
D--80805 Munich, Germany

\end{it}

\bigskip
\bigskip
\end{center}
\vfill 

\par
\noindent
\par
\null
\setcounter{page}{0}
\def\thefootnote{\arabic{footnote}}
\setcounter{footnote}{0}

{\bf Abstract:} \par 
To match the precision of present and future measurements
of $W$-boson production at hadron colliders electroweak radiative 
corrections must be included in the theory predictions.
In this paper we consider their effect on the
transverse momentum ($p_\rT$) distribution of $W$ bosons,
with emphasis on large $p_\rT$. 
We evaluate the full electroweak $\ord(\alpha)$ corrections
to the processes 
$p p \to W + \mathrm{jet}$ and $p \bar p \to W +\mathrm{jet}$ 
including
virtual and real photonic contributions.
We present the explicit expressions in analytical form for the virtual
corrections and provide results for the real corrections, 
discussing in detail the treatment of soft and collinear singularities.
We also provide compact approximate expressions which are valid
in the high-energy region, where the electroweak corrections are strongly
enhanced by logarithms of $\shat/M_W^2$.
These expressions describe the complete asymptotic behaviour at one loop
as well as the leading and next-to-leading logarithms at two loops. 
Numerical results are presented for proton-proton
collisions at $14 \TeV$ and proton-antiproton collisions at $2 \TeV$.
The corrections are negative and their size increases with $p_\rT$.
At the LHC, where transverse momenta of $2 \TeV$ or more can be reached,
the one- and two-loop corrections amount up to
$-40\%$ and $+10\%$, respectively, and will be important for a precise
analysis of $W$ production. At the Tevatron, transverse momenta up to
$300 \GeV$ are within reach. In this case the 
electroweak corrections 
amount up to $-10\%$ and are thus larger than the 
expected statistical error.

\newpage

\section{Introduction}
After the startup of the Large Hadron Collider (LHC) 
hard scattering reactions will be explored with high event
rates and momentum transfers up to several TeV. In order to identify
new phenomena in this region, the predictions of the Standard Model
have to be understood with adequate precision.

The study of gauge-boson production has been among the primary 
goals of hadron colliders, starting with the discovery of the $W$ 
and $Z$ bosons more than two decades ago 
\cite{Arnison:1983rp}.
The investigation of the production dynamics, strictly predicted by 
the electroweak theory, constitutes one of the important tests
of the Standard Model. 
Differential distributions of gauge bosons, in rapidity as well
as in transverse momentum ($p_\rT$), 
have always been the subject of
theoretical and experimental studies.
This allows to search for and set limits on anomalous gauge-boson
couplings, measure the parton distribution functions and, if
understood sufficiently well, use these reactions to calibrate the
luminosity.
For gauge-boson production at large $\pT$ the final state of the 
leading-order process consists of an electroweak
gauge boson plus one recoiling jet. Being, in leading order,
proportional to the strong coupling constant, these reactions could
also lead to a determination of $\alpha_\rS$ in the TeV region.

The high center-of-mass energy at the LHC in
combination
with its enormous luminosity will allow to 
produce gauge bosons with transverse momenta up to \mbox{2 TeV}
or even beyond. 
In this 
kinematic region the electroweak
corrections are strongly enhanced, with the dominant terms 
in $L$-loop approximation being 
leading logarithms (LL) of the form 
$\alpha^L\log^{2L}(\hat{s}/M_W^2)$,
next-to-leading logarithms (NLL) of the form 
$\alpha^L\log^{2L-1}(\hat{s}/M_W^2)$, and so on.
These corrections, also known as electroweak Sudakov logarithms, may
well amount to several tens of 
percent~\cite{Kuhn:2000nn,Jantzen:2005az,Fadin:2000bq,Dittmaier:2001ay,Baur:2001ze,Maina:2004rb,Kuhn:2004em,Kuhn:2005az,Kuhn:2005gv,Baur:2006sn}.
(A recent survey of the literature on electroweak Sudakov logarithms
can be found in \citere{Denner:2006jr}.)
Specifically, the electroweak corrections to the $\pT$-distribution of
photons and $Z$ bosons at hadron colliders were studied in 
\citeres{Maina:2004rb,Kuhn:2004em,Kuhn:2005az,Kuhn:2005gv}.
In \citeres{Kuhn:2004em,Kuhn:2005az,Kuhn:2005gv}, it was found that at transverse momenta of 
$\ord(1\TeV)$ the dominant two-loop contributions to these reactions
amount to several percent and must be included to
match the precision of the LHC experiments.
This is quite different from the production of on-shell gauge bosons 
with  small transverse momenta
\cite{Wackeroth:1996hz}, 
where the electroweak corrections are not 
enhanced by Sudakov logarithms. 
With this motivation in mind we study the electroweak corrections to
hadronic production of $W$ bosons in association with a jet,
$p \pantip \to Wj $, at large $p_\rT$.

As a consequence of the non-vanishing $W$ charge,  QED 
corrections cannot be separated from the purely weak ones and
will thus be included in our analysis. 
Thus, in comparison with 
$Z$-boson production, several 
new aspects arise. 
Real photon emission must be 
included to cancel the infrared divergencies from virtual 
photonic corrections. Collinear singularities, a consequence of radiation 
from massless quarks, must be isolated and absorbed in the parton 
distribution functions (PDFs) in the case of initial-state radiation. 
We regularize soft and collinear singularities in two 
different schemes: 
using small quark  and photon masses which are set to zero 
at the end of the calculation
and, alternatively, dimensional regularization.
In events with real radiation, the $\pT$ of the $W$ boson is balanced 
both by the  $\pT$ of the recoiling parton (quark or gluon) and the photon. 
Configurations involving a small-$\pT$  parton
and a hard photon
are better described as $W\gamma$ final states.
We thus define the  $Wj$ cross section
imposing a lower limit on the jet transverse momentum, 
which is chosen independent of the $W$-boson $\pT$. 
In order to avoid final-state collinear singularities,
we recombine collinear photon-quark final states.

The virtual EW corrections to $Wj$ production are formally connected
with the real emission of $W$ and $Z$ bosons, which leads to $WVj$
final states with $V=W,Z$.  Both contributions are of
$\ord(\alpha^2\alpha_{\mathrm{S}})$.  If integrated over the
full phase space, the real emission of gauge bosons produces large
Sudakov logarithms that partially cancel those resulting from virtual
gauge bosons.  However, in exclusive measurements of $pp\to Wj$, the
available phase space for gauge boson emission is strongly suppressed
by the experimental cuts.  We thus expect that real emission provides
relatively small contributions while the bulk of electroweak effects
originates from virtual corrections.  In fact, for $pp\to Zj$ 
it was shown that, in presence of realistic (and
relatively less exclusive) experimental cuts, the contribution of real
emission is about five times smaller than the virtual corrections
\cite{Baur:2006sn}. Moreover, real emission can be further reduced
with a veto on additional jets, which suppresses multiple-jet events
resulting from the hadronic decay of the radiated gauge bosons.
Therefore we will restrict ourselves to the investigation of virtual
electroweak corrections (and photon bremsstrahlung).  The real
emission of $W$ and $Z$ bosons can be non-negligible and certainly
deserves further detailed studies, however we do not expect a dramatic
impact on our results.

The partonic reactions $\bar{q}q'\rightarrow W^\pm g\, (\gamma)$, 
$q'g\rightarrow W^\pm q\, (\gamma)$ and
{$\bar{q}g\rightarrow W^\pm \bar{q}'\, (\gamma)$}  
with $q=u,d,s,c,b$ are considered. 
All of them are, however, trivially related 
by CP- and crossing-symmetry relations.
Quark-mass effects are neglected throughout, 
which allows to incorporate the effect of quark mixing 
through a simple redefinition of parton distribution functions
(see~\refse{se:hadronicxsection}).
Our conventions for couplings, kinematics and two- as well as three-body phase space are also collected in Sect.~2.
The calculation of the virtual corrections is described in~\refse{se:corr}.
We present 
analytic expressions for the 
one-loop amplitude,
specify the counterterms in the $G_\mu$ renormalization scheme
and isolate the infrared singularities. 
The high-energy limit is studied in detail in \refse{se:helimit}. 
The analytic one-loop result is investigated in the limit $\hat s \gg M_W^2$, 
keeping quadratic and linear logarithms as well as constant terms. 
These results are compared to those derived in the 
NLL approximation \cite{Denner:2001jv}.
In view of their numerical importance we also derive 
the dominant (NLL) two-loop terms, 
using the formalism of \citeres{Denner:2003wi, Melles:2001gw}.
The calculation of the real corrections is performed using the dipole 
subtraction formalism \cite{Dittmaier:1999mb,Catani:1996vz,Catani:2002hc}, 
as discussed in~\refse{se:real}.
The checks which we carried out in order to ensure 
the correctness of 
the results are described in~\refse{se:checks}.

The numerical results are presented in~\refse{se:numerics}.
After convolution with parton distribution functions, we obtain
radiatively corrected predictions for $\pT$-distributions
of $W$ bosons at the LHC and the Tevatron.
The quality of the one-loop NLL and NNLL approximations 
is investigated and the size of the dominant two-loop terms is compared
with the expected statistical precision of the experiments.
Concerning perturbative QCD, our predictions are based on the lowest order.
To obtain realistic absolute cross sections,
higher-order QCD corrections \cite{QCDcorr} must be included.
However, the relative rates for $W^+$, $W^-$ and $Z$
production are expected to be more stable against 
QCD effects. Therefore, the impact of the electroweak corrections 
on these ratios is presented in~\refse{se:numerics}.
Our conclusions and a brief summary can be found in \refse{se:conc}.
Explicit analytic results are collected in the 
Appendices.

A short description of the method of calculation and the main results for LHC have been
given in \citere{Kuhn:2007qc}. After completion of this work,
Hollik, Kasprzik and Kniehl~\cite{Hollik:2007qc} reported results on hadronic
\mbox{$W$-boson} production at 
large $\pT$ qualitatively similar
to those of~\citere{Kuhn:2007qc} and the present paper.

\section{Definitions and conventions}
\label{se:conventions}

\subsection{Hadronic cross section}
\label{se:hadronicxsection}

\newcommand{\pTmin}{\pT^{\mathrm{min}}}
\newcommand{\pTj}{p_{\mathrm{T},\,j}}
\newcommand{\pTk}{p_{\mathrm{T,\,}k}}
\newcommand{\pTW}{p_{\mathrm{T,\,}W}}
\newcommand{\pTminj}{p_{\mathrm{T},\,j}^{\mathrm{min}}}
\newcommand{\pTmink}{p_{\mathrm{T},\,k}^{\mathrm{min}}}
\newcommand{\pTminW}{p^{\mathrm{min}}_{\mathrm{T},\,W}}
\newcommand{\tpphat}{\hat{t}''}
\newcommand{\upphat}{\hat{u}''}
\newcommand{\spphat}{\hat{s}''}
The  $\pT$-distribution of $W$ bosons in the reaction 
$h_1 h_2 \to W^\pm j (\ga)$
is given by 
\newcommand{\pdf}[4]{f_{#1,#2}(#3,#4)}
\newcommand{\pdfmod}[4]{\tilde f_{#1,#2}(#3,#4)}
\newcommand{\tauhmin}{\hat{\tau}_{\rm min}}
\beq
\label{hadroniccs}
\frac{\rd \si^{h_1 h_2}}{\rd \pT}=
\sum_{a,b,k}\int_0^1\rd x_1 \int_0^1\rd x_2
\;
\theta(x_1 x_2- \tauhmin)
\pdf{h_1}{a}{x_1}{\mu^2}
\pdf{h_2}{b}{x_2}{\mu^2}
\frac{\rd \hat{\si}^{a b\to W^\si k (\ga)}}{\rd \pT}
,
\eeq
where $\tauhmin$ depends on the kinematic configuration of the final state and is specified at the end of \refse{sse:kinematics}.
The indices  $a, b$  denote initial-state partons 
and $f_{h_1,a}(x,\mu^2)$, $f_{h_2,b}(x,\mu^2)$
are the corresponding parton distribution functions (PDFs).
$\hat {\si}^{ab\to W^\si k (\ga) }$ is the partonic cross section for 
the subprocess $a b \to W^\si k (\ga) $. 
The sum in \refeq{hadroniccs}
runs over all $a,b,k$ combinations corresponding to the subprocesses
\beqar\label{processespl}
&&\bar d_n u_m \to W^+ g  (\ga),\quad
u_m\bar d_n\to W^+ g  (\ga),\quad
g u_m \to W^+  d_n  (\ga),\quad  \nl&&
u_m g \to W^+  d_n (\ga) ,\quad
\bar d_n g  \to W^+  \bar u_m  (\ga),\quad
g \bar d_n   \to W^+  \bar u_m  (\ga)
,
\eeqar
for $W^+$ production, and  similarly for $W^-$ production. 

The dependence of the partonic cross sections on the family indices $m,n$
amounts to an overall factor $|V_{u_md_n}|^2$. This factor can be easily absorbed by redefining  the 
parton distribution functions as
\beqar\label{redpdfs}
\tilde{f}_{h,{d}_{m}}
&=&
\sum_{n=1}^3|V_{u_m d_n}|^2 f_{h,{d}_{n}}
,\qquad
\tilde{f}_{h,\bar{d}_{m}}
=
\sum_{n=1}^3|V_{u_m d_n}|^2 f_{h,\bar{d}_{n}}
,\nl
\tilde{f}_{h,{u}_{m}}
&=&
{f}_{h,{u}_{m}}
,\qquad
\tilde{f}_{h,\bar{u}_{m}}
=
{f}_{h,\bar{u}_{m}}
,\qquad
\tilde{f}_{h,g}
=
{f}_{h,g}
.
\eeqar
The hadronic cross section \refeq{hadroniccs}
can be computed using the trivial CKM matrix $\tilde V_{u_i d_j}=\de_{ij}$
and the redefined PDFs \refeq{redpdfs}.
Since we do not consider initial or final states involving
(anti-)top quarks, only the contributions of the first two quark families
($m=1,2$) have to be included.
The corresponding redefined PDFs ($\tilde{f}_{h,q}$ with $q=u,d,c,s$)
automatically include the (small) contributions associated with
initial- and final-state bottom quarks.

\subsection{Kinematics} 
\label{sse:kinematics}

For the $2 \to 2$ subprocess 
$a b \to W^\si k$ the Mandelstam variables are defined
in the standard way,
\beq
\shat=(p_a+ p_b)^2 
,\qquad 
\that=(p_a- p_W)^2 
,\qquad 
\uhat=(p_b-p_W)^2
. 
\eeq
The momenta  $p_{a}$, $p_{b}$, $p_{k}$ of the partons are assumed to be massless, whereas 
$p_W^2= M_W^2$. 
In terms of $x_1,x_2,\pT$ and the collider energy $\sqrt{s}$ we have
\beq
\label{tuform}
\shat=x_1 x_2 s,\qquad
\that=\frac{ M_W^2-\shat}{2}(1-\cos\theta),\qquad
\uhat=\frac{ M_W^2-\shat}{2}(1+\cos\theta),
\eeq
with $\cos\theta=\sqrt{1- 4\pT^2 \shat/(\shat- M_W^2)^2}$
corresponding to the cosine of the angle between the momenta $p_a$ and
$p_W$ in the partonic center-of-mass frame.

The $\pT$-distribution 
for the unpolarized partonic subprocess $a b \to W^\si k$
reads
\beqar
\label{partoniccsgen}
\frac{\rd \hat{\si}^{a b\to W^\si k}}{\rd \pT}
&=& 
\mathcal{N}_{ab}
\int \rd  \Phi_2\, \overline{\sum}|\M^{a b\to W^\si k}|^2 \, F_{\mathrm{O}, 2}(\Phi_2),
\eeqar
where
$
\overline{\sum}=
\frac{1}{4}
\sum_{\mathrm{pol}} 
\sum_{\mathrm{col}} 
$
involves the sum over polarization and color as well as the average factor
$1/4$ for initial-state polarization. The factor $\mathcal{N}_{ab}$ is given by
\beq\label{fluxfactor}
\mathcal{N}_{ab}=\frac{(2\pi)^4}{2 \shat N_{ab}},
\eeq
where
$N_{\bar q q'}=N_{q\bar q'}=N_\rc^2$, 
$N_{g q}=N_{q g}=N_{\bar q g}=N_{g\bar q}=N_\rc (N_\rc^2-1)$, 
with $N_\rc=3$, account for the initial-state colour average.
The phase-space measure $\rd \Phi_2$ is given by
\beq
\label{PSMeas2}
   \rd \Phi_2=
   \frac{\rd^3 p_W}{(2\pi)^3 2 p_W^0}
   \frac{\rd^3 p_k}{(2\pi)^3 2 p_k^0}
   \;  \delta^4 \left( p_{a}+p_b-p_W-p_k \right),
\eeq
while the function 
$ F_{\rm O, 2}$
defines the observable of interest, \ie 
the $W$-boson $\pT$-distribution in presence of a cut on the
transverse momentum of the jet,
\beq
\label{fo2}
F_{\mathrm{O}, 2}(\Phi_2)= \delta (\pT - \pTW) 
\theta (\pTj -\pTminj)\,.
\eeq
In the 2-particle phase space the jet is identified with the 
parton $k$ and momentum conservation implies $\pTj=\pTk=\pTW$. 
In practice, since  we always consider the  $\pT$-distribution
in the region $\pT>\pTminW>\pTminj$, the cut on $\pTj$  
in \refeq{fo2} is irrelevant.
The phase-space integral in~\refeq{partoniccsgen} yields two contributions
originating from kinematic configurations in the forward and backward 
hemispheres  with opposite values of  $\cos \theta$ in the center-of-mass frame,
\newcommand{\fwd}{\mathrm{fwd}}
\newcommand{\bkwd}{\mathrm{bkwd}}
\beqar\label{partoniccs1}
\frac{\rd \hat{\si}^{a b\to W^\si k}}{\rd \pT}
&=&
\frac{\rd \hat{\si}_{\fwd}^{a b\to W^\si k}}{\rd \pT}
+
\frac{\rd \hat{\si}_{\bkwd}^{a b\to W^\si k}}{\rd \pT}
,
\eeqar
with 
\beqar\label{partoniccs1b}
\frac{\rd \hat{\si}_{\fwd}^{a b\to W^\si k}}{\rd \pT}
&=&
\frac{\pT}{8\pi N_{a b}\shat|\that-\uhat|}
\overline{\sum}|\M^{a b\to W^\si k}|^2
,\quad
\frac{\rd \hat{\si}_{\bkwd}^{a b\to W^\si k}}{\rd \pT}
=
\frac{\rd \hat{\si}_{\fwd}^{a b\to W^\si k}}{\rd \pT}
\Bigg|_{\that\leftrightarrow \uhat}
.
\eeqar

For the $2 \to 3 $ subprocess $a b \to W^\si k \gamma$  we define the following five
independent invariants
\beqar
\label{3PPSInv}
   \shat=(p_a+ p_b)^2 
   ,\qquad 
   \that&=(p_a - p_W)^2 
   ,\qquad 
   \uhat=&(p_b-p_W)^2 ,
   \nl
   \tphat&=(p_a- p_\ga)^2 
   ,\qquad 
   \uphat=&(p_b -p_\ga)^2  \;,
\eeqar
and the four dependent invariants
\beqar
   \sphat&=(p_k + p_\ga )^2 = \shat + \that + \uhat -M_W^2
   ,\qquad   
   \spphat&=(p_W+ p_k)^2 = \shat + \tphat + \uphat 
   ,\nl
   \tpphat&=(p_a - p_k)^2 =  M_W^2 - \shat -\that -\tphat
   ,\qquad 
   \upphat&=(p_b - p_k)^2 =  M_W^2 - \shat -\uhat -\uphat \,.
\nl
\eeqar 
The $\pT$-distribution for this subprocess reads
\beq
\label{Part:xsec:real}
\frac{\rd \hat{\si}^{a b \to W^\si k \gamma}}{\rd \pT} = \mathcal{N}_{ab}
\int \rd  \Phi_3\, \overline{\sum}|\M^{a b \to  W^\si k \gamma}|^2 \, F_{\mathrm{O}, 3}(\Phi_3)
,
\eeq
where
\beq
\label{PSMeas3}
   \rd \Phi_3=
   \frac{\rd^3 p_W}{(2\pi)^3 2 p_W^0}
   \frac{\rd^3 p_k}{(2\pi)^3 2 p_k^0}
   \frac{\rd^3 p_\ga}{(2\pi)^3 2 p_\ga^0}
   \;  \delta^4 \left( p_a+p_b-p_W-p_k-p_\ga \right) \,
.
\eeq
In the 3-particle phase space, 
the  $W$-boson $\pT$-distribution in $Wj$ production
is defined by the observable function
\beq
\label{fo3}
F_{\mathrm{O}, 3}(\Phi_3)= \delta (\pT - p_{\rT,\,W}) 
\theta (\pTj -\pTminj) \,.
\eeq
The cut on the jet transverse momentum 
rejects events where 
the $W$-boson  $\pT$  is balanced by  an isolated photon
plus a parton with small transverse momentum.
This observable is thus free from singularities associated with
soft and collinear quarks or gluons.
When applying the cut on the jet momentum in the 3-particle phase space, 
care must be taken that the definition of the jet $\pT$ 
is collinear-safe. 
In general the jet cannot be identified with the parton $k$,
since in presence of collinear photon radiation 
the transverse momentum of a charged parton
is not a collinear-safe quantity. 
Thus we identify the jet with the
parton $k$ only if $k$ is a quark
well separated from the photon or a gluon. 
Otherwise, \ie for collinear
quark-photon configurations, 
the recombined momentum of the 
quark and photon is taken as momentum of the jet. 
In practice, we define the separation variable
\beq\label{eq:Rdef}
R(q,\gamma)=\sqrt{(\eta_q-\eta_\gamma)^2+(\phi_q-\phi_\gamma)^2},
\eeq  
where $\eta_i$ is the pseudo-rapidity and $\phi_i$ is the azimuthal
angle of a particle $i$.  
If $R(q,\gamma)<R_{\mathrm{sep}}$, then the
photon and quark momenta are recombined by simple 
\mbox{four-vector} addition
into an effective momentum $p_{j}$ 
and then $\pTj=\sqrt{(\vec  p_{\rT,\,q} +\vec p_{\rT,\,\ga})^2} $, 
otherwise $\pTj=p_{\rT,\, q}$. 
We note that, in the collinear region, lowest-order kinematics implies
$p_{\rT,j}=p_{\rT,q}+p_{\rT,\gamma}=p_{\rT,W}>\pTminj$. This means that the
recombination procedure effectively removes the cut on $p_{\rT,q}$
inside the collinear cone $R(q,\gamma)<R_{\mathrm{sep}}$.
For instance the recombined  $g q' \to W^\si q \gamma$
cross section is given by
\beqar\label{reccs}
\hat \sigma^{g q' \to W^\si q \gamma}_{\mathrm{rec.}}&=&
\int_{R(q,\gamma)<R_\mathrm{sep}} \rd \hat{\si}^{g q' \to W^\si q \gamma}
+
\int_{R(q,\gamma)>R_\mathrm{sep}} 
\theta(p_{\rT,\, q}-\pTminj)\,
\rd \hat{\si}^{g q' \to W^\si q \gamma}
.
\eeqar
In contrast, for the case of final-state gluons, we do not perform
photon-gluon recombination and the cut on $p_{\rT,\, g}$ is imposed in the
entire phase space.

This procedure has the advantage to avoid both collinear-photon and soft-gluon
singularities.  However it implies a different treatment of quark and gluon
final states and can thus be regarded as an arbitrary cut-off prescription for
the final-state collinear singularity.  Moreover, the recombined cross section
\refeq{reccs} has a logarithmic dependence on the cut-off parameter
$R_\mathrm{sep}$.  These aspects are discussed in detail in
\refapp{app:recomb}.
There we compare the recombination procedure with a realistic experimental
definition of exclusive $pp\to Wj$ production, where final-state quarks are
subject to the same cut as final state gluons ($p_{\rT,q}> \pTminj$) within
the entire phase space.
Describing the exclusive $g q' \to W^\si q \gamma$  cross section, 
\beqar\label{exclcs}
\hat \sigma^{g q' \to W^\si q \gamma}_{\mathrm{excl.}}&=&
\int
\theta(p_{\rT,\, q}-\pTminj)\,
\rd \hat{\si}^{g q' \to W^\si q \gamma},
\eeqar
by means of quark fragmentation functions, 
we find that the quantitative difference between the two definitions
\refeq{reccs} and \refeq{exclcs} amounts to less than two permille.
Moreover, we show that the recombined
cross section is extremely stable with respect to variations of the
parameter $R_\mathrm{sep}$.  
This means that the recombination procedure used in our calculation
provides a very good description of exclusive $pp\to Wj$ production.

Another treatment of the singularities, which does not require
recombination and treats quark- and gluon-induced jets uniformly, has
been proposed in \citere{Hollik:2007qc}. There, contributions from
$Wj$ production and $W\gamma$ production to a more inclusive
observable, \ie high-$\pT$ $W$ production, are both calculated.  All
soft and collinear singularities in the final state cancel in the
approach of \citere{Hollik:2007qc} as a result of the more inclusive
observable definition than associated production of the $W$ boson
together with a jet, considered in this work. 
The comparison of our results with those of \citere{Hollik:2007qc}
seems to indicate that these differences in the jet definitions
have a quite small impact on the size of the electroweak corrections.

The quantity  $\tauhmin$ in \refeq{hadroniccs}
is related to the minimum partonic energy 
that is needed to produce final states with $\pTj>\pTminj$ and  $\pTW>\pTminW$,
\beq\label{taudef}
s \tauhmin =\left(\pTminj +\sqrt{(\pTminW)^2+ M_W^2}\right)^2
\,.
\eeq
When we evaluate the $2\to 2$ contributions to the hadronic cross section \refeq{hadroniccs}, after analytic integration of the phase space in \refeq{partoniccsgen},
we can set $\pTminj = \pTminW=\pT$
in \refeq{taudef}.

\subsection{Crossing symmetries}
\label{se:symmetries}
The unpolarized squared matrix elements for the $2\to 2$ processes
in \refeq{processespl} are related by the crossing-symmetry relations
\beqar\label{crossing1}
\overline{\sum}|\M^{g q'\to W^\si q}|^2
&=& 
-\left.\overline{\sum}|\M^{\bar q q'\to W^\si g}|^2 
\right|_{\shat\leftrightarrow \that}
,\nl
\overline{\sum}|\M^{\bar q g\to W^\si \bar q'}|^2
&=& 
-\left.\overline{\sum}|\M^{\bar q q'\to W^\si g}|^2 
\right|_{\shat\leftrightarrow \uhat}
,\nl
\overline{\sum}|\M^{b a\to W^\si k}|^2
&=& 
\left.\overline{\sum}|\M^{a b\to W^\si k}|^2 
\right|_{\that\leftrightarrow \uhat}
.
\eeqar
Moreover, due to CP symmetry, 
the unpolarized partonic cross section for the production of 
positively and negatively charged $W$ bosons are related by
\beq\label{cpsymm}
\overline{\sum}|\M^{\bar d u \to W^+ g}|^2 
=
\overline{\sum}|\M^{d \bar u\to W^- g}|^2 
.
\eeq
Eqs.~\refeq{crossing1} and \refeq{cpsymm} permit to relate the six processes for $W^+$ 
production in  \refeq{processespl} and the six charge conjugate ones to a single process. 
Hence the explicit computation of the unpolarized squared matrix element
needs to be performed only once.
In the following we will present explicit results for the process
$\bar q q'\to W^\si g$.

Similarly, for the unpolarized squared matrix elements for the  $2\to 3$ processes
in \refeq{processespl} we have 
\beqar\label{crossing3}
\overline{\sum}|\M^{g q'\to W^\si q \ga}|^2
&=& 
-\left.\overline{\sum}|\M^{\bar q q'\to W^\si g \ga}|^2 
\right|_{\{\shat\leftrightarrow \upphat, \that \lrar \spphat, \tphat
  \lrar \sphat \}}
,\nl
\overline{\sum}|\M^{\bar q g\to W^\si \bar q' \ga }|^2
&=& 
-\left.\overline{\sum}|\M^{\bar q q'\to W^\si g \ga }|^2 
\right|_{\{\shat\leftrightarrow \tpphat, \uhat \lrar \spphat, \uphat
  \lrar \sphat \}}
,\nl
\overline{\sum}|\M^{b a\to W^\si k \ga }|^2
&=& 
\left.\overline{\sum}|\M^{a b\to W^\si k \ga }|^2 
\right|_{\{\that\leftrightarrow \uhat, \tphat \lrar \uphat \}}
\eeqar
and 
\beq\label{cpsymm2}
\overline{\sum}|\M^{\bar d u \to W^+ g \ga }|^2 
=
\overline{\sum}|\M^{d \bar u\to W^- g \ga}|^2 
.
\eeq
It is thus enough to perform calculations only for the 
$\bar q q' \to
W^\si g \ga$ subprocess.  

\subsection{Couplings and Born matrix element}

For gauge couplings we adopt the conventions of \citere{Pozzorini:rs}.
With this notation the $g q\bar q$ vertex and the $V q'\bar q$ vertices with $V=A,Z,W^\pm$ read 
\beqar
 \hspace{1cm}
\begin{picture}(30.,20.)(0,-3)
\ArrowLine(0,0)(-20,20)
\Text(-25,20)[rt]{$\bar q$}
\Gluon(0,0)(30,0){2.5}{4}
\Text(25,5)[b]{$G^\mu$}
\ArrowLine(-20,-20)(0,0)
\Text(-25,-20)[rb]{$q$}
\Vertex(0,0){1.5}
\end{picture}
&=& - \ri g_\rS  t^a \gamma^{\mu}
, \hspace{2cm}
\begin{picture}(30.,20.)(0,-3)
\ArrowLine(0,0)(-20,20)
\Text(-25,20)[rt]{$\bar q$}
\Photon(0,0)(30,0){2.5}{4}
\Text(25,5)[b]{$V^\mu$}
\ArrowLine(-20,-20)(0,0)
\Text(-25,-20)[rb]{$q'$}
\Vertex(0,0){1.5}
\end{picture}
=\ri e \gamma^{\mu}
\sum_{\lambda=\mathrm{R,L}} \omega_{\lambda}  
I^V_{q_\la {q_\la}^{\hspace{-1.5mm}'}},
\eeqar
where $\omega_{\la}$ are the chiral projectors
\beq\label{projectors}
\omega_{\rR}=\frac{1}{2}(1+\gamma_5)
,\qquad
\omega_{\rL}=\frac{1}{2}(1-\gamma_5),
\eeq
$t^a$ are the Gell-Mann matrices and $I^V$ are matrices in the weak isospin space.
For diagonal matrices such as $I^A$ and $I^Z$ we write $I^V_{q_\la {q_\la}^{\hspace{-1.5mm}'}}=
\de_{q q'} I^V_{q_\la}$.
In terms of the weak isospin $T^3_{q_\la}$ and the weak hypercharge $Y_{q_\la}$
we have
\beqar\label{couplfact0}
I^Z_{q_\la}&=&\frac{\cw}{\sw} T^3_{q_\la}-\frac{\sw}{\cw}\frac{Y_{q_\la}}{2}
,\qquad
I^A_{q_\la}=-Q_{q_\la}= - T^3_{q_\la}-\frac{Y_{q_\la}}{2},
\eeqar
with the shorthands $\cw=\cos{\theta_\rw}$ and  $\sw=\sin{\theta_\rw}$
for the  weak mixing angle $\theta_\rw$. 
The eigenvalues of isospin, 
hypercharge and SU(2) Casimir operators
for left-handed fermions are
\beq\label{eigenvalues}
T^3_{u_\rL} = - T^3_{d_\rL} = \frac 12
,\quad
Y_{u_\rL} = Y_{d_\rL} = \frac 13
,\quad
C_{\mathrm{F}} = \frac 34
,\quad
C_{\mathrm{A}} = 2
.
\eeq
The only non-vanishing components of the generators associated with 
$W$ bosons are
\beq
I^{W^{+}}_{u_\rL d_\rL}=I^{W^{-}}_{d_\rL u_\rL}=\frac{1}{{\sqrt{2}\sw}}.
\eeq
The triple gauge-bosons vertices read
\beqar
 \hspace{1cm}
\begin{picture}(30.,20.)(0,-3)
\Photon(0,0)(-20,20){2.5}{4}
\Text(-25,20)[rt]{$V_a^{\mu_1}$}
\Photon(0,0)(30,0){2.5}{4}
\Text(25,5)[b]{$V_c^{\mu_3}$}
\Photon(-20,-20)(0,0){2.5}{4}
\Text(-25,-20)[rb]{$V_b^{\mu_2}$}
\Vertex(0,0){1.5}
\end{picture}
&=& \frac{e}{\sw}\varepsilon^{V_a V_b V_c}
[
g^{\mu_1\mu_2}(k_1-k_2)^{\mu_3}
+g^{\mu_2\mu_3}(k_2-k_3)^{\mu_1}
\nl&&{}
+g^{\mu_3\mu_1}(k_3-k_1)^{\mu_2}
],
\eeqar
where the totally anti-symmetric tensor $\varepsilon^{V_1 V_2V_3}$ 
is defined through the commutation relations
\beqar\label{commrel}
\left[I^{V_1},
I^{V_2}
\right]
=\frac \ri\sw \sum_{V_3=A,Z,W^\pm}
\varepsilon^{V_1V_2V_3}
I^{\bar{V}_3},
\eeqar
and has components
$\varepsilon^{Z W^+W^-}=-\ri \cw$ and
$\varepsilon^{A W^+W^-}=\ri \sw$.

To lowest order in $\alpha$ and $\alpha_\rS$,
the unpolarized squared matrix element 
for the $\bar q q'\to W^\si g$ process
reads 
\beq\label{generalamplitude}
\overline{\sum}|\M_0^{\bar q q'\to W^\si g}|^2=
8 \pi^2 \alpha \alpha_\rS (N_\rc^2-1)
\left(I^{W^{-\si}}_{q_\rL q_\rL'}\right)^2
\frac{\that^2+\uhat^2+2 M_W^2 \shat}{\that\uhat} 
,
\eeq
where $\alpha=e^2/(4 \pi)$ and $\alpha_\rS=g_\rS^2/(4 \pi)$ 
are the electromagnetic and the strong coupling constants.

\section{Virtual corrections}
\label{se:corr}

\newcommand{\diagtI}[1]{
\begin{picture}(94.,90.)(-47,-45)
\ArrowLine(0,20)(-40,20)
\Vertex(0,20){1.5}
\Photon(0,20)(40,20){2.5}{6}
\ArrowLine(-40,-20)(0,-20)
\Vertex(0,-20){1.5}
\Gluon(0,-20)(40,-20){-2.5}{6}
\ArrowLine(0,-20)(0,20)
\Text(0,-35)[t]{#1}
\end{picture}}

\newcommand{\diagtII}[1]{
\begin{picture}(94.,90.)(-47,-45)
\ArrowLine(0,20)(-40,20)
\Vertex(0,20){1.5}
\Gluon(0,20)(40,20){2.5}{6}
\ArrowLine(-40,-20)(0,-20)
\Vertex(0,-20){1.5}
\Photon(0,-20)(40,-20){-2.5}{6}
\ArrowLine(0,-20)(0,20)
\Text(0,-35)[t]{#1}
\end{picture}}

\newcommand{\diagcI}[1]{
\begin{picture}(94.,90.)(-47,-45)
\ArrowLine(0,20)(-40,20)
\Photon(0,20)(40,20){2.5}{6}
\ArrowLine(-40,-20)(0,-20)
\Vertex(0,-20){1.5}
\Gluon(0,-20)(40,-20){-2.5}{6}
\ArrowLine(0,-20)(0,0)\ArrowLine(0,0)(0,20)
\Text(0,-35)[t]{#1}
\BCirc(0,0){4}\Line(-2.83,-2.83)(2.83,2.83)\Line(-2.83,2.83)(2.83,-2.83)
\end{picture}}

\newcommand{\diagcII}[1]{
\begin{picture}(94.,90.)(-47,-45)
\ArrowLine(0,20)(-40,20)
\Vertex(0,20){1.5}
\Gluon(0,20)(40,20){2.5}{6}
\ArrowLine(-40,-20)(0,-20)
\Vertex(0,-20){1.5}
\Photon(0,-20)(40,-20){-2.5}{6}
\ArrowLine(0,-20)(0,0)\ArrowLine(0,0)(0,20)
\Text(0,-35)[t]{#1}
\BCirc(0,0){4}\Line(-2.83,-2.83)(2.83,2.83)\Line(-2.83,2.83)(2.83,-2.83)
\end{picture}}

\newcommand{\diagcIII}[1]{
\begin{picture}(94.,90.)(-47,-45)
\ArrowLine(0,20)(-40,20)
\Photon(0,20)(40,20){2.5}{6}
\ArrowLine(-40,-20)(0,-20)
\Vertex(0,-20){1.5}
\Gluon(0,-20)(40,-20){-2.5}{6}
\ArrowLine(0,-20)(0,20)
\Text(0,-35)[t]{#1}
\BCirc(0,-20){4}\Line(-2.83,-22.83)(2.83,-17.18)\Line(-2.83,-17.18)(2.83,-22.83)
\end{picture}}

\newcommand{\diagcIV}[1]{
\begin{picture}(94.,90.)(-47,-45)
\ArrowLine(0,20)(-40,20)
\Vertex(0,20){1.5}
\Gluon(0,20)(40,20){2.5}{6}
\ArrowLine(-40,-20)(0,-20)
\Vertex(0,-20){1.5}
\Photon(0,-20)(40,-20){-2.5}{6}
\ArrowLine(0,-20)(0,20)
\Text(0,-35)[t]{#1}
\BCirc(0,20){4}\Line(-2.82,17.18)(2.83,22.82)\Line(-2.83,22.83)(2.83,17.18)
\end{picture}}

\newcommand{\diagcV}[1]{
\begin{picture}(94.,90.)(-47,-45)
\ArrowLine(0,20)(-40,20)
\Photon(0,20)(40,20){2.5}{6}
\ArrowLine(-40,-20)(0,-20)
\Vertex(0,-20){1.5}
\Gluon(0,-20)(40,-20){-2.5}{6}
\ArrowLine(0,-20)(0,20)
\Text(0,-35)[t]{#1}
\BCirc(0,20){4}\Line(-2.82,17.18)(2.83,22.82)\Line(-2.83,22.83)(2.83,17.18)
\end{picture}}

\newcommand{\diagcVI}[1]{
\begin{picture}(94.,90.)(-47,-45)
\ArrowLine(0,20)(-40,20)
\Vertex(0,20){1.5}
\Gluon(0,20)(40,20){2.5}{6}
\ArrowLine(-40,-20)(0,-20)
\Vertex(0,-20){1.5}
\Photon(0,-20)(40,-20){-2.5}{6}
\ArrowLine(0,-20)(0,20)
\Text(0,-35)[t]{#1}
\BCirc(0,-20){4}\Line(-2.83,-22.83)(2.83,-17.18)\Line(-2.83,-17.18)(2.83,-22.83)
\end{picture}}

\newcommand{\diagsI}[1]{
\begin{picture}(94.,90.)(-47,-45)
\ArrowLine(0,20)(-40,20)
\Vertex(0,20){1.5}
\Photon(0,20)(40,20){2.5}{6}
\ArrowLine(-40,-20)(0,-20)
\Vertex(0,-20){1.5}
\Gluon(0,-20)(40,-20){-2.5}{6}
\PhotonArc(0,0)(10,90,270){-2.5}{4}\Text(-15,0)[r]{$\scriptstyle{V}$}
\ArrowLine(0,-20)(0,-10)\Vertex(0,-10){1.5}\ArrowLine(0,-10)(0,10)\Vertex(0,10){1.5}\ArrowLine(0,10)(0,20)
\Text(0,-35)[t]{#1}
\end{picture}}

\newcommand{\diagsII}[1]{
\begin{picture}(94.,90.)(-47,-45)
\ArrowLine(0,20)(-40,20)
\Vertex(0,20){1.5}
\Gluon(0,20)(40,20){2.5}{6}
\ArrowLine(-40,-20)(0,-20)
\Vertex(0,-20){1.5}
\Photon(0,-20)(40,-20){-2.5}{6}
\PhotonArc(0,0)(10,90,270){-2.5}{4}\Text(-15,0)[r]{$\scriptstyle{V}$}
\ArrowLine(0,-20)(0,-10)\Vertex(0,-10){1.5}\ArrowLine(0,-10)(0,10)\Vertex(0,10){1.5}\ArrowLine(0,10)(0,20)
\Text(0,-35)[t]{#1}
\end{picture}}

\newcommand{\diagvI}[1]{
\begin{picture}(94.,90.)(-47,-45)
\ArrowLine(0,20)(-40,20)
\Vertex(0,20){1.5}
\Photon(0,20)(40,20){2.5}{6}
\ArrowLine(-40,-20)(-15,-20)
\Vertex(-15,-20){1.5}
\Vertex(15,-20){1.5}
\Gluon(15,-20)(40,-20){-2.5}{3.5}
\ArrowLine(-15,-20)(15,-20)
\Photon(-15,-20)(0,5){2.5}{4.5}\Text(-13,-5)[r]{$\scriptstyle{V}$}\ArrowLine(15,-20)(0,5)
\Vertex(0,5){1.5}\ArrowLine(0,5)(0,20)
\Text(0,-35)[t]{#1}
\end{picture}}

\newcommand{\diagvII}[1]{
\begin{picture}(94.,90.)(-47,-45)
\ArrowLine(-15,20)(-40,20)
\Vertex(-15,20){1.5}
\Vertex(15,20){1.5}
\Gluon(15,20)(40,20){2.5}{3.5}
\ArrowLine(15,20)(-15,20)
\Photon(0,-5)(-15,20){2.5}{4.5}\Text(-13,5)[r]{$\scriptstyle{V}$}\ArrowLine(0,-5)(15,20)
\Vertex(0,-5){1.5}\ArrowLine(0,-20)(0,-5)
\ArrowLine(-40,-20)(0,-20)
\Vertex(0,-20){1.5}
\Photon(0,-20)(40,-20){-2.5}{6}
\Text(0,-35)[t]{#1}
\end{picture}}

\newcommand{\diagvIII}[1]{
\begin{picture}(94.,90.)(-47,-45)
\ArrowLine(-15,20)(-40,20)
\Vertex(-15,20){1.5}
\Vertex(15,20){1.5}
\Photon(15,20)(40,20){2.5}{3.5}
\ArrowLine(15,20)(-15,20)
\Photon(0,-5)(-15,20){2.5}{4.5}\Text(-13,5)[r]{$\scriptstyle{V}$}\ArrowLine(0,-5)(15,20)
\Vertex(0,-5){1.5}\ArrowLine(0,-20)(0,-5)
\ArrowLine(-40,-20)(0,-20)
\Vertex(0,-20){1.5}
\Gluon(0,-20)(40,-20){-2.5}{6}
\Text(0,-35)[t]{#1}
\end{picture}}

\newcommand{\diagvIV}[1]{
\begin{picture}(94.,90.)(-47,-45)
\ArrowLine(0,20)(-40,20)
\Vertex(0,20){1.5}
\Gluon(0,20)(40,20){2.5}{6}
\ArrowLine(-40,-20)(-15,-20)
\Vertex(-15,-20){1.5}
\Vertex(15,-20){1.5}
\Photon(15,-20)(40,-20){-2.5}{3.5}
\ArrowLine(-15,-20)(15,-20)
\Photon(-15,-20)(0,5){2.5}{4.5}\Text(-13,-5)[r]{$\scriptstyle{V}$}\ArrowLine(15,-20)(0,5)
\Vertex(0,5){1.5}\ArrowLine(0,5)(0,20)
\Text(0,-35)[t]{#1}
\end{picture}}

\newcommand{\diagvV}[1]{
\begin{picture}(94.,90.)(-47,-45)
\ArrowLine(-15,20)(-40,20)
\Vertex(-15,20){1.5}
\Vertex(15,20){1.5}
\Photon(15,20)(40,20){2.5}{3.5}
\Photon(15,20)(-15,20){-2.5}{4.5}\Text(0,25)[b]{$\scriptstyle{V_1}$}
\ArrowLine(0,-5)(-15,20)\Photon(0,-5)(15,20){2.5}{4.5}\Text(13,5)[l]{$\scriptstyle{V_2}$}
\Vertex(0,-5){1.5}\ArrowLine(0,-20)(0,-5)
\ArrowLine(-40,-20)(0,-20)
\Vertex(0,-20){1.5}
\Gluon(0,-20)(40,-20){-2.5}{6}
\Text(0,-35)[t]{#1}
\end{picture}}

\newcommand{\diagvVI}[1]{
\begin{picture}(94.,90.)(-47,-45)
\ArrowLine(0,20)(-40,20)
\Vertex(0,20){1.5}
\Gluon(0,20)(40,20){2.5}{6}
\ArrowLine(-40,-20)(-15,-20)
\Vertex(-15,-20){1.5}
\Vertex(15,-20){1.5}
\Photon(15,-20)(40,-20){-2.5}{3.5}
\Photon(-15,-20)(15,-20){-2.5}{4.5}\Text(0,-25)[t]{$\scriptstyle{V_2}$}
\ArrowLine(-15,-20)(0,5)\Photon(15,-20)(0,5){2.5}{4.5}\Text(13,-5)[l]{$\scriptstyle{V_1}$}
\Vertex(0,5){1.5}\ArrowLine(0,5)(0,20)
\Text(0,-35)[t]{#1}
\end{picture}}

\newcommand{\diagbI}[1]{
\begin{picture}(94.,90.)(-47,-45)
\ArrowLine(-15,20)(-40,20)
\Vertex(-15,20){1.5}
\Vertex(15,20){1.5}
\Photon(15,20)(40,20){2.5}{3.5}
\ArrowLine(15,20)(-15,20)
\ArrowLine(-40,-20)(-15,-20)
\Vertex(-15,-20){1.5}
\Vertex(15,-20){1.5}
\Gluon(15,-20)(40,-20){-2.5}{3.5}
\ArrowLine(-15,-20)(15,-20)
\Photon(-15,-20)(-15,20){2.5}{5}\Text(-20,0)[r]{$\scriptstyle{V}$}
\ArrowLine(15,-20)(15,20)
\Text(0,-35)[t]{#1}
\end{picture}}

\newcommand{\diagbII}[1]{
\begin{picture}(94.,90.)(-47,-45)
\ArrowLine(-15,20)(-40,20)
\Vertex(-15,20){1.5}
\Vertex(15,20){1.5}
\Gluon(15,20)(40,20){2.5}{3.5}
\ArrowLine(15,20)(-15,20)
\ArrowLine(-40,-20)(-15,-20)
\Vertex(-15,-20){1.5}
\Vertex(15,-20){1.5}
\Photon(15,-20)(40,-20){-2.5}{3.5}
\ArrowLine(-15,-20)(15,-20)
\Photon(-15,-20)(-15,20){2.5}{5}\Text(-20,0)[r]{$\scriptstyle{V}$}
\ArrowLine(15,-20)(15,20)
\Text(0,-35)[t]{#1}
\end{picture}}

\newcommand{\diagbIII}[1]{
\begin{picture}(94.,90.)(-47,-45)
\ArrowLine(-15,20)(-40,20)
\Vertex(-15,20){1.5}
\Vertex(15,20){1.5}
\Photon(15,20)(40,20){2.5}{3.5}
\Photon(15,20)(-15,20){-2.5}{4.5}\Text(0,25)[b]{$\scriptstyle{V_1}$}
\ArrowLine(-40,-20)(-15,-20)
\Vertex(-15,-20){1.5}
\Vertex(15,-20){1.5}
\Gluon(15,-20)(40,-20){-2.5}{3.5}
\ArrowLine(-15,-20)(15,-20)
\Photon(-15,-20)(15,20){2.5}{7}\Text(13,0)[lb]{$\scriptstyle{V_2}$}
\ArrowLine(15,-20)(-15,20)
\Text(0,-35)[t]{#1}
\end{picture}}

\newcommand{\diagqqzct}{
\begin{picture}(40.,20.)(0,-3)
\ArrowLine(0,0)(-20,20)
\Photon(0,0)(30,0){2.5}{4}
\ArrowLine(-20,-20)(0,0)
\BCirc(0,0){4}\Line(-2.83,-2.83)(2.83,2.83)\Line(-2.83,2.83)(2.83,-2.83)
\end{picture}}

\newcommand{\diagqqgct}{
\begin{picture}(40.,20.)(0,-3)
\ArrowLine(0,0)(-20,20)
\Gluon(0,0)(30,0){2.5}{4}
\ArrowLine(-20,-20)(0,0)
\BCirc(0,0){4}\Line(-2.83,-2.83)(2.83,2.83)\Line(-2.83,2.83)(2.83,-2.83)
\end{picture}}

\newcommand{\diagqqct}{
\begin{picture}(40.,10.)(0,-3)
\ArrowLine(0,0)(-30,0)
\ArrowLine(30,0)(0,0)
\BCirc(0,0){4}\Line(-2.83,-2.83)(2.83,2.83)\Line(-2.83,2.83)(2.83,-2.83)
\end{picture}}

In this section we present the virtual electroweak corrections to the 
$\qbar q' \to W^\si g$ process.
The algebraic reduction to gauge-coupling structures, standard matrix elements
and one-loop scalar integrals is described in \refse{se:reduction}.
The renormalization of ultraviolet divergences and the subtraction of infrared
singularities originating from soft and collinear virtual photons are
discussed in \refse{se:renormalization}. and \refse{se:masssing},
respectively.
In \refse{se:results} we summarize the one-loop result for the
unpolarized squared matrix element.

\subsection{Preliminaries}
\label{se:preliminaries}
As discussed in the previous section, the twelve different processes relevant 
for $Wj$ production are related by CP and crossing symmetries. 
It is thus sufficient to consider only one of these processes.
{\unitlength 1pt \small
\begin{figure}
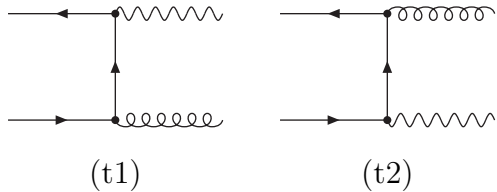

\begin{center}
\diagtI{(t1)}
\diagtII{(t2)}
\end{center}
 \caption{Tree-level Feynman diagrams for the process $\bar q q' \to W^\si g$.}
 \label{fig:treediags}
 \end{figure}}
In the following we derive the one-loop corrections for the $\qbar q' \to W^\si g$ process.
The matrix element 
\beqar\label{oneloopsplitting}
\Mqq_1&=&\Mqq_0+\delta \Mqq_{1}
\eeqar
is expressed as a function of the Mandelstam invariants 
\beq
\shat=(p_{\bar q}+ p_{q'})^2 
,\qquad 
\that=(p_{\bar q}- p_W)^2 
,\qquad 
\uhat=(p_{q'}-p_W)^2.
\eeq
The Born contribution  $\Mqq_0$ results from the $t$- and $u$-channel diagrams 
of \reffi{fig:treediags}.
The loop and counterterm 
diagrams contributing to the corrections, 
\beqar\label{CTsplitting}
\delta \Mqq_1&=&
\delta \Mqq_{1,\mathrm{loops}}+
\delta \Mqq_{1,\mathrm{CT}},
\eeqar
are depicted in \reffi{fig:loopdiags} and \reffi{fig:ctdiags}, respectively.
{\unitlength 1pt \small
\begin{figure}
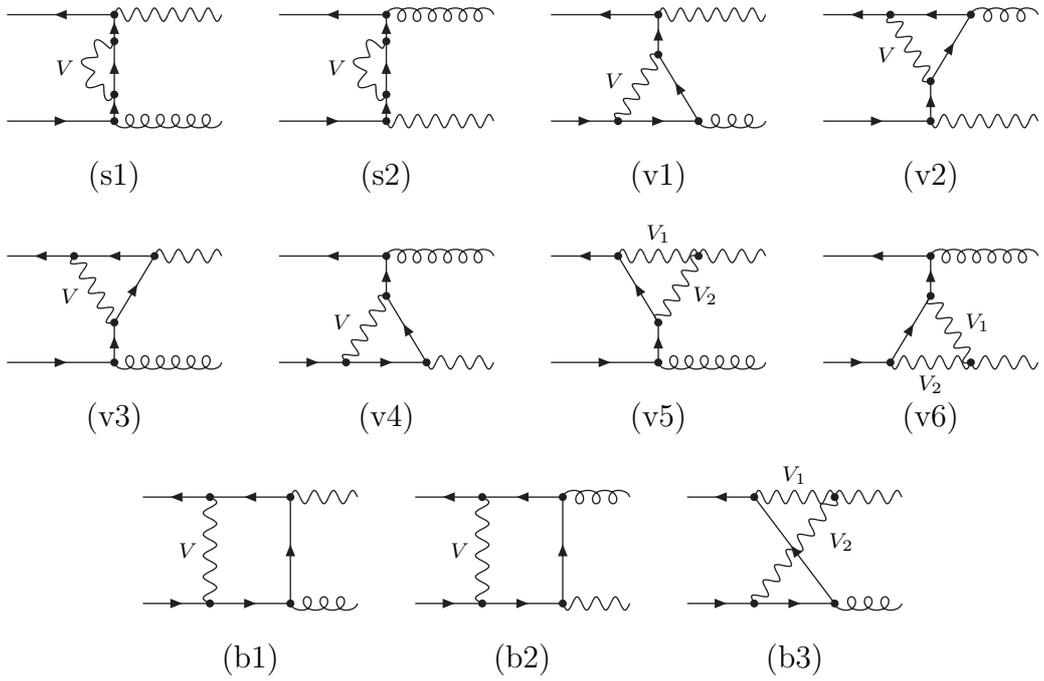

\begin{center}
\diagsI{(s1)}
\diagsII{(s2)}
\diagvI{(v1)}
\diagvII{(v2)}
\diagvIII{(v3)}
\diagvIV{(v4)}
\diagvV{(v5)}
\diagvVI{(v6)}
\diagbI{(b1)}
\diagbII{(b2)}
\diagbIII{(b3)}
\end{center}
 \caption{
One-loop Feynman diagrams for the process $\bar q q' \to W^\si g$.
The diagrams s1, s2, v1 and v2 receive contributions
from neutral and charged gauge bosons, $V=A,Z,W^\pm$.
The diagrams  v3, v4, b1 and b2  involve only 
neutral gauge bosons, $V=A,Z$.
The remaining diagrams, v5, v6 and b3 involve two contributions 
with one charged and one neutral gauge boson:
$(V_1,V_2)=(V,W^\pm)$ 
and
$(W^\pm,V)$ with  $V=A,Z$.
}
 \label{fig:loopdiags}
 \end{figure}}
{\unitlength 1pt \small
\begin{figure}
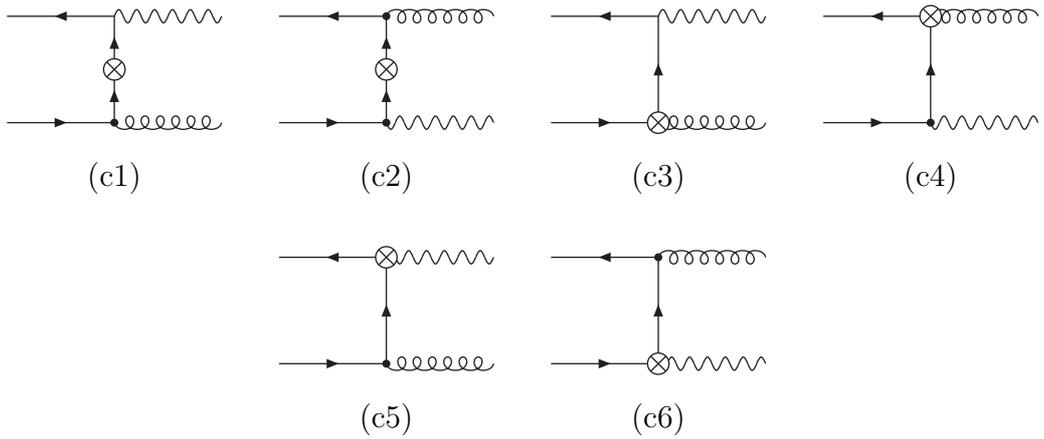

\begin{center}
\diagcI{(c1)}
\diagcII{(c2)}
\diagcIII{(c3)}
\diagcIV{(c4)}
\diagcV{(c5)}
\diagcVI{(c6)}
\end{center}
 \caption{Counterterm diagrams for the process $\bar q q' \to W^\si g$.}
 \label{fig:ctdiags}
 \end{figure}}

The quarks that are present in the loop diagrams of \reffi{fig:loopdiags} are
treated as massless, and the regularization of the collinear singularities 
that arise in this limit is discussed in \refse{se:masssing}.
The only quark-mass effects that we take into account are the $m_t$-terms 
that enter the counterterms through gauge-boson self-energies.

Our calculation has been performed at the matrix-element level 
and provides full control over polarization effects.
However, at this level, the analytical expressions are too large to be 
published.
Explicit results will thus be presented only for the 
unpolarized squared matrix element.

\subsection{Algebraic reduction}
\label{se:reduction}
The matrix element \refeq{oneloopsplitting} has the general form 
\beqar\label{spinors}
\Mqq_1
&=&  \ri\,e\, g_\rS\,t^a\,  
\bar{v}(p_{\qbar}) 
\M_1^{\rL,\mu\nu} \omega_\rL u(p_{q'})\, 
\varepsilon^*_\mu(p_W) \varepsilon^*_\nu(p_g).
\eeqar
Since we neglect quark masses,
$\M_1^{\rL,\mu\nu}$ consists of terms involving an odd number of matrices 
$\gamma^\rho$ with $\rho=0,\dots,3$. The $\gamma^5$-terms are isolated in the 
chiral projector 
$\omega_{\rL}$ defined in \refeq{projectors}.
The polarization dependence of the quark spinors 
and gauge-boson polarization vectors is implicitly understood.
In analogy to \refeq{oneloopsplitting} and \refeq{CTsplitting} we write
\beq
\M_1^{\rL,\mu\nu}=\M_0^{\rL,\mu\nu}+\de\M_1^{\rL,\mu\nu}
,\qquad
\de\M_1^{\rL,\mu\nu}=\de\M_{1,\mathrm{loops}}^{\rL,\mu\nu}+\de\M_{1,\mathrm{CT}}^{\rL,\mu\nu}
.
\eeq

Following the approach adopted in \citere{Kuhn:2005az}, we isolate the
SU(2)$\times$U(1) couplings that appear in the Feynman diagrams and reduce the
one-loop amplitude to a sum of contributions associated with independent
coupling structures.  
As we will see,
besides an abelian and a non-abelian contribution that are related to
the ones found for $Zj$ production \cite{Kuhn:2005az}, for $Wj$ production we
have two additional coupling structures.

The coupling structure of the Born amplitude
is trivial and consists simply of the $q_\rL q'_\rL$ component of the 
SU(2) generator,
\beqar\label{bornampli}
\M_0^{\rL,\mu\nu}
&=& I^{W^{-\si}}_{q_\rL q'_\rL}
\smel^{\mu\nu}_0
=
\frac{\smel^{\mu\nu}_0}
{\sqrt 2 \sw}
,\qquad
\smel^{\mu\nu}_0=
\frac{\gamma^\mu(\ps_W-\ps_\qbar)\gamma^\nu}{\that}+
\frac{\gamma^\nu(\ps_g-\ps_\qbar)\gamma^\mu}{\uhat}
.
\eeqar
The contribution of the loop diagrams of \reffi{fig:loopdiags} 
can be written as
\beqar\label{diagramsandcouplings}
\delta \M_{1,\mathrm{loops}}^{\rL,\mu\nu}
&=& \frac{\alpha}{4\pi}
\Biggl\{\sum_{V=\mathrm{A,Z,W^\pm}}\Biggl[
\left(I^{W^{-\si}} I^{V} I^{\bar V}\right)_{q_\rL q'_\rL} 
D_1^{\mu\nu}(M_V^2)
+
\left(I^{V} I^{\bar V} I^{W^{-\si}}\right)_{q_\rL q'_\rL} 
D_2^{\mu\nu}(M_V^2)
\Biggr]
\nl&&{}
+\sum_{V=\mathrm{A,Z}}\Biggl[
\left(I^{V} I^{W^{-\si}} I^{V}\right)_{q_\rL q'_\rL} 
D_3^{\mu\nu}(M_V^2)
+
\frac{\ri}{\sw} \varepsilon^{W^{\si}V W^{-\si}} 
\left(I^{V} I^{W^{-\si}}\right)_{q_\rL q'_\rL} 
\nl&&{}\times
D_4^{\mu\nu}(M_V^2,M_W^2)
+\frac{\ri}{\sw} \varepsilon^{V W^{\si} W^{-\si}} 
\left(I^{W^{-\si}} I^{V}\right)_{q_\rL q'_\rL} 
D_4^{\mu\nu}(M_W^2,M_V^2)
\Biggr]
\Biggr\}
.
\eeqar
In the following, treating the electroweak gauge couplings as
isospin matrices and using group-theoretical identities
(see App.~B of \citere{Pozzorini:rs}), 
we express the above amplitude in terms of 
the eigenvalues of isospin, hypercharge and SU(2) Casimir operators
for left-handed fermions \refeq{eigenvalues}.

The tensors $D_1^{\mu\nu}(M_V^2)$ and $D_2^{\mu\nu}(M_V^2)$ 
in \refeq{diagramsandcouplings} describe the contributions
of the diagrams s1, v1 and  s2, v2, respectively.
These diagrams may involve charged or neutral virtual bosons. 
In the former case $(V=W^\pm)$, the corresponding couplings read%
\footnote{The following identities have to be understood as matrix identities,
where the $q_\rL q'_\rL$ indices of the SU(2) generators are implicitly understood.}
\beqar
\label{couplings1a}
\sum_{\rho=\pm}I^{W^{-\si}} I^{W^\rho} I^{W^{-\rho}} 
&=& \sum_{\rho=\pm} I^{W^\rho} I^{W^{-\rho}} I^{W^{-\si}}=
\frac{C_{\mathrm{F}}-(T^3)^2}{\sw^2}
\,
I^{W^{-\si}}
.
\eeqar
In the latter case ($V=A,Z$) the coupling factors read 
\beqar
\label{couplings1c}
I^{W^{-\si}} I^{V} I^{V} 
&=&
\biggl[\de^{\mathrm{SU}(2)}_{VV}\frac{(T^3)^2}{\sw^2}
+X_V T^{3}Y
+\de^{\mathrm{U}(1)}_{VV}\frac{Y^2}{4\cw^2} \biggr]I^{W^{-\si}}
,
\nl
I^{V} I^{V} I^{W^{-\si}}
&=&
\biggl[
\de^{\mathrm{SU}(2)}_{VV}\frac{(T^3)^2}{\sw^2}
-X_V T^{3}Y
+\de^{\mathrm{U}(1)}_{VV}\frac{Y^2}{4\cw^2}\biggr]I^{W^{-\si}}
.
\eeqar
Here
\beq\label{mixingcouplings}
\de^{\mathrm{SU}(2)}_{VV}=(U_{VW^3})^2
,\qquad
X_V=\frac{U_{VW^3}U_{VB}}{\sw\cw}
,\qquad
\de^{\mathrm{U}(1)}_{VV}=(U_{VB})^2,
\eeq
where $U$ is the electroweak mixing matrix.
For $V=A,Z$ we have
$\de^{\mathrm{SU}(2)}_{AA}=\sw^2$,
$X_A=-1$,
$\de^{\mathrm{U}(1)}_{AA}=\cw^2$,
and 
$\de^{\mathrm{SU}(2)}_{ZZ}=\cw^2$,
$X_Z=1$,
$\de^{\mathrm{U}(1)}_{ZZ}=\sw^2$.
We note that 
\beqar\label{mixmatrelat}
\sum_{V=A,Z}\de^\mathrm{SU(2)}_{VV}
&=&\sum_{V=A,Z}\de^\mathrm{U(1)}_{VV}
=1
,\qquad
\sum_{V=A,Z}X_{V}=0.
\eeqar
The tensor $D_3^{\mu\nu}(M_V^2)$ in \refeq{diagramsandcouplings}
corresponds to the diagrams v3, v4, b1 and b2.
These diagrams receive contributions from 
neutral virtual gauge bosons only ($V=A,Z$). For the corresponding couplings 
we have 
\beqar
\label{couplings3}
&&I^{V} I^{W^{-\si}} I^{V} 
= \biggl[
\de^{\mathrm{SU}(2)}_{VV}\frac{C_{\mathrm{F}}-C_{\mathrm{A}}/2}{\sw^2}
+\de^{\mathrm{U}(1)}_{VV}\frac{Y^2}{4\cw^2}\biggr]I^{W^{-\si}}
.
\eeqar
Finally,  $D_4^{\mu\nu}(M_{V_1}^2,M_{V_2}^2)$ represents the diagrams v5, v6 and b3.
These diagrams  involve a neutral gauge boson ($V=A,Z$) and a $W$ boson.
The coupling factors yield
\beqar
\label{couplings2a}
\frac{\ri}{\sw} \varepsilon^{W^{\si}V W^{-\si}} I^{V} I^{W^{-\si}}
&=& \biggl[
\de^{\mathrm{SU}(2)}_{VV}\frac{C_{\mathrm{A}}}{4\sw^2}
-X_V T^{3}Y
\biggr]I^{W^{-\si}}
,\nl
\frac{\ri}{\sw} \varepsilon^{V W^{\si} W^{-\si}} I^{W^{-\si}} I^{V}
&=& \biggl[
\de^{\mathrm{SU}(2)}_{VV}\frac{C_{\mathrm{A}}}{4\sw^2}
+X_V T^{3}Y
\biggr]I^{W^{-\si}}
.
\eeqar
Using the above identities we express the one-loop amplitude \refeq{diagramsandcouplings} 
for $W$-boson production
in a form that is analogous to the one adopted in \citeres{Kuhn:2005az,Kuhn:2005gv} 
to describe 
the production of neutral gauge bosons.
To this end we define%
\footnote{In our notation we emphasize the dependence of the form factors 
$\delta \A^{\mu\nu}_{1,\mathrm{I}}$
on $M_V$, whereas the dependence on the external momenta as well as the
$M_W$-dependence (for $\delta \A^{\mu\nu}_{1,\mathrm{N}}$ and $\delta \A^{\mu\nu}_{1,\mathrm{Y}}$) 
is implicitly understood.
}
\beqar\label{formfactors}
\delta \A^{\mu\nu}_{1,\mathrm{A}}(M^2_V)&=&D^{\mu\nu}_1(M^2_V)+D^{\mu\nu}_2(M^2_V)+D^{\mu\nu}_3(M^2_V)
,\nl
\delta \A^{\mu\nu}_{1,\mathrm{N}}(M^2_V)&=&\frac{1}{2}\left[D^{\mu\nu}_4(M^2_V,M_W^2)+D^{\mu\nu}_4(M_W^2,M^2_V)\right]-D^{\mu\nu}_3(M^2_V)
,\nl
\delta \A^{\mu\nu}_{1,\mathrm{X}}(M^2_V)&=&D^{\mu\nu}_1(M^2_V)+D^{\mu\nu}_2(M^2_V)
,\nl
\delta \A^{\mu\nu}_{1,\mathrm{Y}}(M^2_V)&=&D^{\mu\nu}_4(M^2_V,M_W^2)-D^{\mu\nu}_4(M_W^2,M^2_V)
+D^{\mu\nu}_2(M^2_V)-D^{\mu\nu}_1(M^2_V).
\eeqar
The tensor $\delta \A^{\mu\nu}_{1,\mathrm{A}}(M^2_V)$ 
is identical to the
abelian tensor defined in \citere{Kuhn:2005az},
and $\delta \A^{\mu\nu}_{1,\mathrm{N}}(M^2_V)$
is equal to the non-abelian tensor of \citere{Kuhn:2005az}
for $M^2_V=M_W^2$.
The remaining two tensors, 
$\delta \A^{\mu\nu}_{1,\mathrm{X}}(M^2_V)$ and $\delta \A^{\mu\nu}_{1,\mathrm{Y}}(M^2_V)$,
are new.
Using \refeq{couplings1a}--\refeq{formfactors} we can write the one-loop amplitude  \refeq{diagramsandcouplings} as
\beqar\label{ymsplitting}
\delta \M_{1,\mathrm{loops}}^{\rL,\mu\nu}
&=& \frac{\alpha}{4\pi\sqrt{2}\sw}
\Biggl\{\sum_{V=A,Z}\Biggl[
\left(\de^{\mathrm{SU}(2)}_{VV}\frac{{C_{\mathrm{F}}}}{\sw^2}+\de^{\mathrm{U}(1)}_{VV}\frac{Y_{q_\rL}^2}{4\cw^2} \right)
\,\delta \A_{1,\mathrm{A}}^{\mu\nu}(M^2_V)
\nl&&{}
-{\de^{\mathrm{SU}(2)}_{VV}}\frac{C_{\mathrm{F}}-(T_{q_\rL}^3)^2}{\sw^2}
\,\delta \A_{1,\mathrm{X}}^{\mu\nu}(M^2_V)
+\de^{\mathrm{SU}(2)}_{VV}\frac{C_{\mathrm{A}}}{2\sw^2}
\,\delta \A_{1,\mathrm{N}}^{\mu\nu}(M^2_V)
\nl&&{}
-X_V T_{q_\rL}^{3}Y_{q_\rL}
\delta \A_{1,\mathrm{Y}}^{\mu\nu}(M^2_V)
\Biggr]
+\frac{C_{\mathrm{F}}-(T_{q_\rL}^3)^2}{\sw^2}
\,\delta \A_{1,\mathrm{X}}^{\mu\nu}(M_W^2)
\Biggr\}
.
\eeqar
This amplitude has been reduced algebraically using 
the Dirac equation, the identity
$p^\mu \varepsilon_\mu(p)=0$ for gauge-boson polarization vectors
and Dirac algebra.
Moreover, tensor loop integrals have been reduced to scalar ones 
by means of the Passarino-Veltman technique~\cite{pass-velt}.
The result has been expressed in the form 
\beqar\label{algebraicred}
\delta \A_{1,\mathrm{I}}^{\mu\nu}(M^2_V)
=\sum_{i=1}^{10}  
\sum_{j}
\F_{\mathrm{I}}^{ij}(M_V^2)
\smel_{i}^{\mu\nu} 
\loops_{j}(M_V^2)
,
\eeqar
for I=A,N,X,Y. 
The quantities $\F_{\mathrm{I}}^{ij}(M_V^2)$ 
are rational functions of Mandelstam invariants and masses.
Explicit expressions for the tensors 
$\smel_{i}^{\mu\nu}$ 
and the scalar loop integrals $\loops_{j}(M_V^2)$
are provided in \refapp{app:smel} and \refapp{app:loopint}.

\subsection{Renormalization}
\label{se:renormalization}
\newcommand{\Deltamsbar}{\bar \Delta_{\mathrm{UV}}}
While the tensors
$\delta \A_{1,\mathrm{X}}^{\mu\nu}$
and $\delta \A_{1,\mathrm{Y}}^{\mu\nu}$
are ultraviolet finite, the abelian and the nonabelian tensors
give rise to the ultraviolet singularities
\beq\label{UVsing}
\left.\delta \A_{1,\mathrm{A}}^{\mu\nu}(M^2_V)\right|_{\mathrm{UV}}=
\Deltamsbar\, \smel_{0}^{\mu\nu} 
,\qquad
\left.\delta \A_{1,\mathrm{N}}^{\mu\nu}(M^2_V)\right|_{\mathrm{UV}}=
2 \Deltamsbar\, \smel_{0}^{\mu\nu}, 
\eeq
where $\smel_{0}^{\mu\nu}$ is the tensor structure of the Born amplitude 
\refeq{bornampli}, and
\beq\label{msbarsubt}
\Deltamsbar = 
\left(\frac{4\pi \mu^2}{M_Z^2}\right)^\varepsilon
\frac{\Gamma(1+\varepsilon)}{\varepsilon}
=
\frac 1 \varepsilon -\gamma_{\mathrm{E}} +\ln(4\pi)+\ln\left(\frac{\mu^2}{M_Z^2}\right)
+\ord{(\varepsilon)}
\eeq
in $D=4-2\varepsilon$ dimensions. 
These singularities are cancelled by the 
counterterm diagrams depicted in \reffi{fig:ctdiags}
and the results are independent 
of the scale
$\mu$ of dimensional regularization.
The counterterms
 that are responsible for the contributions
of diagrams c1, c2, c3 and c4 read
\beq
\hspace{2.5cm}
\diagqqct
= 
\ri \ps \,
\omega_{\rL} \delta Z_{q_{\rL}}
,\hspace{2cm}
\diagqqgct
= 
-\ri g_\rS  t^a \gamma^{\mu} 
\omega_{\rL} \delta Z_{q_{\rL}}
.
\eeq
Since there is no $\ord(\alpha)$ contribution to the renormalization 
of the strong coupling constant $g_\rS$,
these counterterms depend only on the wave-function renormalization constants 
for left-handed quarks, $\delta Z_{q_{\rL}}$. 
Their combined contribution to the $\qbar q'\to W^\si g$ process,
\ie the sum of the diagrams  c1, c2, c3 and c4, vanishes.
The renormalization of the $\qbar q'\to W^\si g$ process is thus 
provided by the diagrams c5 and c6, which originate from 
the $W\qbar q'$ counterterm, 
\beqar\label{qqzct}
\diagqqzct
&=& \ri e \gamma^{\mu}\omega_{\rL} 
 I^{W^{-\si}}_{q_{\rL}q'_{\rL}}
\left[\delta C^{\mathrm{A}}+\delta C^{\mathrm{N}}\right],
\eeqar
with 
\beqar\label{deCCts}
\delta C^{\mathrm{A}} 
&=&
\frac{1}{2} \left(\delta Z_{u_{\rL}}+\delta Z_{d_{\rL}} \right)
,\qquad
\delta C^{\mathrm{N}} 
=
\frac{1}{2} \left( \de Z_W+\frac{\delta g_2^2}{g_2^2} \right)
,
\eeqar
and yields
\beqar\label{Ctcontrib}
\de\M_{1,\mathrm{CT}}^{\rL,\mu\nu}
&=& 
\left(\delta C^{\mathrm{A}}+\delta C^{\mathrm{N}}\right)
\M_0^{\rL,\mu\nu}
.
\eeqar
The wave-function renormalization constants of
massless left-handed quarks and on-shell $W$ bosons are related to the
corresponding self-energies by 
\beq\label{wfcts}
\delta Z_{q_{\rL}} =- \mathrm{Re}\left(\Sigma^{q,{\rL}}(0)\right)
,\qquad
\delta Z_{W} 
= 
-\mathrm{Re}
\left(
\frac{\partial
\Sigma^{{W}}_{\rT}(p^2) 
}{\partial p^2} 
\right)\Bigg|_{p^2=M_W^2}
,
\eeq
and have been evaluated using the explicit results of \citere{Denner:1991kt}.

For the definition and the renormalization of the 
SU(2) coupling constant,
\beq\label{g2ct}
g_2^2=\frac{4\pi\alpha}{\sw^2}
,\qquad
\frac{\de g_2^2}{g_2^2} =
\frac{\de \alpha}{ \alpha}- \frac{\de\sw^2}{\sw^2} 
,
\eeq
we adopt the $G_\mu$-scheme, where 
the 
electromagnetic 
coupling  constant $\alpha$ 
is expressed in
terms of the Fermi constant $G_\mu$,
and 
the weak mixing angle is related to the
on-shell masses $M_Z$, $M_W$ of the gauge bosons,
\beq 
\alpha = \frac{\sqrt{2}G_\mu M_W^2 \sw^2}{\pi}
,\qquad
\sw^2=1-\cw^2=1-M_W^2/M_Z^2
.
\eeq
The counterterm $\delta \alpha/\alpha$ in the $G_\mu$-scheme can
be  derived from the on-shell counterterm $\de\alpha(0)/\alpha(0)$ 
for the fine-structure constant in the Thompson limit.
Using the one-loop relation 
$\alpha 
=
\alpha(0)\left[1+\Delta r\right]
$
and requiring $\alpha+\de\alpha=\alpha(0)+\de\alpha(0)
$ we have
\beqar\label{alphacts}
\frac{\de\alpha}{\alpha} =
\frac{\de\alpha(0)}{\alpha(0)} - \Delta r.
\eeqar
Combining the relations \refeq{g2ct}--\refeq{alphacts} and using the
explicit one-loop expression for $\Delta r$ 
\cite{deltar, Bohm:2001yx},
we obtain
\beqar
\frac{\de g_2^2}{g_2^2} =
\mathrm{Re}\,\left[
\frac{\Sigma^{W}_{\rT}(M_W^2)-\Sigma^{W}_{\rT}(0) }{M_W^2}
\right]-
\frac{\alpha}{\pi\sw^2}\left[
\Deltamsbar
+\frac 14
\left(
6+\frac{7-
12\sw^2
}{2\sw^2}\ln\left(\frac{M_W^2}{M_Z^2}\right)
\right)
\right].
\nl
\eeqar
The above conterterms yield the ultraviolet singularities
\beqar\label{UVcounterterms}
\left.\delta C^{\mathrm{A}}\right|_{\mathrm{UV}} 
&=&-\frac{\alpha}{8\pi}\,\Deltamsbar
 \sum_{q=u,d}\sum_{V=\mathrm{A,Z,W^\pm}}
\left( I^V I^{\bar{V}} \right)_{q_{\rL}}
=-\frac{\alpha}{4\pi}\,\Deltamsbar
\left(\frac{{C_{\mathrm{F}}}}{\sw^2}+\frac{Y_{q_\rL}^2}{4\cw^2} \right)
,\nl
\left.\delta C^{\mathrm{N}}\right|_{\mathrm{UV}} 
&=&-\frac{\alpha}{2\pi\sw^2}\Deltamsbar\,
.
\eeqar
Using \refeq{mixmatrelat}
one can easily verify that these singularities cancel those 
resulting from the loop diagrams
[see \refeq{ymsplitting} and \refeq{UVsing}].

\subsection{Soft and collinear singularities}
\label{se:masssing}
\newcommand{\irfin}{fin}

Loop diagrams and wave-function renormalization constants involve 
singularities originating from 
soft and collinear virtual photons 
(for brevity denoted in the following as IR singularities).  
In order to isolate these singularities and check 
that they are cancelled by corresponding ones originating from
real photon bremsstrahlung, we split the
wave-function renormalization constants and the photon contributions 
to \refeq{ymsplitting},
\ie the terms $\delta \A_{1,\mathrm{I}}^{\mu\nu}(M_A^2)$, 
in IR-singular (IR) and 
IR-finite (\irfin) parts:
\beqar\label{splittingdef1}
\delta Z_{q_{\rL}}
&=&
\delta Z^{\mathrm{IR}}_{q_{\rL}}+
\delta Z^{\mathrm{\irfin}}_{q_{\rL}}
,\nl
\delta Z_{W}
&=&
\delta Z^{\mathrm{IR}}_{W}+
\delta Z^{\mathrm{\irfin}}_{W}
,\nl
\delta \A_{1,\mathrm{I}}^{\mu\nu}(M_A^2)
&=&
\delta \A^{\mathrm{IR},\mu\nu}_{1,\mathrm{I}}
+
\delta \A^{\mathrm{\irfin},\mu\nu}_{1,\mathrm{I}}
.
\eeqar
The singular parts depend on the scheme adopted to regularize IR singularities. The remaining parts are scheme-independent
and free 
from IR singularities, but can contain ultraviolet poles.
For the regularization of  IR singularities we use, alternatively, 
two different schemes:
\begin{itemize}
\item 
In the first scheme, which we denote as mass-regularization scheme (MR),
we use infinitesimal quark  masses $m$ and a 
photon-mass regulator, 
$M_A=\la$ with $0<\la \ll m$. 
Since the quark-mass dependence disappears in the final result,
we perform the computation using the same mass $m$ for all quarks.
To denote quantities evaluated in this scheme 
we use the label MR;

\item 
In the second scheme we perform the calculation using massless fermions and
photons, $M_A=m=0$, and we evaluate IR singularities in dimensional regularization (DR).
To denote quantities evaluated in this scheme  we use the label DR.
\end{itemize}
The singular parts of the wave-function renormalization constants read
\newcommand{\wfsing}[2]{h^\mathrm{IR}_{#1 #2}}
\newcommand{\loopsing}[2]{f^\mathrm{IR}_{#1 #2}}
\beqar\label{irwfrc}
\delta Z^{\mathrm{IR}}_{q_{\rL}}&=&
\frac{\alpha}{4\pi} 
\left(\frac{4\pi\mu^2}{M_W^2}\right)^\varepsilon\Gamma(1+\varepsilon)
Q_q^2 \wfsing{q}{}
,\nl
\delta Z^{\mathrm{IR}}_{W}&=&
\frac{\alpha}{4\pi} 
\left(\frac{4\pi\mu^2}{M_W^2}\right)^\varepsilon\Gamma(1+\varepsilon)
\wfsing{W}{}
,
\eeqar
with 
\beqar\label{msCtcontrib1}
\wfsing{q}{,\mathrm{MR}}
&=&
-\ln\left(\frac{M_W^2}{m^2}\right)
-2\ln\left(\frac{\la^2}{m^2}\right)-4
,\qquad
\wfsing{W}{,\mathrm{MR}}
=
-2\ln\left(\frac{\la^2}{M_W^2}\right),
\eeqar
in the MR scheme and
\beqar\label{msCtcontrib2}
\wfsing{q}{,\mathrm{DR}}
&=&
\frac{1}{\varepsilon}
,\qquad
\wfsing{W}{,\mathrm{DR}}
=
-\frac{2}{\varepsilon},
\eeqar
in the DR scheme.
The splitting  of the loop contributions $\delta
\A^{\mu\nu}_{1,\mathrm{I}}(M^2_A)$ into IR-singular and IR-finite parts is performed 
at the level of the 
scalar loop integrals $\loops_{i}(M_A^2)$:
\beq\label{splittingdef2}
\loops_{i}(M_A^2)
= 
\loops_{i}^{\mathrm{IR}}
+
\loops_{i}^{\mathrm{\irfin}}
.
\eeq
Explicit expression for the IR-singular and IR-finite parts of
individual loop integrals are presented in \refapp{app:IRsing}.
Combining all singular contributions $\loops_{i}^{\mathrm{IR}}$ we obtain
\beqar\label{factorization}
\delta \A^{\mathrm{IR},\mu\nu}_{1,\mathrm{I}}
&=&
\left(\frac{4\pi\mu^2}{M_W^2}\right)^\varepsilon\Gamma(1+\varepsilon)
f_{\mathrm{I}}^{\mathrm{IR}}\smel_{0}^{\mu\nu},
\eeqar
\ie the IR singularities factorize%
\footnote{To be precise, the tensors
$\delta \A^{\mu\nu}_{1,\mathrm{X}}(M_A^2)$ and
$\delta \A^{\mu\nu}_{1,\mathrm{N}}(M_A^2)$ 
contain also non-factorizable IR divergences.
However these  non-factorizable singularities are related by
\beqar
\delta \A^{\mathrm{non-fact},\mu\nu}_{1,\mathrm{X}}(M_A^2)
=
2 \delta \A^{\mathrm{non-fact},\mu\nu}_{1,\mathrm{N}}(M_A^2),
\nonumber
\eeqar
and due to the identity
$C_{\mathrm{F}}-(T_{q_\rL}^3)^2=C_{\mathrm{A}}/4$,
which relates the coupling structures associated with the X- and N-terms in 
\refeq{ymsplitting}, they cancel.}
 with respect to the Born
amplitude \refeq{bornampli}. 
The IR-singular part of the renormalized amplitude can be expressed
in terms of the electromagnetic charges of the external particles as
\newcommand{\CLa}{1}
\newcommand{\CLb}{2}
\newcommand{\CLc}{3}
\beqar\label{ymsplitting3}
\delta \M_{1,\mathrm{IR}}^{\rL,\mu\nu}
=
\frac{\alpha}{4\pi}\left(\frac{4\pi\mu^2}{M_W^2}\right)^\varepsilon\Gamma(1+\varepsilon)
\Biggl[
-Q_q Q_{q'}  \loopsing{\CLa}{}
+\si Q_q    \loopsing{\CLb}{}
-\si Q_{q'} \loopsing{\CLc}{}
\Biggr]
\M_0^{\rL,\mu\nu}
,
\eeqar
where $\si=\pm 1$ is the charge of the $W$ boson and
\beqar\label{chargeff}
\loopsing{\CLa}{}
&=&
-\loopsing{\mathrm{A}}{}
-\wfsing{q}{}
,\nl
\loopsing{\CLb}{}
&=&
-\loopsing{\mathrm{A}}{}
-\loopsing{\mathrm{N}}{}
+\frac 12
\left(
\loopsing{\mathrm{X}}{}
-\loopsing{\mathrm{Y}}{}
-\wfsing{q}{}
-\wfsing{W}{}
\right)
,\nl
\loopsing{\CLc}{}
&=&
-\loopsing{\mathrm{A}}{}
-\loopsing{\mathrm{N}}{}
+\frac 12
\left(\loopsing{\mathrm{X}}{}
+\loopsing{\mathrm{Y}}{}
-\wfsing{q}{}
-\wfsing{W}{}
\right).
\eeqar
In the MR scheme we obtain
\beqar\label{massivesing2}
\loopsing{\CLa}{,\mathrm{MR}}
&=& 
-2\ln\left(\frac{\la^2}{M_W^2}\right)
\ln\left(\frac{-\shat}{m^2}\right)
-\ln^2\left(\frac{m^2}{M_W^2}\right)
+3\ln\left(\frac{m^2}{M_W^2}\right)
+2\ln\left(\frac{\la^2}{m^2}\right)
,\nl
\loopsing{\CLb}{,\mathrm{MR}}
&=& 
\ln\left(\frac{\la^2}{M_W^2}\right)
\left[\ln\left(\frac{m^2}{M_W^2}\right)-2\ln\left(1-\frac{\that}{M_W^2}\right)\right]
-\frac 12 \ln^2\left(\frac{m^2}{M_W^2}\right)
\nl&&{}
+\frac 12 \ln\left(\frac{m^2}{M_W^2}\right)
+2\ln\left(\frac{\la^2}{M_W^2}\right)
,\nl
\loopsing{\CLc}{,\mathrm{MR}}
&=& \loopsing{\CLb}{,\mathrm{MR}}
\Big|_{\that \to \uhat},
\eeqar
and in the DR scheme 
\beqar\label{dimsing}
\loopsing{\CLa}{,\mathrm{DR}}
&=& 
\frac{2}{\varepsilon^2}
-\frac{1}{\varepsilon}
\left[2 \ln\left(\frac{-\shat}{M_W^2}\right)
-3\right]+4
,\nl
\loopsing{\CLb}{,\mathrm{DR}}
&=& 
\frac{1}{\varepsilon^2} 
-\frac{1}{\varepsilon} 
\left[2
\ln\left(1-\frac{\that}{M_W^2}\right)
-\frac 52
\right]+2
,\nl
\loopsing{\CLc}{,\mathrm{DR}}
&=& \loopsing{\CLb}{,\mathrm{DR}}
\Big|_{\that \to \uhat}
.
\eeqar
The splitting \refeq{splittingdef1} has been performed in such a way that 
in the high-energy limit ($\shat,|\that|,|\uhat|\gg M_W^2$) the IR-finite 
part of the amplitude has the same logarithmic behaviour 
as the virtual corrections regularized by a photon mass $M_A=M_W$.
Indeed the IR-singular parts $\loopsing{i}{}$ correspond exactly
 to the contribution called purely electromagnetic
 in \citere{Denner:2001jv}.  This implies that, up to terms that are
 not logarithmically enhanced at high energies, the IR-finite part
 of the corrections corresponds to the symmetric electroweak
 contribution of \citere{Denner:2001jv}, which is constructed by
 setting the photon mass equal to $M_W$.  
This property is evident in the asymptotic high-energy 
expressions \refeq{heres6} for the IR-finite part 
of the diagrams involving virtual photons.

\subsection{Result}
\label{se:results}
Let us summarize our 
result for the unpolarized squared matrix element
for the $\qbar q'\to W^\sigma g$ process.
To $\ord(\alpha^2\alpha_\rS)$,
\beqar\label{NLOunpol}
\overline{\sum}|\Mqq_{1}|^2 =
\overline{\sum}|\Mqq_{0}|^2 
+
2 \mathrm{Re}\left[ \overline{\sum}\left( \Mqq_0\right)^* \de \Mqq_1\right]
.
\eeqar
Using \refeq{spinors} and summing over the polarizations we 
can express the interference term as
\beqar\label{polsumid}
\lefteqn{
2 \mathrm{Re}\left[ \overline{\sum}\left( \Mqq_0\right)^* \de \Mqq_1\right]
=
2 \pi^2 \alpha  \alpha_\rS (N_c^2-1)
}\quad&&\nl&&{}\times
\mathrm{Re}\left[\mathrm{Tr}\left(
\ps_{q'} \overline{\M}_0^{\rL,\mu\nu}  \ps_{\qbar} \de \M_1^{\rL,\mu'\nu'}\right)\right]
g_{\nu\nu'} \left( g_{\mu\mu'}-\frac{p_{W\mu} p_{W\mu'}}{p_W^2} \right),
\eeqar
where $ \overline{\M}=\gamma^0 \M^\dagger\gamma^0$.
Combining the contributions of
the bare one-loop diagrams \refeq{ymsplitting} and the counterterms \refeq{Ctcontrib}
yields
\beqar\label{generalresult}
\overline{\sum}
|\Mqq_{1}|^2 
&=&
\left[1+2\, \mathrm{Re}\left(\de C^{\mathrm{A}}+\de C^{\mathrm{N}}\right)\right]
\overline{\sum}
|\Mqq_{0}|^2 
+
\frac{ 2 \pi \alpha^2 \alpha_\rS}{\sw^2} (N_\mathrm{c}^2-1)
\nl&&{}\times\mathrm{Re}\,
\Biggl\{\sum_{V=\mathrm{A,Z}}\Biggl[
\left(\de^{\mathrm{SU}(2)}_{VV}\frac{{C_{\mathrm{F}}}}{\sw^2}+\de^{\mathrm{U}(1)}_{VV}\frac{Y_{q_\rL}^2}{4\cw^2} \right)
H_1^{\mathrm{A}}(M^2_V)
\nl&&{}
-{\de^{\mathrm{SU}(2)}_{VV}}\frac{C_{\mathrm{F}}-(T_{q_\rL}^3)^2}{\sw^2}
H_1^{\mathrm{X}}(M^2_V)
+\de^{\mathrm{SU}(2)}_{VV}\frac{C_{\mathrm{A}}}{2\sw^2}
H_1^{\mathrm{N}}(M^2_V)
\nl&&{}
-X_V T_{q_\rL}^{3}Y_{q_\rL}
H_1^{\mathrm{Y}}(M^2_V)
\Biggr]
+\frac{C_{\mathrm{F}}-(T_{q_\rL}^3)^2}{\sw^2}
H_1^{\mathrm{X}}(M_W^2)
\Biggr\}.  
\eeqar 
The unpolarized Born contribution is given in
\refeq{generalamplitude}, the counterterms  $\de C^{\mathrm{A}}$ and $\de
C^{\mathrm{N}}$ are presented in \refse{se:renormalization},
and the coupling factors are specified by \refeq{eigenvalues}
and \refeq{mixingcouplings}.
The functions $H_1^{\mathrm{I}}(M^2_V)$
represent the contributions 
resulting from the loop diagrams of \reffi{fig:loopdiags}.
They are related to the tensors 
$\delta \A_{1,\mathrm{I}}^{\mu\nu}(M^2_V)$
in \refeq{ymsplitting}--\refeq{algebraicred} by
\beqar\label{unpolfunct}
   H_1^{\mathrm{I}}(M_V^2) &=& 
\frac{1}{8}
\mathrm{Tr}\left[ \,
\ps_{q'} \, \overline{\smel}_0^{\mu\nu} \, \ps_{\qbar} \, 
\delta \A_{1,\mathrm{I}}^{\mu'\nu'}(M^2_V)
\, 
\right] 
g_{\nu\nu'} \left( g_{\mu\mu'}-\frac{p_{W\mu} p_{W\mu'}}{p_W^2} \right)
.
\eeqar
The couplings associated with
$H_1^{\mathrm{A}}$,
$H_1^{\mathrm{N}}$
and
$H_1^{\mathrm{X}}$
are the same for $q=u$ and $q=d$.
Thus the crossing and CP symmetry relations \refeq{crossing1} and \refeq{cpsymm} imply that these 
functions are symmetric with respect to the transformation 
$\that\leftrightarrow \uhat$.
In contrast, $H_1^{\mathrm{Y}}$  is antisymmetric with respect to
$\that\leftrightarrow \uhat$ exchange
since the corresponding coupling is proportional to $T^3_{q_{\rL}}$
and has thus opposite signs for $q=u$ and $q=d$.  
The functions $H_1^{\mathrm{I}}(M_V^2)$ are presented in 
\refapp{app:coeff} as linear combinations of 
scalar loop integrals.
We note that, in contrast to the definition adopted in
the case of $Zj$ production \cite{Kuhn:2005az}, here 
we do not include the contributions of the fermionic wave-function 
renormalization constants in $H_1^{\mathrm{A}}(M_V^2)$.

For the IR-singular part of the renormalized one-loop correction 
we obtain 
\beqar\label{msinggeneralresult2}
\overline{\sum}|\Mqq_{1,\mathrm{IR}}|^2 
&=& 
\frac{\alpha}{2\pi}
\mathrm{Re}\, \Biggl[
-Q_q Q_{q'}  \loopsing{\CLa}{}
+\si Q_q    \loopsing{\CLb}{}
-\si Q_{q'} \loopsing{\CLc}{}
\Biggr]
\nl&&{}\times
\left(\frac{4\pi\mu^2}{M_W^2}\right)^\varepsilon\Gamma(1+\varepsilon)
\,\overline{\sum}|\M_{0}^{\bar q q'\to W^\si g}|^2 
.
\eeqar
The IR-singular functions $ \loopsing{i}{}$ in the MR and DR schemes
are presented in \refse{se:masssing}.

\section{High-energy limit}
\label{se:helimit}
In this section we provide compact analytic expressions that describe the
behaviour of the IR-finite part of the virtual electroweak corrections
in the limit $M_W^2/\shat\to 0$ with 
$\that/\shat$ and  $\uhat/\shat$ constant.
In this limit, 
which is applicable for transverse momenta of 
$\ord(100 \GeV)$ or beyond,
the electroweak corrections
are dominated by logarithmic contributions of the
type $\ln(\shat/M_W^2)$.
In \refse{se:NNLL} we present the asymptotic expansion of the one-loop
corrections, including leading and next-to-leading logarithms, as well as
terms that are not logarithmically enhanced at high energies.
In \refse{se:twoloops} we present the two-loop corrections to next-to-leading
logarithmic accuracy.

\subsection{Next-to-next-to-leading approximation at one loop}
\label{se:NNLL}
In this section we discuss the high-energy behaviour of the IR-finite
part of the one-loop corrections to the $\qbar q'\to W^\si g$ process, 
obtained by subtracting the
IR divergence \refeq{msinggeneralresult2} from the 
renormalized one-loop result \refeq{generalresult},
\beqar\label{remres}
\overline{\sum}|\Mqq_{1,\mathrm{\irfin}}|^2 
&=& 
\overline{\sum}|\Mqq_{1}|^2 
-
\overline{\sum}|\Mqq_{1,\mathrm{IR}}|^2
.
\eeqar
In the following we present  explicit asymptotic expressions for
the unrenormalized loop contributions, \ie 
for the IR-finite parts $H^{\mathrm{I,\irfin}}_1$
of the functions $H^{\mathrm{I}}_1$ in \refeq{generalresult}.
Using the general results of \citere{Roth:1996pd}, we evaluate the functions
 $H^{\mathrm{I,\irfin}}_1$ to next-to-next-to-leading logarithmic (NNLL)
accuracy.
This approximation accounts for all
contributions that are not suppressed by powers of $M_W^2/\shat$.
It includes double and single logarithms as well as terms that
are not logarithmically enhanced in the high-energy limit.
To simplify non-logarithmic functions of the ratio  $M_Z/M_W$ 
we have performed an expansion
in $\sw^2=1-M_W^2/M_Z^2$, keeping only terms up to the first order\footnote{In practice we find that all terms of $\ord(\sw^2)$
cancel in the result.}
in $\sw^2$. 
The NNLL expansion of $H^{\mathrm{I,\irfin}}_1 (M_V^2)$
has the general form
\beq\label{nnllstracture}
H^{\mathrm{I,\irfin}}_1 (M_V^2)\NNLLa
\mathrm{Re }\,\left[
g_0^{\mathrm{I}}(M_V^2)\,
\frac{ \that^2+\uhat^2}{\that\uhat}
+g_1^{\mathrm{I}}(M_V^2)\,
\frac{ \that^2-\uhat^2}{\that\uhat}
+g_2^{\mathrm{I}}(M_V^2)
\right].
\eeq
It involves the rational function
$(\that^2+\uhat^2)/\that\uhat$,
which has the same angular behaviour as the 
squared Born amplitude \refeq{generalamplitude}
in the high-energy limit,
and two other rational functions,
which describe different angular dependencies.
The functions $g_i^{\mathrm{I}}$ consist of logarithms 
of the kinematical variables and constants.
The loop diagrams involving $Z$ and $W$ bosons, with mass
$M_V=M_Z,M_W$, yield
\beqar\label{heres1}
 g_0^{\mathrm{N}}(M_V^2)&=&
2\left[\Deltamsbar +\ln\left(\frac{M_Z^2}{M_W^2}\right)+\ln\left(\frac{M_V^2}{M_W^2}\right)\right]
+\ln^2\left(\frac{-\shat}{M_V^2}\right)
-\frac{1}{2}\Biggl[
\ln^2\left(\frac{-\that}{M_V^2}\right)
\nl&&{}
+
\ln^2\left(\frac{-\that}{M_W^2}\right)
+
\ln^2\left(\frac{-\uhat}{M_V^2}\right)
+
\ln^2\left(\frac{-\uhat}{M_W^2}\right)
\Biggr]
+\ln^2\left(\frac{\that}{\uhat}\right)
-\frac{3}{2}\Biggl[
\ln^2\left(\frac{\that}{\shat}\right)
\nl&&{}
+\ln^2\left(\frac{\uhat}{\shat}\right)
\Biggr]
-\frac{20\pi^2}{9}
-\frac{2\pi}{\sqrt{3}}
+4 
,\nl
 g_1^{\mathrm{N}}(M_V^2)&=&
\frac{1}{2}\Biggl[
\ln^2\left(\frac{\uhat}{\shat}\right)
-
\ln^2\left(\frac{\that}{\shat}\right)
\Biggr]
,\nl
 g_2^{\mathrm{N}}(M_V^2)&=&
-2\Biggl[
\ln^2\left(\frac{\that}{\shat}\right)
+\ln^2\left(\frac{\uhat}{\shat}\right)
+\ln\left(\frac{\that}{\shat}\right)
+\ln\left(\frac{\uhat}{\shat}\right)
\Biggr]
+2\ln\left(\frac{M_V^2}{M_W^2}\right)-4{\pi^2}
,
\nl
 g_0^{\mathrm{A}}(M_V^2)&=&
-\ln^2\left(\frac{-\shat}{M_V^2}\right)
+3\ln\left(\frac{-\shat}{M_V^2}\right)
+\frac{3}{2}\Biggl[
\ln^2\left(\frac{\that}{\shat}\right)
+\ln^2\left(\frac{\uhat}{\shat}\right)
+\ln\left(\frac{\that}{\shat}\right)
\nl&&{}
+\ln\left(\frac{\uhat}{\shat}\right)
\Biggr]
+\frac{7\pi^2}{3}-\frac{5}{2}
+ g_0^{\mathrm{A},\mathrm{UV}}(M_V^2)
,\nl
 g_1^{\mathrm{A}}(M_V^2)&=&
- g_1^{\mathrm{N}}(M_W^2)+
\frac{3}{2}\Biggl[
\ln\left(\frac{\uhat}{\shat}\right)
-
\ln\left(\frac{\that}{\shat}\right)
\Biggr]
,\nl
 g_2^{\mathrm{A}}(M_V^2)&=&
- g_2^{\mathrm{N}}(M_W^2)
,\nl
g_0^{\mathrm{X}}(M_V^2)&=&0
,\nl
g_1^{\mathrm{X}}(M_V^2)&=&0
,\nl
g_2^{\mathrm{X}}(M_V^2)&=&
-2\left[
2\ln\left(\frac{-\shat}{M_V^2}\right)
+\ln\left(\frac{\that}{\shat}\right)
+\ln\left(\frac{\uhat}{\shat}\right)
-3
\right]
,\nl
 g_0^{\mathrm{Y}}(M_V^2)&=&
\ln^2\left(\frac{-\that}{M_W^2}\right)
-\ln^2\left(\frac{-\that}{M_V^2}\right)
-\ln^2\left(\frac{-\uhat}{M_W^2}\right)
+\ln^2\left(\frac{-\uhat}{M_V^2}\right)
,\nl
g_1^{\mathrm{Y}}(M_V^2)&=&0
,\nl
g_2^{\mathrm{Y}}(M_V^2)&=&
2\ln\left(\frac{\that}{\uhat}\right)
,
\eeqar
where $\Deltamsbar$ is defined in \refeq{msbarsubt} and, in order to
facilitate the comparison with \citere{Kuhn:2005az}, we have isolated the 
term
\beq
g_0^{\mathrm{A},\mathrm{UV}}(M_V^2)
=
\Deltamsbar +\ln\left(\frac{M_Z^2}{M_V^2}\right)-\frac{1}{2}
.
\eeq
If we included the fermionic wave-function renormalization constants in
the definition of the function $H^{\mathrm{A}}_1$, 
as we had done for the case of $Zj$ production in \citere{Kuhn:2005az},
this term would cancel and the function $g_0^{\mathrm{A}}(M_V^2)$ would
be identical to the one obtained in \citere{Kuhn:2005az}.

For the loop diagrams involving photons ($M_V=M_A$), 
after subtraction of the IR-singular parts, we obtain
\beqar\label{heres6}
 g_0^{\mathrm{N}}(M_A^2)&=&
 g_0^{\mathrm{N}}(M_W^2)
-\frac{7\pi^2}{9}+\frac{2\pi}{\sqrt{3}}
,\nl
 g_0^{\mathrm{A}}(M_A^2)&=&
 g_0^{\mathrm{A}}(M_W^2)
+\pi^2
,\nl
g_0^{\mathrm{I}}(M_A^2)&=&g_0^{\mathrm{I}}(M_W^2)
\qquad\mbox{for}\quad \mathrm{I=X,Y}
,\nl
g_1^{\mathrm{I}}(M_A^2)&=&g_1^{\mathrm{I}}(M_W^2)
\qquad\mbox{for}\quad\mathrm{I=A,N,X,Y}
,\nl
g_2^{\mathrm{I}}(M_A^2)&=&g_2^{\mathrm{I}}(M_W^2)
\qquad\mbox{for}\quad\mathrm{I=A,N,X,Y}
.
\eeqar

The contribution of the counterterms
$\de C^{\mathrm{A}}$ and $\de C^{\mathrm{N}}$ to the IR-finite part of the renormalized result \refeq{remres} is obtained 
by subtracting from \refeq{deCCts}
the IR-divergent part  of the wave-function
renormalization constants \refeq{irwfrc}. This contribution, consisting of on-shell self-energies 
and their derivatives, does not depend on the scattering
energy. Therefore we evaluate the IR-finite parts 
of the counterterms 
in numerical form without applying any approximation.
Using the input parameters specified in \refse{se:numerics} we obtain
\beqar\label{deCCtsb}
\delta C^{\mathrm{A,\irfin}} 
&=&
\frac{1}{2} \left(\delta Z^{\mathrm{\irfin}}_{u_{\rL}}+\delta Z^{\mathrm{\irfin}}_{d_{\rL}} \right)
=\left.\delta C^{\mathrm{A}}\right|_{\mathrm{UV}} 
+5.57
\times 10^{-4}
,\nl
\delta C^{\mathrm{N,\irfin}} 
&=&
\frac{1}{2} \left( \de Z^{\mathrm{\irfin}}_W+\frac{\delta g_2^2}{g_2^2} \right)
=\left.\delta C^{\mathrm{N}}\right|_{\mathrm{UV}} 
-1.49
\times 10^{-3}
.
\eeqar
The UV divergences $\delta C^{\mathrm{A,N}}|_{\mathrm{UV}}$
[see \refeq{UVcounterterms}]
cancel against the  $\Deltamsbar$-terms in
 \refeq{heres1}--\refeq{heres6}.

The results \refeq{heres1}--\refeq{heres6}, for the 
 $\qbar q'\to W^\si g$ process, are valid for arbitrary values 
of the Mandelstam invariants and can easily be 
translated to all other processes in \refeq{processespl}
by means of the relations \refeq{crossing1}--\refeq{cpsymm}.
Logarithms with negative arguments 
in \refeq{heres1}--\refeq{heres6} 
are defined through the 
usual $\ri\varepsilon$ prescription, 
$\rhat\to \rhat +\ri\varepsilon$
for  $\rhat=\shat,\that,\uhat$.

In next-to-leading logarithmic (NLL) approximation,
\ie retaining only double and single logarithms that grow with energy,
the above results assume a particularly compact form.
In this approximation the counterterms 
do not contribute,
\beq\label{nllapprox2}
\delta C^{\mathrm{A}}\NLLa
\delta C^{\mathrm{N}}\NLLa
0,
\eeq
and for the functions  $H^{\mathrm{I,\irfin}}_1 (M_V^2)$,
neglecting logarithms of $M_Z/M_W$, we obtain
\beqar\label{nllapprox}
H_1^{\mathrm{N,\irfin}}(M_V^2)&\NLLa&
-\left[
\ln^2\left(\frac{|\that|}{M_W^2}\right)
+\ln^2\left(\frac{|\uhat|}{M_W^2}\right)
-\ln^2\left(\frac{|\shat|}{M_W^2}\right)
\right] 
\frac{ \that^2+\uhat^2}{\that\uhat}
,\nl
H_1^{\mathrm{A,\irfin}}(M_V^2)&\NLLa&
-\left[\ln^2\left(\frac{|\shat|}{M_W^2}\right)
-3\ln\left(\frac{|\shat|}{M_W^2}\right)
\right] 
\frac{ \that^2+\uhat^2}{\that\uhat}
,\nl
H_1^{\mathrm{X,\irfin}}(M_V^2)&\NLLa&
-4\ln\left(\frac{|\shat|}{M_W^2}\right)
,\nl
H_1^{\mathrm{A,\irfin}}(M_V^2)&\NLLa&0
,
\eeqar
for $V=A,Z,W$.
We note that, owing to $H_1^{\mathrm{X,\irfin}}(M_Z)\NLLa
H_1^{\mathrm{X,\irfin}}(M_A)$ and \refeq{mixmatrelat},
the NLL contribution of the function  $H_1^{\mathrm{X,\irfin}}(M_V)$
cancels in \refeq{generalresult}. 
Thus the NLL corrections \refeq{nllapprox} are proportional to the rational function $(\that^2+\uhat^2)/\that\uhat$,
which describes the angular dependence of the Born cross section.

\subsection{Next-to-leading logarithms up to two loops}
\label{se:twoloops}
\newcommand{\logar}[2]{\mathrm{L}^{#1}_{#2}}
Let us now present our results for the  NLL asymptotic behaviour of the
electroweak corrections  up to two loops.
For a discussion of the calculation we refer to 
\citere{Kuhn:2004em}, where the same class of corrections 
has been computed for $Zj$ production.
The results have been obtained in the $M_Z=M_W$ approximation.
As in the previous section, we present results for the 
IR-finite part of the electroweak 
corrections, obtained after subtraction of IR singularities.
As discussed in \refse{se:masssing}, at one loop this subtraction is performed
in such a way that, to NLL accuracy, the IR-finite part corresponds to the complete electroweak
correction regularized with a fictitious photon mass 
$M_A=M_W$. The same prescription is adopted at the two-loop level.

The unpolarized squared matrix element for  $\qbar q'\to W^\si g$,
including NLL terms 
up to the two-loop level, has the general form 
\beq\label{generalamplitudetwo}
\overline{\sum}|\Mqq_2|^2=
8 \pi^2
{\alpha\alpha_\rS}
(N_\rc^2-1)
\frac{\that^2+\uhat^2}{\that\uhat} 
\left[
A^{(0)}
+\left(\frac{\alpha}{2\pi}\right)A^{(1)}
+\left(\frac{\alpha}{2\pi}\right)^2 A^{(2)}
\right]
.
\eeq
The Born contribution reads 
\beq\label{bornresult}
 A^{(0)}=
\frac 1 {2\sw^2}
. 
\eeq
At one loop, the NLL part consists of double- and single-logarithmic
terms and reads
\beq\label{oneloopresult}
A^{(1)}\NLLa - 
\frac 1 {2\sw^2}
\left[\cew_{q_\rL}\left(\logar{2}{\shat}-3\logar{}{\shat}\right)
+
\frac{C_{\mathrm{A}}}{2 \sw^2}
\left(\logar{2}{\that}+\logar{2}{\uhat}-\logar{2}{\shat}\right)
\right]
.
\eeq
Here we used the shorthand  $\logar{k}{\rhat}=\ln^k(|\rhat|/ M_W^2)$
for the logarithms and 
$\cew_{q_\rL}=Y_{q_\rL}^2/(4\cw^2)+C_{\mathrm{F}}/\sw^2$
are the eigenvalues of the electroweak Casimir operator 
for left-handed quarks.
This expression is consistent with the process-independent results of
\citere{Denner:2001jv} as well as with the 
NLL part  of the one-loop asymptotic expressions presented in 
\refse{se:NNLL}.
At two loops we obtain
\beqar\label{twolooplogs}
A^{(2)}&\NLLa&
\frac 1 {2\sw^2}
\Biggl\{
\frac{1}{2}
\left(\cew_{q_\rL}+\frac{C_{\mathrm{A}}}{2\sw^2}\right)
\Biggl[\cew_{q_\rL}\left(\logar{4}{\shat}-6\logar{3}{\shat}\right)
+
\frac{C_{\mathrm{A}}}{2\sw^2}
\left(\logar{4}{\that}+\logar{4}{\uhat}-\logar{4}{\shat}\right) 
\Biggr]
\nl&&{}
+\frac{1}{6}
\Biggl[
\frac{b_1}{\cw^2}\left(\frac{Y_{q_\rL}}{2}\right)^2
+\frac{b_2}{\sw^2} \left(
C_{\mathrm{F}}
+\frac{C_{\mathrm{A}}}{2}\right)
\Biggr]\logar{3}{\shat}
\Biggr\},
\eeqar
where $b_1=-41/(6\cw^2)$ and $b_2=19/(6\sw^2)$ are the one-loop $\beta$-function coefficients associated with the U(1) and SU(2) couplings, respectively.
The LLs as well as the angular-dependent subset of the NLLs in \refeq{twolooplogs}, \ie all contributions of the form $\logar{4}{\rhat}$
with $\rhat=\shat,\that,\uhat$, have been derived from \citere{Denner:2003wi}.
There, by means of a diagrammatic two-loop calculation
in the spontaneously broken electroweak theory, it was shown that such two-loop terms result from the exponentiation of the corresponding one-loop corrections. 
The additional NLLs of the form  $\logar{3}{\shat}$ in \refeq{twolooplogs}
have been obtained via a fixed-order expansion
of the process-independent resummed expression proposed  in \citere{Melles:2001gw}.
This resummation \cite{Melles:2001gw} relies on the 
assumption that effects from spontaneous breaking of the
SU(2)$\times$U(1) symmetry  can be neglected in the high-energy limit.

Our NNLO predictions include only the LL and NLL terms. 
Thus they are affected by a potentially
large theoretical uncertainty, due to missing subleading contributions
of order $\alpha^2\ln^k(\shat/M_W^2)$ with $k=2,1,0$.
For four-fermion scattering it was found that, at $\shat\sim 1\TeV^2$,
the two-loop logarithmic expansion has an oscillating behaviour
characterized by large cancellations between leading and subleading
terms \cite{Jantzen:2005az}. In this case the subleading terms play a
very important role and the NLL approximation yields misleading results.
In contrast, in the case of $Wj$ production, the relative weight
of the LL, NLL and NNLL contributions at one loop indicates a
fairly good convergence of the logarithmic expansion.
Indeed, as can be seen from our numerical results in
\refse{se:numerics}, the one-loop corrections are clearly dominated by
the negative LL contributions, while the NLL terms are relatively
small and the NNLL contributions almost negligible.
A similar convergence is expected also at two-loops, owing to the
exponentiation property of the logarithmic corrections. Thus our NLL
two-loop predictions can be regarded as a plausible estimate of the
size of the two-loop electroweak effects in high-$\pT$ $W$-boson
production.

\section{Real corrections}
\label{se:real}

\newcommand{\ise}{{initial-state emitter}}
\newcommand{\iss}{{initial-state spectator}}
\newcommand{\fse}{{final-state emitter}}
\newcommand{\fss}{{final-state spectator}}
\newcommand{\mM}{{{\sf M}}}

In order to cancel IR singularities from the 
virtual-photon 
corrections, real emission corrections need to be calculated.
As discussed in \refse{se:symmetries},
all relevant partonic reactions are related to
the $\qbar q' \to W^\si g \gamma$  process
through crossing and CP symmetry.
\begin{figure}
  \begin{center}
\epsfig{file=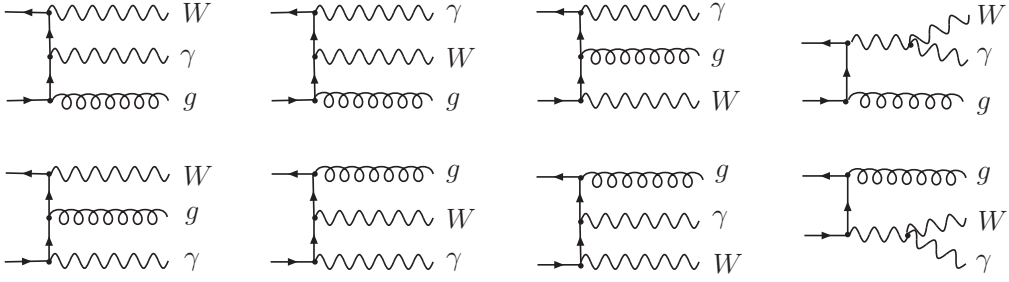, angle=0, width=0.9\textwidth}
\end{center}
\vspace*{-2mm}
\caption{Tree-level diagrams for the process $\bar q q' \to W g \gamma$.}
\label{fig:real_diag}
\end{figure}
The tree-level diagrams for this process are shown in \reffi{fig:real_diag}. 

The squared matrix element  for the  $2\to3$ process 
$a b \to W^\si k \gamma$, summed over polarization
and colour as well as averaged over initial-state
polarization, can be written in a general form~\cite{NLOrealcorrexp}
\beqar
\label{MReal}
    \overline{\sum}| \M^{a b \to W^\si k \gamma}_{0}|^2
    &=&4\pi\alpha \left[  
    -Q_q Q_{q'} H^{1,ab}_\rr (\shat,\that,\uhat,\tphat,\uphat) \right.\nl 
    &+& \left.\si Q_q    H^{2,ab}_\rr (\shat,\that,\uhat,\tphat,\uphat)
    -\si Q_{q'} H^{3,ab}_\rr (\shat,\that,\uhat,\tphat,\uphat)
    \right],
\eeqar
with the kinematical invariants defined in~\refse{sse:kinematics}.
In the limit of 
soft and/or collinear photon emission, the squared
matrix element \refeq{MReal} exhibits IR singularities. 
To combine these singularities with those originating from virtual corrections
we have to extract them in analytic form.
This is done with the help of the dipole subtraction formalism~\cite{Dittmaier:1999mb,Catani:1996vz,Catani:2002hc}.
Within this framework the partonic differential cross section can be
schematically written as 
\beq
\label{dip:gen}
\frac{\rd \hat{\si}^{a b \to W^\si k \gamma}}{\rd \pT} = \mathcal{N}_{ab}
\int \rd \Phi_3  
\left[ \mM^{a  b} (\Phi_3)
-  \mM^{a b}_{\rm sub} (\Phi_3) \right]  
+ \frac{\rd \hat{\si}^{a b}_{\rm A}}{\rd \pT}, 
\eeq
with $\mathcal{N}_{ab}$ given in \refeq{fluxfactor}.
The quantity  $\mM^{a  b}$ reads
\beq
 \mM^{a  b}(\Phi_3) = \overline{\sum} | \M^{a b \to W^\si k \gamma}_{0}|^2 F_{\rm O,3} (\Phi_3) \,.
\eeq
The auxiliary function $\mM^{a b}_{\rm sub}$ is chosen such that it has the same
singular  behaviour as $\mM^{a  b}$
in the soft and collinear limits. This ensures that the difference 
$\mM^{a  b} -  \mM^{a b}_{\rm sub}$ can be integrated numerically. 
To compensate for the subtraction, the
integral of the auxiliary function  $\mM^{a b}_{\rm sub}$, denoted
here 
${\rd \hat{\si}^{a b}_{\rm A}}/{\rd \pT}$, 
is then
added back. The analytical form of 
${\rd \hat{\si}^{a b}_{\rm A}}/{\rd \pT}$ 
is obtained after
performing integration over the subspace of the radiated photon.
The result of this one-particle subspace integration 
contains singular contributions  which must be combined with those in
virtual corrections.
The algorithms for constructing the auxiliary subtraction function and
its integrated counterpart have been developed both for the case
of photon radiation off massless or massive
fermions~\cite{Dittmaier:1999mb} and QCD radiation off massless~\cite{Catani:1996vz} or massive
partons~\cite{Catani:2002hc}. 
In \refses{se:massreg} and \ref{se:dimreg} we discuss the application of both
formalisms to calculate the $\ord(\al)$ real corrections to the $Wj$ production process. 
In both approaches we use expressions 
for the emission off a massive fermion to describe the emission off a $W$ boson, 
since 
only soft singularities are present in this case and they depend only on 
the charge of the external particle and not on its spin. 

After adding the real and virtual corrections, collinear
singularities remain. 
Final-state singularities are avoided by recombining collinear
photon-quark configurations as discussed in \refse{sse:kinematics}.
Initial-state singularities are absorbed in the definition of PDFs 
using the $\MSBAR$ scheme.

\subsection{Mass regularization}
\label{se:massreg}

The formalism of \citere{Dittmaier:1999mb} employs small photon and
fermion masses to regularize soft and collinear singularities.
The subtraction term for the squared matrix element is constructed from the 
appropriate dipole factors. Keeping the original notation 
of \citere{Dittmaier:1999mb}
we can write for the process $ab \to W^\si k \ga$ (where $a$ ($b$) can be $\qbar,q',g$)
\newcommand{\gsub}[2]{g^{\mathrm{sub}{#1}}_ {#2}}
\newcommand{\TPHI}[1]{\tilde{\Phi}_{#1}}
\beqar
\label{MDip}
    \mM^{a b}_{\rm sub} (\Phi_3) =
    -4\pi\alpha \sum_{\tau=\pm}\Bigg\{ \Bigg(
    &Q_a Q_b &
    \gsub{}{ab,\tau}(p_a,p_b,p_\ga)\;  \mM_0^{a'b'}\!\left(\TPHI{2,ab}\right)
    \nl 
    - &Q_a \si &
    \gsub{}{a W,\tau}(p_a,p_W,p_\ga)\; \mM_0^{a'b'}\!\left(\TPHI{2,aW} \right)    \nl 
    -&\si Q_a &
    \gsub{}{W a,\tau}(p_W,p_a,p_\ga)\; \mM_0^{a'b'}\!\left(\TPHI{2,Wa} \right)
    \nl  
    - &Q_a Q_k & 
    \gsub{}{ak,\tau}(p_a,p_k,p_\ga)\; \mM_0^{a'b'}\!\left(\TPHI{2,ak}\right)
    \nl
    - &Q_k Q_a & 
    \gsub{}{ka,\tau}(p_k,p_a,p_\ga)\; \mM_0^{a'b'}\!\left(\TPHI{2,ka}\right)
    +  (a\lrar b)
    \Bigg)\Bigg| _{\{a'=a,\;b'=b\}} 
    \nl 
    + &Q_k \si &
    \gsub{}{k W,\tau}(p_k,p_W,p_\ga)\; \mM_0^{a b}\!\left(\TPHI{2,kW} \right)  
    \nl 
    + &\si Q_k &
    \gsub{}{W k,\tau}(p_W,p_k,p_\ga)\; \mM_0^{a b }\!\left(\TPHI{2,Wk}
    \right) \Bigg\}
    \;,
\eeqar
with 
\beq
\label{m0fo2}
\mM^{ab}_0 (\TPHI{2,nm}) =  
\overline{\sum} |\M_0^{a b \to W^\si k} (\TPHI{2,nm})|^2 F_{\rm O,2}(\TPHI{2,nm}) \,. 
\eeq
Due to $Q_g=0$ the dipole terms with gluon
indices do not contribute to  
\refeq{MDip}
and for each subprocess
the subtraction term $\mM^{a b}_{\rm sub}$ is constructed from six dipole 
terms, characterised by the $\gsub{}{}$ functions. Expressions for these
functions are taken directly from~\citere{Dittmaier:1999mb}. 
In \refapp{app:diptables} (see Table~\ref{table:massreg})
we list all the functions which are used to calculate~(\ref{MDip}), together
with the corresponding equation numbers in~\citere{Dittmaier:1999mb}.

For each subprocess the six dipole terms fall into three groups,
each containing two dipole terms and coming with a 
specific charge 
combination, either $-Q_q Q_{q'}$ or $\si Q_q$, or $-\si Q_{q'}$. The
subtraction term $\mM^{a b}_{\rm sub}$ has then the same structure as 
$\mM^{ab}$ in~\refeq{MReal} and the IR-singular part of the virtual
corrections~(\ref{msinggeneralresult2}). 
Thus the cancellation of singularities can be analyzed for
each charge combination separately.

The construction of the reduced phase space $\TPHI{2,nm}$ follows the
prescriptions of~\citere{Dittmaier:1999mb}. 
Generally $\TPHI{2,nm}$ is a mapping from the 
3-particle phase space into a 2-particle phase space. The
mapping respects all mass shell conditions. 
For different types of dipoles, different
mappings are necessary. 
In Table~\ref{table:massreg} we list numbers 
of equations in~\citere{Dittmaier:1999mb} 
which we used to perform mapping for the
dipole terms appearing in our calculations. In particular, the observable-defining 
function  $F_{\rm O, 2} (\TPHI{2,nm})$  in~(\ref{m0fo2}) is then
\beq
\label{fo2T}
F_{\mathrm{O}, 2}(\TPHI{2,nm})= \delta (\pT - \tilde p_{\rT, W}) 
\theta (\tilde \pTj -\pTminj)
\,,
\eeq
with $\tilde p_{\rT, W}$ and $\tilde \pTj$ belonging to $\TPHI{2,nm}$.

\newcommand{\Gsub}[2]{G^{\mathrm{sub}}_{\mathrm{#1}}(#2)}
\newcommand{\cGsub}[2]{\mathcal{G}^{\mathrm{sub}}_{\mathrm{#1}}(#2)}
\newcommand{\rI}{\mathrm{I}}
\newcommand{\rFM}{\mathrm{FM}}
\newcommand{\sigpt}{\frac{ \rd \hat \sigma^{a b\to W^\si k}}{ \rd \pT}}
\newcommand{\sigptP}{\frac{ \rd \hat \sigma^{a b\to W^\si k}_{\fwd}}{ \rd \pT}}
\newcommand{\sigpta}{\frac{ \rd \hat \sigma^{a b, \,\rm asym}_0}{ \rd \pT}}

The expression for the subtraction term integrated over the
phase space of the photon reads
\newcommand{\tauh}{\hat{\tau}}
\newcommand{\tauzero}{0}
\beqar
\label{MiDip}
\frac{\rd \hat{\si}^{a b}_{\rm A}}{\rd \pT} 
=  -\frac{\alpha}{2\pi} 
     \Bigg \{
     2 Q_a Q_b &\Bigg[& 
     \Gsub{I,I}{\rhat_{ab}} \; \sigptP \left(\shat,\pT\right)
     \nl &+&
     \int_{\tauzero}^1 \rd x 
     \left[ \cGsub{I,I}{\rhat_{ab},x} \right]_+ 
     \; \sigptP \left(x \shat,\pT\right) + (a \lrar b)
     \Bigg]  
     \nl - Q_a \si &\Bigg[ &  
     \Gsub{I,FM}{\rhat_{aW}} \; \sigptP (\shat,\pT) 
     \nl &+&
     \int_{\tauzero}^1 \rd x  
     \left[ \cGsub{I,FM}{\rhat_{aW}(x),x} \right]_+ 
     \; \sigptP(x \shat,\pT) + (a \lrar b)
     \Bigg]  
     \nl - Q_b \si &\Bigg[ &  
     \Gsub{I,FM}{\rhat_{bW}} \; \sigptP (\shat,\pT) 
     \nl &+&
     \int_{\tauzero}^1 \rd x  
     \left[ \cGsub{I,FM}{\rhat_{bW}(x),x} \right]_+ 
     \; \sigptP(x \shat,\pT) + (a \lrar b)
     \Bigg]  
     \nl - Q_a Q_k &\Bigg[ & 
     \Gsub{I,F}{\rhat_{ak}} \; \sigptP(\shat,\pT)   
     \nl &+&
     \int_{\tauzero}^1 \rd x 
     \left[ \cGsub{I,F}{\rhat_{ak}(x),x} \right]_+ 
     \; \sigptP(x \shat,\pT)
     + (a \lrar b) \Bigg]
     \nl - Q_b Q_k &\Bigg[ & 
     \Gsub{I,F}{\rhat_{bk}} \; \sigptP(\shat,\pT)   
     \nl &+&
     \int_{\tauzero}^1 \rd x 
     \left[ \cGsub{I,F}{\rhat_{bk}(x),x} \right]_+ 
     \; \sigptP(x \shat,\pT)
     + (a \lrar b) \Bigg]  
     \nl +  Q_k \si &\Bigg[& 
     \Gsub{F,FM}{\rhat_{kW}} \; \sigptP (\shat,\pT)
     + (a \lrar b) \Bigg]      
     \Bigg\} \,.
\eeqar
The relevant invariants in~(\ref{MiDip}) are defined as 
$\rhat_{ab} = (p_a+p_b)^2 = \shat$, $\rhat_{kW} = (p_k + p_W)^2 =
\shat$, 
and
\beqar
\label{KiniDip}
   \rhat_{aW}(x)&=&(x p_a-p_W)^2,
   \nl
   \rhat_{ak}(x)&=&(x p_a-p_k)^2,
   \nl
   \rhat_{nl}&=&\rhat_{nl}(1).
\eeqar
The terms proportional to 
$\rd\hat\si^{ab\to W^\si k}_\fwd/\rd \pT$ in \refeq{MiDip}
represent the contributions originating from the forward hemisphere 
in the 2-particle phase space [see \refeq{partoniccs1}-\refeq{partoniccs1b}].
The $(a \lrar b)$ terms are the 
contributions from the backward hemisphere, and
\beqar\label{partoniccsfb}
\frac{\rd \hat{\si}_{\fwd}^{a b\to W^\si k}}{\rd \pT}
\Bigg|_{a\leftrightarrow b}
=
\frac{\rd \hat{\si}_{\bkwd}^{a b\to W^\si k}}{\rd \pT}
.
\eeqar
Note that the first argument of 
$\rd\hat\si^{ab\to W^\si k}_\fwd/\rd \pT$ in \refeq{MiDip} directly 
indicates the $x$-dependence 
of the actual values of the $\that$, $\uhat$ invariants defined in~\refeq{tuform}. 
The plus-distributions appearing in eq.~(\ref{MiDip}) are
evaluated according to the prescription
\beqar
\lefteqn{\int_{\tauzero}^1 \rd x  
   \left[ \cGsub{}{\rhat(x),x} \right]_+
     \frac{\rd\hat\si}{\rd\pT}\left(x \shat,\pT\right)
   =}\quad &&\nl&=& 
\int_0^1 \rd x  \Bigg[ \cGsub{}{\rhat(x),x}
     \frac{\rd\hat\si}{\rd\pT}\left(x \shat,\pT\right) 
     \theta\left( x-\tauh \right)
    - 
   \cGsub{}{\rhat(1),x} 
     \frac{\rd\hat\si}{\rd\pT}\left(\shat,\pT\right)
   \Bigg],
\nl   
\eeqar
where 
$ \tauh=\left(\pT +\sqrt{\pT^2+ M_W^2}\right)^2\!\Big/ \shat $ guarantees the minimal 
center-of-mass energy to produce the final state. 
The expressions for the functions $G^\mathrm{sub}$ and
$\mathcal{G}^\mathrm{sub}$ follow directly%
\footnote{The function $\Gsub{\rm F,FM}{\rhat}$ has been derived from eq.~(4.10) in \citere{Dittmaier:1999mb} 
by taking the limit of an infinitesimal quark mass. }
 from the results in \citere{Dittmaier:1999mb}.
For the functions  $G^\mathrm{sub}$ they read
\beqar
\label{Gfcts}
\Gsub{\rI,\rI}{\rhat} &=& \frac{1}{2} \Bigg[
\mathrm{Re}\left(\loopsing{\CLa}{,\mathrm{MR}}\right)
   +\ln^2\left(\frac{\rhat}{M_W^2}\right) - 3\ln\left(\frac{\rhat}{M_W^2}\right)
   -\frac{2}{3}\pi^2+4 \Bigg],
   \\
   \Gsub{\rI,\rFM}{\rhat} &=&
   \mathrm{Re}\left(\loopsing{\CLb}{,\mathrm{MR}}\right)\Big|_{\that \leftrightarrow \rhat} 
   +2\ln^2\left(1-\frac{\rhat}{M_W^2}\right)
   -\ln^2\left(2-\frac{\rhat}{M_W^2}\right)
   \nl &&
   +\ln\left(1-\frac{\rhat}{M_W^2}\right)        
   \left( \frac{M_W^4}{\rhat^2}-\frac{3M_W^2}{\rhat} -3 \right)
   -2{\rm Li}_2\left(\frac{M_W^2}{2 M_W^2-\rhat}\right)
   \nl && 
   +2{\rm Li}_2\left(\frac{\rhat}{2M_W^2-\rhat}\right)
   -2{\rm Li}_2\left(\frac{-\rhat}{2M_W^2-\rhat}\right)
   +\frac{M_W^2}{\rhat}+\frac{\pi^2}{6}+\frac{1}{2} ,
   \\
   \Gsub{\rI,\rF}{\rhat} &=& 
   \mathrm{Re}\left(\loopsing{\CLa}{,\mathrm{MR}}\right)\Big|_{\shat \leftrightarrow \rhat} 
   +\ln^2\left(\frac{-\rhat}{M_W^2}\right) - 3\ln\left(\frac{-\rhat}{M_W^2}\right)
   -\frac{\pi^2}{3}+\frac{1}{2},
   \\
   \Gsub{\rF,\rFM}{\rhat} &=& 
   \mathrm{Re}\left(\loopsing{\CLb}{,\mathrm{MR}}\right)\Big|_{\that \leftrightarrow \rhat} 
   +\ln^2\left(\frac{\rhat}{M_W^2}-1\right) 
   +\ln^2\left(1-\frac{M_W^2}{\rhat}\right)
   \nl &&
   +\frac{1}{2}\ln\left(\frac{\rhat}{M_W^2}\right)
   -\frac{7}{2}\ln\left(\frac{\rhat}{M_W^2}-1\right)
   -\frac{3}{2}\ln\left(\frac{\sqrt{\rhat}-M_W}{\sqrt{\rhat}+M_W}\right)
   \nl &&+
   4{\rm Li}_2\left(\frac{M_W^2}{\rhat}\right) -4{\rm Li}_2\left(\sqrt{\frac{M_W^2}{\rhat}}\right)
   +\frac{M_W^2}{2\rhat}-\frac{2\pi^2}{3}+3
   .
\eeqar
The IR-singular functions $\loopsing{i}{,\mathrm{MR}}$
cancel against those in the virtual corrections, cf. \refse{se:masssing}. 
The explicit 
forms of the functions $\mathcal{G}^\mathrm{sub}$ are
\newcommand{\Pff}[1]{\frac{1+#1^2}{1-#1}}
\beqar
\label{cGfcts}
   \cGsub{\rI,\rI}{\rhat,x} &=&
   \chi(x) + \Pff{x}\left\{ \ln\left(\frac{\rhat}{\mu_{\rm QED}^2}\right)+2\ln\left(1-x\right) \right\}+1-x  ,
   \\
   \cGsub{\rI,\rFM}{\rhat,x} &=&
   \chi(x) + \Pff{x}\Bigg\{ \ln\left(\frac{M_W^2-\rhat}{x \mu_{\rm QED}^2 }\right) 
   +\ln\Big( (1-x)(1-z_1(\rhat,x)) \Big) \Bigg\}
   \nl &&
   +\frac{z_1(\rhat,x)-1}{2(1-x)} \left(3+z_1(\rhat,x)-\frac{4M_W^2 x}{(\rhat-M_W^2)(1-x)} \right)+1-x,
   \\
   \cGsub{\rI,\rF}{\rhat,x} &=& 
   \chi(x) + \Pff{x}\left\{ \ln\left(\frac{-\rhat}{x \mu_{\rm QED}^2}\right)+\ln\left(1-x\right) \right\}
   \nl &&
   -\frac{3}{2(1-x)} +1-x ,
\eeqar
with 
\beq
\label{z1}
z_1(\rhat,x)=\frac{M_W^2 x}{M_W^2-(1-x)\rhat }
\eeq
and
\beq
\label{chifct}
   \chi(x)= \frac{1+x^2}{1-x} \left\{
   \ln\left(\frac{\mu_{\rm QED}^2}{m^2}\right)-2\ln\left(1-x\right)-1 
   \right\},
\eeq
where $\mu_{\rm QED}$ is the factorization scale and $m$ stands for
the quark-mass regulator.
The functions $\chi(x)$ are singular. These singularities are related to
the collinear photon radiation off an initial-state quark and are absorbed
in the definition of the PDFs, yielding the hadronic cross section
finite. The procedure bears complete
analogy to absorbing collinear QCD singularities into the
definition of the PDFs. In the \msbar~factorization scheme,
the redefinition is achieved by 
replacing~\cite{Dittmaier:2001ay}
\beq
\label{pdfnlo}
   f_{h,q} (x,\mu^2_{\rm QCD}) \to
   f_{h,q} (x,\mu^2_{\rm QCD},\mu^2_{\rm QED})
   -\frac{\alpha}{2\pi} Q_q^2 
   \int_x^1 \frac{\rd z}{z} \; f_{h,q}(\frac{x}{z},
    \mu^2_{\rm QCD},\mu^2_{\rm QED}) \; \left[ \chi(z) \right]_+ \;.
\eeq

\subsection{Dimensional regularization}
\label{se:dimreg}

In an independent calculation we used the results of
\citeres{Catani:1996vz,Catani:2002hc} to evaluate 
the dipole subtraction terms and
their integrated counterparts. The formalism
of \cite{Catani:1996vz,Catani:2002hc} is concerned with QCD radiation and 
expressions for dipoles are given as matrices in
colour and helicity space. Since we consider photon emission
off a fermion line, the colour and helicity structure disappears
and the dipole matrices reduce to simple expressions.
More precisely, to adapt the formalism~\citeres{Catani:1996vz,Catani:2002hc} for
the calculation of  
QED corrections, we make use of expressions describing 
gluon radiation off a fermion line in~\citeres{Catani:1996vz,Catani:2002hc} and 
replace
\beq
\label{QED:replace}
\alpha_\rS\to \alpha,\  {\bf T}_i \to \si_i Q_i, \ \ C_\mathrm{F}\to Q_i^2, \ \ T_\mathrm{R}\to 1,\ \
C_\mathrm{A} \to 0 \;,
\eeq
where ${\bf T}_i$ 
indicates the colour of the emitting parton, $Q_i$ is the 
electric charge in units of the positron charge for this parton,
and $\si_i=+1$ $(-1)$ for incoming (outgoing) partons.
Adopting notation analogous to~\citeres{Catani:1996vz,Catani:2002hc}, the subtraction term
for the process $ ab \to W^\si k \ga$ can be then
written
\newcommand{\Cdip}[2]{\cD^{#1}_{{#2}\;{\rm QED}}}
\beqar
\label{MDip:dimreg}
\mM^{a b}_{\rm sub}(\Phi_3)  &=& 
\left[  
\Cdip{a \ga,b}{}
+
\Cdip{a \ga}{W,}
+ \Cdip{a}{\ga W,}
+
\Cdip{a \ga}{k,}
+ \Cdip{a}{\ga k,}
+ (a \lrar b) 
\right]\nl
&&{}+
\Cdip{}{\ga k, W,}
+ \Cdip{}{\ga W,k,}
\,,
\eeqar
where 
\beq
\label{dip:def:dr}
\Cdip{I}{F,}= 
F_{\rm O,2} (\TPHI{2,nm}) \,\cD^{I}_{F}
(p_W,\,p_k,\,p_\ga;\, p_a,\, p_b) 
\Big|_{\tiny{\rm replacements~of~eq.~(\ref{QED:replace})}}.
\eeq
It is understood in eq.~(\ref{MDip:dimreg}) that dipole subtraction
terms with a gluon index do not contribute to $\mM^{a b}_{\rm sub}$.
In a complete analogy to eq.~(\ref{MDip}), for any initial state $ab$ 
the expression for $\mM^{a b}_{\rm sub}$ is
constructed from six dipole subtraction terms  $\Cdip{I}{F,}$, each
associated with one of the three possible charge combinations $-Q_q
Q_{q'}$, $\si Q_q$ or $-\si Q_{q'}$.  
The dipole subtraction functions  $\cD^{I}_{F}$ are taken directly
from~\citeres{Catani:1996vz,Catani:2002hc}. 
A list of the functions $\cD^{I}_{F}$ used to
calculate the subtraction term $\mM^{a b}_{\rm sub}$
in~(\ref{MDip:dimreg}), 
together with the corresponding equation numbers 
in~\citeres{Catani:1996vz,Catani:2002hc}, 
is presented in Table~\ref{table:dimreg}, \refapp{app:diptables}. 
Additionally, for each dipole subtraction
term appearing in~(\ref{MDip:dimreg}) we include a
description of its type.
The 
mappings from $\Phi_3$ to $\TPHI{2,nm}$
agree between the formalism of~\citeres{Catani:1996vz,Catani:2002hc}
and~\cite{Dittmaier:1999mb}. 
However, for the sake of completeness, Table~\ref{table:dimreg}
contains  numbers of equations which provide mapping formulae
in~\citeres{Catani:1996vz,Catani:2002hc}. 
The function $F_{\rm O,2}$ in~\refeq{dip:def:dr} is given by expression~\refeq{fo2T}.

Moreover, apart from the \fse, \fss~case, 
\ie 
the dipoles
$\Cdip{}{\ga k, W,}$ and $\Cdip{}{\ga W,k,}$,
there 
is a direct correspondence between the dipole
subtraction terms in the two formalisms of the form 
\beqar
&& 
\Cdip{a \ga, b}{}
\stackrel{{\veps \to 0}}{\to}  
-Q_a Q_b \; 4 \pi \al \sum_{\tau=\pm}
\gsub{}{a b,\tau } (p_a,p_b,p_\ga) \mM_0^{ab}(\TPHI{2, ab}),
\nl  
&& 
\Cdip{a \ga}{W,}
\stackrel{{\veps \to 0}}{\to}  
Q_a \si \; 4 \pi \al  \sum_{\tau=\pm}
\gsub{}{a W, \tau} (p_a, p_W, p_\ga) \mM_0^{ab}(\TPHI{2, aW}),
\nl  
&& 
\Cdip{a}{\ga W,}
\stackrel{{\veps \to 0}}{\to}  
Q_a \si \; 4 \pi \al \sum_{\tau=\pm} 
\gsub{}{W a,\tau }  (p_W, p_a, p_\ga) 
 \mM_0^{ab}(\TPHI{2,W a}), 
\nl
&& 
\Cdip{a \ga}{k,}
\stackrel{{\veps \to 0}}{\to}  
Q_a Q_k \; 4 \pi \al \sum_{\tau=\pm}
\gsub{}{a k, \tau} (p_a, p_k, p_\gamma) \mM_0^{ab}(\TPHI{2,ak}),
\nl  
&& 
\Cdip{a}{\ga k,}
\stackrel{{\veps \to 0}}{\to}  
Q_a Q_k \; 4 \pi \al \sum_{\tau=\pm}
\gsub{}{k a, \tau} (p_k, p_a, p_\gamma) \mM_0^{ab}(\TPHI{2,ka}).
\eeqar
\newcommand{\I}{\mathcal{I}}
\newcommand{\cK}{\mathcal{K}}
\newcommand{\cP}{\mathcal{P}}
\newcommand{\shatnm}{\shat_{nm}}
The subtraction term integrated over the photon phase space is constructed
according to 
\beqar
\frac {\rd \hat \si^{a b}_{\rm A}}{\rd \pT} 
&=&  
\frac{\al}{2 \pi} 
\Bigg[ \I^{ab}(\shat,\pT) \; +
\int_{\tauzero}^1 \rd x\; \Big(\cK^{ab}(x,\shat,\pT)+\cP^{ab}(x,\shat,\pT)\Big) \Bigg]
,
\eeqar
where the expressions for $\I,\cK$ and $\cP$ functions follow from
results for the integrated dipole functions in~\citeres{Catani:1996vz,Catani:2002hc} after
performing replacements of eq.~(\ref{QED:replace}). 
For the photonic corrections to any of the subprocesses
$ab \to W^\si k$ we can write
\beqar
\label{idipI:dimreg}
\I^{a b}(\shat,\pT) &=&
Q_a Q_b \left[ \tilde \I(\shat_{ab}) \sigptP (\shat,\pT)  + (a \lrar b) \right]
\nl
&&{}- Q_a \si \left[ \I' (\shat_{aW})\sigptP (\shat,\pT) + (a \lrar b) \right]
\nl
&&{}- Q_b \si \left[ \I' (\shat_{bW})\sigptP (\shat,\pT) + (a \lrar b) \right]
\nl
&&{}- Q_a Q_k \left[ \tilde \I (\shat_{ak}) \sigptP (\shat,\pT)+(a \lrar b) \right]
\nl
&&{}- Q_b Q_k \left[ \tilde \I (\shat_{bk}) \sigptP (\shat,\pT)+(a \lrar b) \right] 
\nl
&&{}+ Q_k \si \left[ \I'(\shat_{kW}) \sigptP (\shat,\pT)+(a \lrar b) \right]  
\,,
\eeqar
where  $\shatnm=2 p_n  p_m$, and we make use of $Q_g=0$. 
As in \refeq{MiDip}, the terms proportional to 
$\rd\hat\si^{ab\to W^\si k}_\fwd/\rd \pT$ 
originate from the forward hemisphere
and the $(a \lrar b)$ terms from the backward hemisphere.
The integrated dipole functions
in~(\ref{idipI:dimreg}) read
\beqar 
\label{idipI:detail}
\tilde \I(\shatnm)&=&
-\left( \frac{4 \pi \mu^2}{M_W^2}\right)^\veps \Gamma(1+\veps)\,
\mathrm{Re}\left( \loopsing{\CLa}{,\mathrm{DR}}\right)\bigg|_{\shat=\shatnm}
- 
\ln ^2\left(\frac{\shatnm}{M_W^2}\right)+3 \ln 
 \left(\frac{\shatnm}{M_W^2}\right)
\nl &&
+\frac{4 \pi ^2}{3}-6 \;,
\nl
\I'(\shatnm) &=& -\left( \frac{4 \pi \mu^2}{M_W^2}\right)^\veps
\Gamma(1+\veps)\,
\mathrm{Re}\left( \loopsing{\CLb}{,\mathrm{DR}}\right)\bigg|_{\that=M_W^2+\si_n \si_m \shatnm}
-\ln^2\left(\frac{\shatnm}{M_W^2}\right)
\nl
&&{}
-\ln \!
   \left(\frac{\shatnm}{M_W^2}\right) \!\! \left(\ln \!
   \left(\frac{\shatnm}{M_W^2+\shatnm}\right) \!\! -3\right)
-\ln \!
   \left(\frac{\shatnm}{M_W^2+\shatnm}\right)
\nl
&&{}
 + \ln \!
   \left(\frac{M_W^2}{M_W^2+\shatnm}\right) \ln \!
   \left(\frac{\shatnm}{M_W^2+\shatnm}\right)
+3 \ln \!
   \left(1-\sqrt{\frac{M_W^2}{M_W^2+\shatnm}}\right)
\nl
&&{}
+\frac{M_W^2}{\shatnm}
\ln \! \left(\frac{M_W^2}{M_W^2+\shatnm}\right)
+ 2 \mathrm{Li}_2\left(\frac{\shatnm}{M_W^2+\shatnm}\right)
+\frac{3 M_W}{\sqrt{\shatnm+M_W^2}+M_W}
\nl
&&{}
+\pi ^2-6 
\,.
\eeqar
The structure of 
the singular terms $\loopsing{i}{,\mathrm{DR}}$
in~(\ref{idipI:detail}) is
kept the same as in~(\ref{msinggeneralresult2}) to manifestly show cancellation of
singularities between virtual and real corrections.
 For the $x$-dependent functions we have 
\beqar
\label{kpfunctions:gen}
\cK^{a b}(x,\shat, \pT) =&& 
(Q_a^2 + Q_b^2)\,\left[   \bar{\cK}(x) \sigptP (x \shat, \pT) +(a \lrar b) \right]
\nl
&+& 2\, Q_a Q_b \,\left[   \tilde{\cK}(x)\sigptP (x \shat, \pT) +(a \lrar b) \right]
\nl
&-& Q_a \si \left[  \cK'(x,\shat_{aW}) \sigptP  (x \shat, \pT) +(a \lrar b)\right]
\nl
&-&
 Q_b \si \left[  \cK'(x,\shat_{bW}) \sigptP  (x \shat, \pT)  +(a \lrar b)\right]
\nl
& -& ( Q_a Q_k + Q_b Q_k) \, \left[  \cK''(x) \sigptP (x \shat, \pT) +(a \lrar b)\right] \,,
\nl
\cP^{ab}(x, \shat, \pT) =&& 
2 Q_a Q_b \,\left[  \cP(x,\shat_{ab}) \sigptP (x \shat, \pT)  +(a \lrar b)\right]
\nl
&-& Q_a \si\left[  \cP(x,\shat_{aW})\sigptP  (x \shat, \pT)  +(a \lrar b)\right]
\nl
&-& Q_b \si \left[ \cP(x,\shat_{bW})\sigptP  (x \shat, \pT)  +(a \lrar b)\right]
\nl 
&-& Q_a Q_k \left[ \cP(x,\shat_{ak})\sigptP  (x \shat, \pT)  +(a \lrar b)\right]
\nl
&-& Q_b Q_k \left[ \cP(x,\shat_{bk})\sigptP  (x \shat, \pT)  +(a \lrar b)\right]
\,,
\eeqar
with 
\beqar
\bar{\cK}(x) &=& 
P_{\rm reg}(x) \ln \left(\frac {1-x}{x} \right)+  (1-x) 
+ \left(  \frac{2}{1-x}\ln\left(\frac{1-x}{x}\right)\right)_+ 
\nl
&& -\delta(1-x) (5 -\pi^2) \;,\nl
\tilde{\cK}(x)&=& 
- P_{\rm reg}(x) \ln (1-x) - \left[ 2  \left(
\frac{\ln(1-x)}{1-x}\right)_+ - \frac{\pi^2}{3}  \delta(1-x)\right] \;,\nl
\cK'(x,\shatnm) &=& 
- 2 \left( \frac{\ln(1-x)}{1-x}\right)_+
+ 2  \frac{\ln(2-x)}{1-x} \nl
&&-
\left( \frac{1-x}{2\left(1-x+M_W^2/\shatnm\right)^2} - \frac{2}{1-x}
  \left[ 1+ \ln\left( 1-x + \frac{M_W^2}{\shatnm}\right)\right]\right)_+ \nl
&&-
\left(\frac{2}{1-x}\right)_+ 
\left[
\ln \left(2-x+ \frac{M_W^2}{\shatnm} \right) +\ln \left( \frac{(2-x) \shatnm}{(2-x)
  \shatnm+M_W^2} \right) 
\right] \nl
&&-
P_{\rm reg}(x) \ln\left(  \frac{(1-x) \shatnm}{(1-x) \shatnm+M_W^2} \right) \nl
&&-
 \delta(1-x)
\left[ -\frac{3}{2} + \frac{M_W^2}{\shatnm} \ln \left( \frac {M_W^2}{\shatnm+M_W^2} \right)
+ \frac{3 M_W}{ \sqrt{\shatnm+M_W^2} +M_W} 
\right. \nl
&&+
\left. 
 \frac{3}{2} \ln \left( \frac{\shatnm - 2 M_W\sqrt{\shatnm+M_W^2}
    +2M_W^2}{\shatnm} \right)
+\frac{1}{2}\frac {M_W^2}{\shatnm+M_W^2} 
\right],
\nl
\cK''(x) &=& \frac{3}{2} \left[ \left(\frac{1}{1-x}\right)_+ +
  \delta(1-x) \right] \;, \nl
\cP(x,\shatnm) &=& \left(\frac{1+x^2}{1-x}\right)_+ \ln \left( \frac{\mu_{\rm
    QED}^2}{x \shatnm} \right) \;,
\eeqar
and
\beq
P_{\rm reg}(x)= \Bigg(\frac{1+x^2}{1-x}\Bigg)_+ -
\Bigg(\frac{2}{1-x}\Bigg)_+ -\frac{3}{2}\delta(1-x) \;.
\eeq
Note that in contrast to eq.~(\ref{idipI:dimreg}), the quantity  $\shatnm$ in
(\ref{kpfunctions:gen}) can be
implicitly dependent on the fraction $x$. More precisely, 
it is the case if $\shatnm$ involves the momentum of a final-state particle. 
The final-state momentum belongs then to the phase space for which the 
squared center-of-mass
energy is $x \shat = 2 x p_a p_b$
\cite{Catani:2002hc}. 

The evaluation of the terms involving the plus-distribution is carried out
as indicated in~\citere{Catani:2002hc}, \ie according to
\beqar
\lefteqn{\int_0^1 \rd x   \left[ {\cal{R}} {(x,\shatnm(x))} \right]_+
     \frac{\rd\hat\si}{\rd\pT}\left(x \shat,\pT\right)
   =}\quad&&\nl&=& 
   \int_0^1 \rd x  \Bigg[
   {\cal{R}} {(x,\shatnm(x))}
    \frac{\rd\hat\si}{\rd\pT}\left(x \shat,\pT\right) 
    \theta(x-\tauh)
   -
      {\cal{R}} {(x,\shatnm(1))}
    \frac{\rd\hat\si}{\rd\pT}\left( \shat,\pT\right) \Bigg], 
\nl
\eeqar 
with 
$ \tauh=\left(\pT +\sqrt{\pT^2+ M_W^2}\right)^2\!\Big/ \shat $. 

In the formalism of~\citeres{Catani:1996vz,Catani:2002hc} the collinear counterterms 
associated with PDF renormalization
are included in the expressions for integrated dipole functions, \ie the
final results which we use are free 
from collinear singularities. 
The expressions presented here are calculated using the
\msbar~factorization scheme.

As can be seen from the presented formulae, the explicit expressions
for the integrated dipole functions in the two
formalisms are different.  
In particular, the expressions for the end-point
contributions have different forms due to specific conventions 
\wrt calculating the plus-distribution terms in the two formalisms.
However we have checked that, after subtraction of the IR singularities,
for each charge combination  apart from $\sigma
Q_k$ the integrated dipole contributions to 
$\rd \si^{ab}_{\rm A}/\rd \pT$
in the two formalisms are equivalent.

\section{Checks}
\label{se:checks}
Every part of the presented calculation has been performed in two 
completely independent ways. 
The algebraic reductions were done using two different 
\Mathematica~\cite{mathematica} codes. 
For the numerical evaluation we have implemented the results in two
independent {\sc Fortran} programs. 
Comparing the results at numerical level we find agreement within 
the statistical errors.

Furthermore, in order to control the correctness of our results
we performed various consistency checks. On the side of virtual corrections 
we have verified that the one-loop corrections \refeq{algebraicred}
satisfy the Ward Identity
\beqar\label{wardid}
\varepsilon^*_\mu(p_W)\,
p_{g\nu}\,
\bar{v}(p_{\qbar}) 
\left[\delta \A_{1,\mathrm{I}}^{\mu\nu}(M^2_V)
\omega_\la \right]
u(p_q)
&=&0
\quad
\mbox{for I=A,N,X,Y}. 
\eeqar
A similar Ward identity holds for the lowest-order amplitude%
\footnote{We note that the abelian one-loop contribution satisfies two additional 
Ward identities%
\beqar
p_{W\mu}\,
\varepsilon^*_\nu(p_g) \,
\bar{v}(p_{\qbar}) 
\left[\delta \A_{1,\mathrm{A}}^{\mu\nu}(M^2_V)
\omega_\la \right]
u(p_q)
=0
,\quad
p_{W\mu} \,p_{g\nu}
\,
\bar{v}(p_{\qbar}) 
\left[\delta \A_{1,\mathrm{A}}^{\mu\nu}(M^2_V)
\omega_\la \right]
u(p_q)
=0
.\nonumber
\eeqar
Similar identities for the N-, X- and Y- form factors exist but are less trivial 
due to the non-vanishing contributions from would-be Goldstone bosons on the 
right-hand side.
This means that the calculation of the 
unpolarized cross section requires the use of the exact expression for the 
$W$-boson polarization sum. Instead, owing to \refeq{wardid},
the gluon polarization sum can be implemented as $-g_{\nu\nu'}$.
}.
The cancellation of the ultraviolet divergencies has been 
verified analytically and numerically. 
For the numerical evaluation of the loop integrals we use a set of
routines by A. Denner and, alternatively, the {\sc FF} library 
\cite{vanOldenborgh:1990yc}. 
The NLL approximation that was derived from the
full one-loop calculation, has been checked against results 
from the general derivation of NLL terms \cite{Denner:2001jv}.
Also the IR-singular contributions in the high-energy limit 
have been reproduced within this framework.

The squared matrix element for the real corrections was checked
numerically against {\sc MadGraph}~\cite{Stelzer:1994ta}. 
The cancellation of IR singularities between real and virtual corrections
was done analytically using the dipole formalism. 
The subtraction terms were derived and implemented in two different 
ways, using the mass regularization of IR singularities 
and the dimensional regularization. 
The phase-space integration for the real corrections was performed
with adaptive Monte-Carlo integration using 
{\sc VEGAS} \cite{vegas}.
Detailed comparisons at analytical and numerical level
were performed, and  
the agreement between the predictions generated 
within 
two different regularization
schemes  provided a strong check on the calculation of the real
corrections. 

\section{Numerical results}
\label{se:numerics}
In this section we present numerical predictions for the large-$\pT$
production of $W$ bosons at the hadron colliders
LHC and Tevatron.
The following input parameters are used:
 $G_\mu=1.16637\times10^{-5}\GeV^{-2}$,
$M_W=80.39\GeV$,
$M_Z=91.19\GeV$,
\mbox{$m_t=171.4\GeV$},
$M_H=120\GeV$.
For the numerical values of elements in the CKM quark
mixing matrix we refer to~\cite{Eidelman:2004wy}. 

The hadronic cross sections are obtained using LO MRST2001 PDFs~\cite{Martin:2002dr}.
We choose $\mu^2_{\rm QCD}=p^2_{{\rm T}}$ as the factorization scale
and, similarly, as the scale at which the 
strong coupling constant
is evaluated%
\footnote{Note that when calculating the contribution to the hadronic
cross section coming from the subtraction term in the real
corrections, we take the transverse momentum of the $W$ boson in the
reduced phase space, $\tilde p_{\rT,W}$, as the  factorization scale and the argument of 
$\al_\rS$.}.
We also adopt, in agreement with the value used in
the PDF analysis, the value $\al_\rS(M_Z^2)=0.13$ 
and use the one-loop running expression for $\al_\rS(\mu^2_\mathrm{QCD})$. 
In our calculations of the real corrections we choose the 
$\MSBAR$ factorization scheme with the scale $\mu^2_{\rm QED}=M_W^2$.
We note that in order to consistently include $\Oa$ corrections in a
calculation of a hadronic cross section, PDFs that are used in the
calculation need to take into account QED effects. Such PDF
analysis has been performed in~\cite{Martin:2004dh} and the $\Oa$ 
effects are known to be small for $\mu_{\mathrm{QED}}\lsim 100\GeV$, 
both concerning the change in the
quark distribution functions (below ${\cal O}(1\%)$~\cite{Roth:2004ti})
and the size of the photon distribution function. 
Moreover, the currently available PDFs incorporating $\Oa$ corrections,
MRST2004QED~\cite{Martin:2004dh}, include QCD effects at the NLO in
$\al_\rS$. Since our calculations are of the lowest order in QCD, and QED effects
on PDFs are estimated to be small for $\mu_{\mathrm{QED}}\lsim 100\GeV$,
we prefer to use a LO QCD PDF set without QED corrections
incorporated, rather than MRST2004QED, and we set
$\mu_{\mathrm{QED}}=M_W$\footnote{
The use of different factorization scales, $\mu_\mathrm{QCD}=\pT$ and
$\mu_\mathrm{QED}=M_W$, is due to the fact that $\mu_\mathrm{QCD}$ and
$\mu_\mathrm{QED}$ play a different role in our calculation.
The dependence on $\mu_\mathrm{QCD}$ is due to the LO evolution of the
PDFs and represents an effect of
$\mathcal{O}(\alpha_S\ln(\mu_\mathrm{QCD}/\mu_0))$, where $\mu_0$ is
the scale at which the PDF evolution starts. This dependence would be
compensated by NLO QCD contributions of
$\mathcal{O}(\alpha_S\ln(\pT/\mu_\mathrm{QCD}))$ and, although QCD
corrections are not included in our calculation, choosing
$\mu_\mathrm{QCD}=\pT$ we can absorb large NLO QCD logarithms of the
scale $\pT$ in the LO PDF evolution.
In contrast, the $\mu_\mathrm{QED}$ dependence of our predictions is
due to $\mathcal{O}(\alpha\ln(\pT/\mu_\mathrm{QED}))$ terms in the
photon bremsstrahlung corrections. This dependence is not compensated
by the PDF evolution since we use a PDF set that does not include QED
effects, assuming that these effects are negligible.
This approach makes sense only if the scale $\mu_\mathrm{QED}$ is
chosen in such a way that the (potential) impact of QED effects on the
PDFs is very small.  In \citere{Roth:2004ti} it was shown that the
QED corrections to the PDFs grow with $\mu_\mathrm{QED}$ but do not
exceed one percent for $\mu_\mathrm{QED}\lsim 100\GeV$.  This
motivates our choice $\mu_\mathrm{QED}=M_W$ for the QED factorization
scale.}. 
Moreover we do not include photon-induced contributions, which are
parametrically suppressed by a factor $\alpha / \alpha_{\mathrm{S}}$.
However, in the concurrent to this paper (and subsequent to
\citere{Kuhn:2007qc}), work of \citere{Hollik:2007qc}, it has been
reported that photon-induced contributions are of numerical
significance for large $\pT$ $W$-boson production at the
LHC. Estimates of the exact size of these effects are obscured by
large theoretical uncertainty on the photon's PDF, as demonstrated in
\citere{Hollik:2007qc}.

We choose the following values of the $\pT$-cuts: 
$\pTminj=100\GeV$ 
for LHC and $\pTminj=50\GeV$ for Tevatron. 
The value of the separation parameter
below which the recombination procedure is applied is taken to be 
$R_{\mathrm{sep}}=0.4$. 
The dependence of our predictions on $R_{\mathrm{sep}}$
is negligible. 
We have verified that the shift of the transverse-momentum
distribution induced by variations of this parameter in the range
$0.1 \le R_{\mathrm{sep}} \le 1.0$ does not exceed a few permille. 

Our lowest-order (LO) predictions result from \refeq{generalamplitude}. 
The next-to-leading order predictions (NLO)
include the LO$+$virtual contributions \refeq{generalresult} 
and the real bremsstrahlung \refeq{dip:gen}.
We also study the relative importance of the IR-finite parts of the
virtual ($\mathrm{NLO}_{\mathrm{virt}}$)
and real ($\mathrm{NLO}_{\mathrm{real}}$) contributions.
These IR-finite parts are constructed by subtracting the IR divergence
\refeq{msinggeneralresult2} from the virtual corrections and adding it to the real ones.
The next-to-leading-logarithmic (NLL) and  next-to-next-leading-logarithmic (NNLL)
predictions%
\footnote{For details concerning the treatment of angular-dependent logarithms at the NLL level we refer to \citere{Kuhn:2004em}. } 
are obtained adding to the LO the approximations
\refeq{oneloopresult} and \refeq{nnllstracture}
for the $\mathrm{NLO}_{\mathrm{virt}}$ part and neglecting
the $\mathrm{NLO}_{\mathrm{real}}$ part of the corrections.
As we will demonstrate, for the case of fully inclusive photon radiation
neglecting this piece provides a good approximation of the complete 
calculation.
The next-to-next-to-leading order predictions (NNLO) 
include the full NLO results plus the two-loop NLL 
corrections \refeq{twolooplogs}.

\begin{figure}[p]
\vspace*{2mm}
  \begin{center}
\epsfig{file=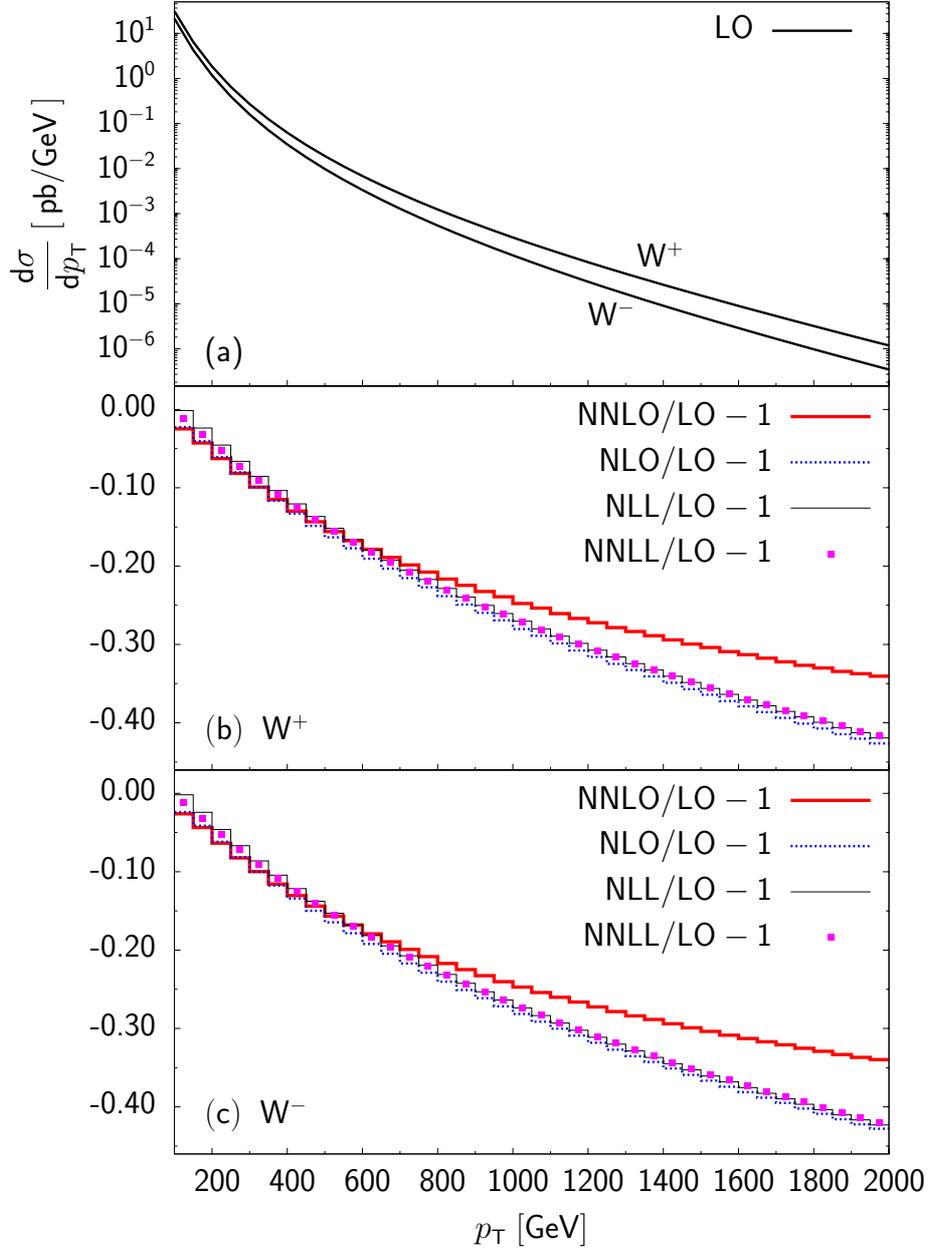, angle=0, width=12cm}
\end{center}
\vspace*{-2mm}
\caption{Transverse-momentum distribution for $W$-boson production at
the LHC.
(a) LO distribution for $pp\rar W^+ j$ and $pp\rar W^- j$. 
(b) Relative NLO (dotted), NLL (thin solid), NNLL (squares) and NNLO (thick solid)
electroweak correction \wrt the LO distribution for  $pp\rar W^+ j$.
(c) Relative NLO (dotted), NLL (thin solid), NNLL (squares) and NNLO (thick solid)
electroweak correction \wrt the LO distribution for  $pp\rar W^- j$.
}
\label{fig:ptwplusminuslhc}
\end{figure}
The LO transverse-momentum distributions for $p p \rightarrow W^+ j$
and $pp\rightarrow W^- j$ at the LHC are shown in
\reffi{fig:ptwplusminuslhc}a. In~\reffi{fig:ptwplusminuslhc}b and
~\reffi{fig:ptwplusminuslhc}c we plot the relative size of the NLO,
one-loop NLL, one-loop NNLL and NNLO corrections \wrt the LO
predictions for $W^+$ and $W^-$ production, respectively.
The behaviour of the relative corrections to $W^+$ and $W^-$
production is very similar.  As expected, the importance of the NLO
contribution increases significantly with $\pT$ and leads to a
negative correction ranging from $-15\%$ at \mbox{$\pT=500\GeV$} to
$-43\%$ at $\pT=2\TeV$.  We also observe that the one-loop NLL and
NNLL approximations are in good agreement (at the 1-2$\%$ level) with
the full NLO result for $\pT \ge 100\GeV$.
The difference between NLO and NNLO curves is significant.  The
two-loop terms are positive and amount to $+3\%$ at $\pT=1\TeV$ and
$+9\%$ at $\pT=2\TeV$. This shifts the
relative corrections for $W^+$ production up to $-25\%$ at $\pT=1\TeV$
and $-34\%$ at $\pT=2\TeV$.

\begin{figure}[p]
  \begin{center}
\epsfig{file=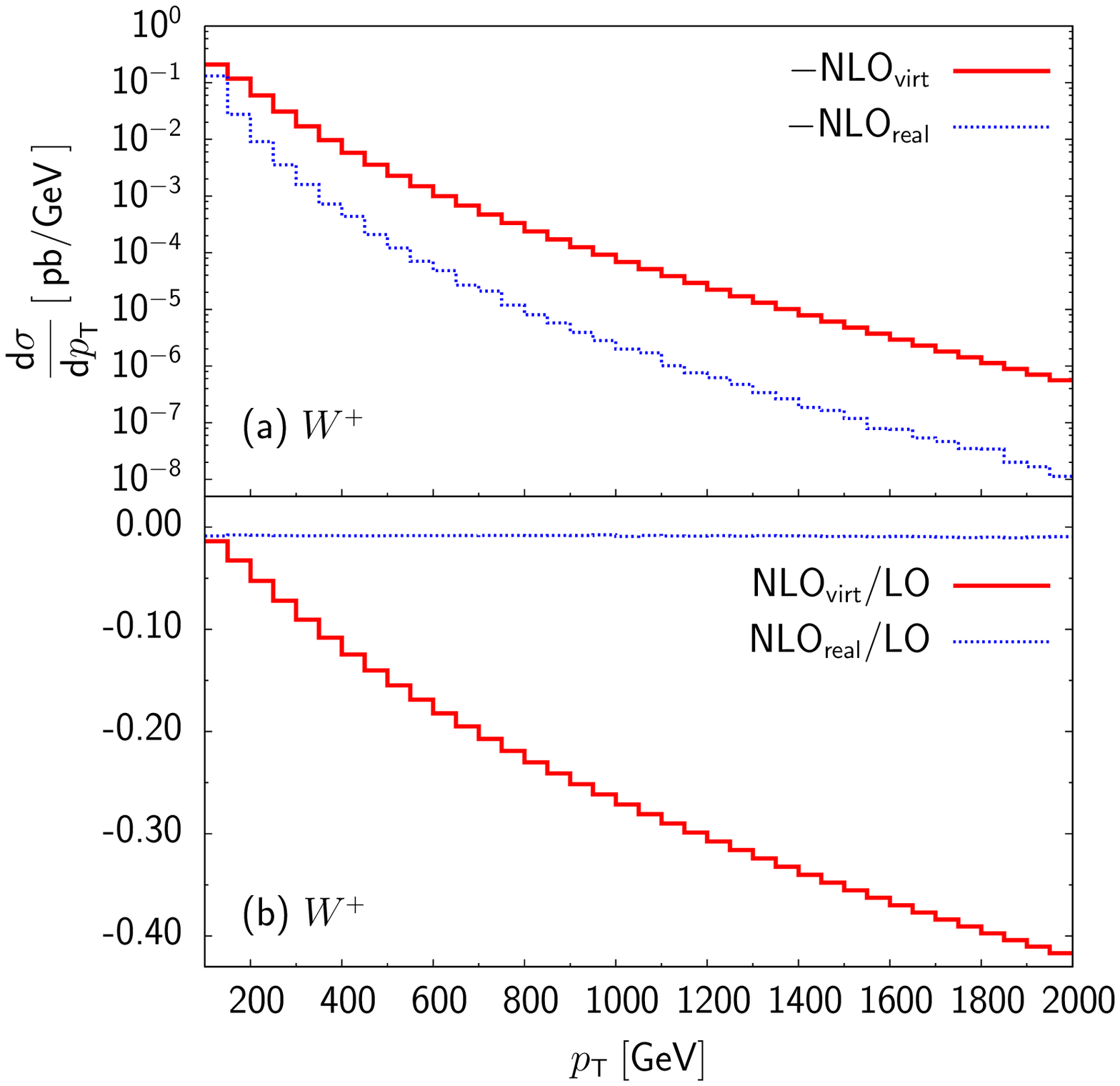, angle=0, width=12cm}
\end{center}
\vspace*{-2mm}
\caption{IR-finite parts of the 
virtual ($\mathrm{NLO}_{\mathrm{virt}}$)
and real ($\mathrm{NLO}_{\mathrm{real}}$) contributions
to the $\pT$-distribution of $W$ bosons in the process $pp\rar W^+
  j$ at $\sqrt{s}= 14\TeV$.} 
\label{fig:ptrealvirtlhc}
\end{figure}
The IR-finite parts of the virtual
($\mathrm{NLO}_{\mathrm{virt}}$) and real
($\mathrm{NLO}_{\mathrm{real}}$) corrections to $W^+$ production at
the LHC are shown separately in~\reffi{fig:ptrealvirtlhc}a.
\reffi{fig:ptrealvirtlhc}b shows the relative size of the
$\mathrm{NLO}_{\mathrm{virt}}$ and $\mathrm{NLO}_{\mathrm{real}}$
corrections \wrt the LO predictions.
The $\mathrm{NLO}_{\mathrm{virt}}$
contribution dominates the full NLO correction and amounts up to
$-42\%$ at $\pT=2\TeV$.
The $\mathrm{NLO}_{\mathrm{real}}$ part contributes with a smaller and
nearly constant correction of about $-1\%$ in the entire $\pT$-range.
This means that, for the case of fully inclusive photon
radiation, the $\mathrm{NLO}_{\mathrm{virt}}$ part represents a good
approximation of the full NLO correction.

\begin{figure}[p]
  \begin{center}
\epsfig{file=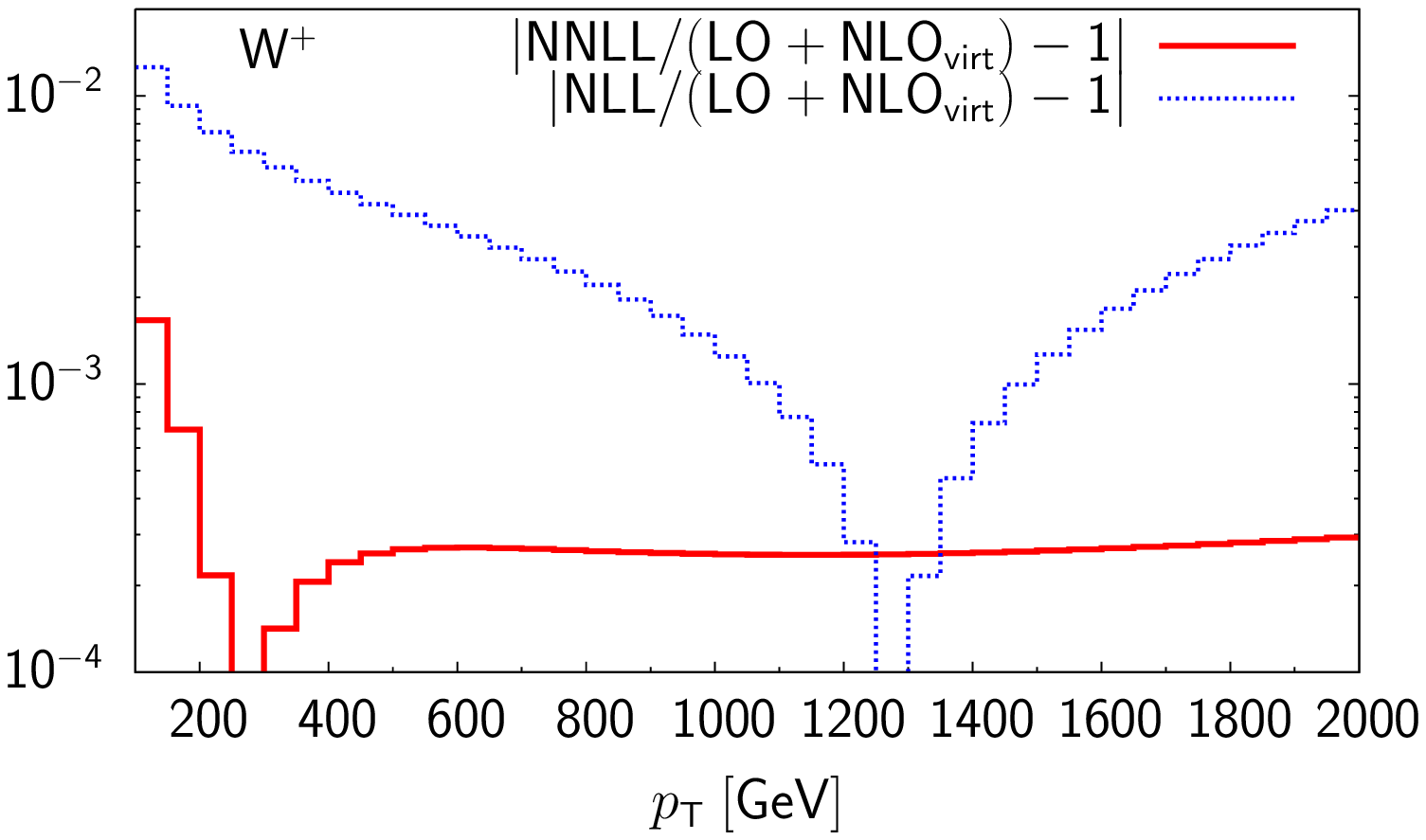, angle=0, width=12cm}
\end{center}
\vspace*{-2mm}
\caption{Relative precision of the high-energy approximations at one loop 
in the process $pp\rar W^+ j$ at $\sqrt{s}= 14\TeV$
as a function of $\pT$: NNLL (solid) and NLL (dashed) \wrt the 
IR-finite part of the exact one-loop result (LO$+\mathrm{NLO}_{\mathrm{virt}}$).} 
\label{fig:ptqualNNLLlhc}
\end{figure}
The high-energy behaviour of the $\mathrm{NLO}_{\mathrm{virt}}$ part is 
described by the compact NLL and NNLL approximations presented in 
\refse{se:helimit}.
The quality of these approximations is shown in  \reffi{fig:ptqualNNLLlhc}.
We observe that the NLL approximation works well differing from the exact
$\mathrm{NLO}_{\mathrm{virt}}$ result by
less than $1\%$ for $\pT \ge 200 \GeV$. 
The quality of the NNLL approximation 
is of the order of one permille or better in the entire  $\pT$-range.

\begin{figure}[p]
\begin{center}
\epsfig{file=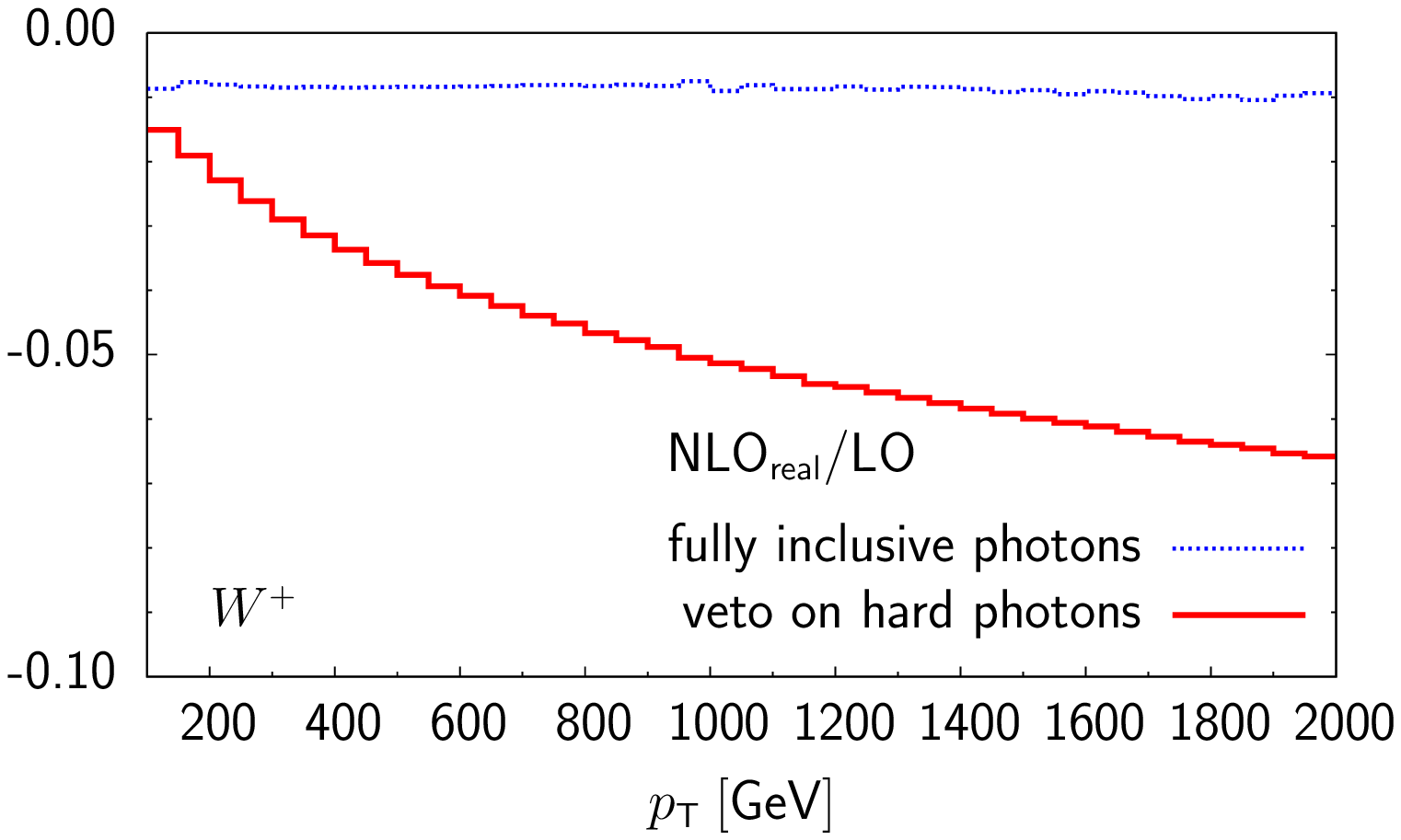, angle=0, width=12cm}
\end{center}
\vspace*{-2mm}
\caption{
Relative size of the real correction ($\mathrm{NLO}_{\mathrm{real}}$)
\wrt the $\mathrm{LO}$ for fully inclusive photon radiation
and the case where visible photons are rejected, plotted as a function of 
$\pT$ for the process $pp\rar W^+ j$ at $\sqrt{s}= 14\TeV$.
}
\label{fig:photoncutslhc}
\end{figure}
For less inclusive observables 
where a veto on hard photons is imposed, 
the  $\mathrm{NLO}_{\mathrm{real}}$ contribution can become important.
\reffi{fig:photoncutslhc} shows the relative $\mathrm{NLO}_{\mathrm{real}}$ corrections  for $W^+$ production.
We compare the fully inclusive photon radiation with the case where visible 
photons with $p_{\rT,\gamma}>10\GeV$ and $R(\gamma,j)>0.4$ are rejected. 
This veto leads to a significant enhancement of the
(absolute size of the) $\mathrm{NLO}_{\mathrm{real}}$ part,
which can exceed $-5\%$ for $\pT \ge 1 \TeV$. 

\begin{figure}[p]
  \begin{center}
\epsfig{file=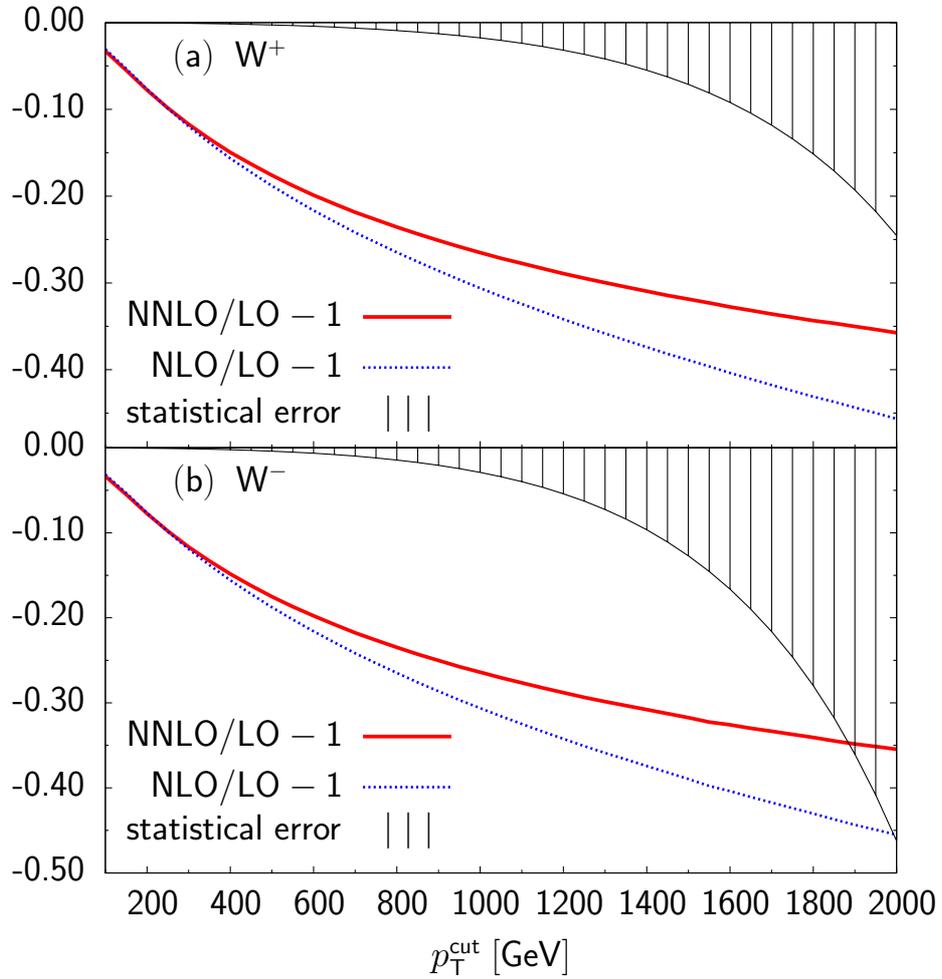, angle=0, width=12cm}
\end{center}
\vspace*{-2mm}
\caption{Relative NLO (dotted) and NNLO (solid) electroweak corrections \wrt the LO and statistical
  error (shaded area) for the unpolarized integrated cross section for 
(a) $p p\rar
  W^+j$ at $\sqrt{s}= 14\TeV$ and (b)  $p p\rar
  W^-j$ at $\sqrt{s}= 14\TeV$ as a function of $\pTcut(W)$.
}
\label{fig:cswplusminuslhc}
\end{figure}
To underline the relevance of the large electroweak corrections for
$W$ production at the LHC, 
in \reffi{fig:cswplusminuslhc}a and  \reffi{fig:cswplusminuslhc}b we present 
the relative NLO and NNLO 
corrections to the $W^+$ and $W^-$ cross sections
integrated over $\pT$ starting from $\pT = \pTcut$, as a function of
$\pTcut$. This is compared with the statistical error, defined as
$\Delta \si_{\rm stat} / \si = 1 /\sqrt N$ with $N= \calL \times {\rm
  BR} \times \si_{\rm LO}$. The branching ratio $\rm{BR}=2/9$ accounts for
the full efficiency of $W$-detection in the 
$e \bar{\nu_e}$ and $\mu \bar{\nu_\mu}$ modes
(for this estimate we ignore experimental efficiencies and cuts) and
we assume a total integrated luminosity $\mathcal{L}=300\mathrm{fb}^{-1}$ 
for the LHC \cite{LHClum}. It is clear that the size of the NLO corrections 
is much bigger than the statistical error. 
Indeed, already for $\mathcal{L}=3 \mathrm{fb}^{-1}$ and $\pT \lsim 800\GeV$ 
they correspond to a 
two standard deviation effect.
Also the difference between
the NNLO and NLO corrections, due to two-loop logarithmic effects, is
significant. In terms of the estimated statistical error, these two-loop 
contributions amount to 1--3 standard deviations for $\pT$ of $\ord(1\TeV)$.

\begin{figure}[p]
\vspace*{2mm}
  \begin{center}
\epsfig{file=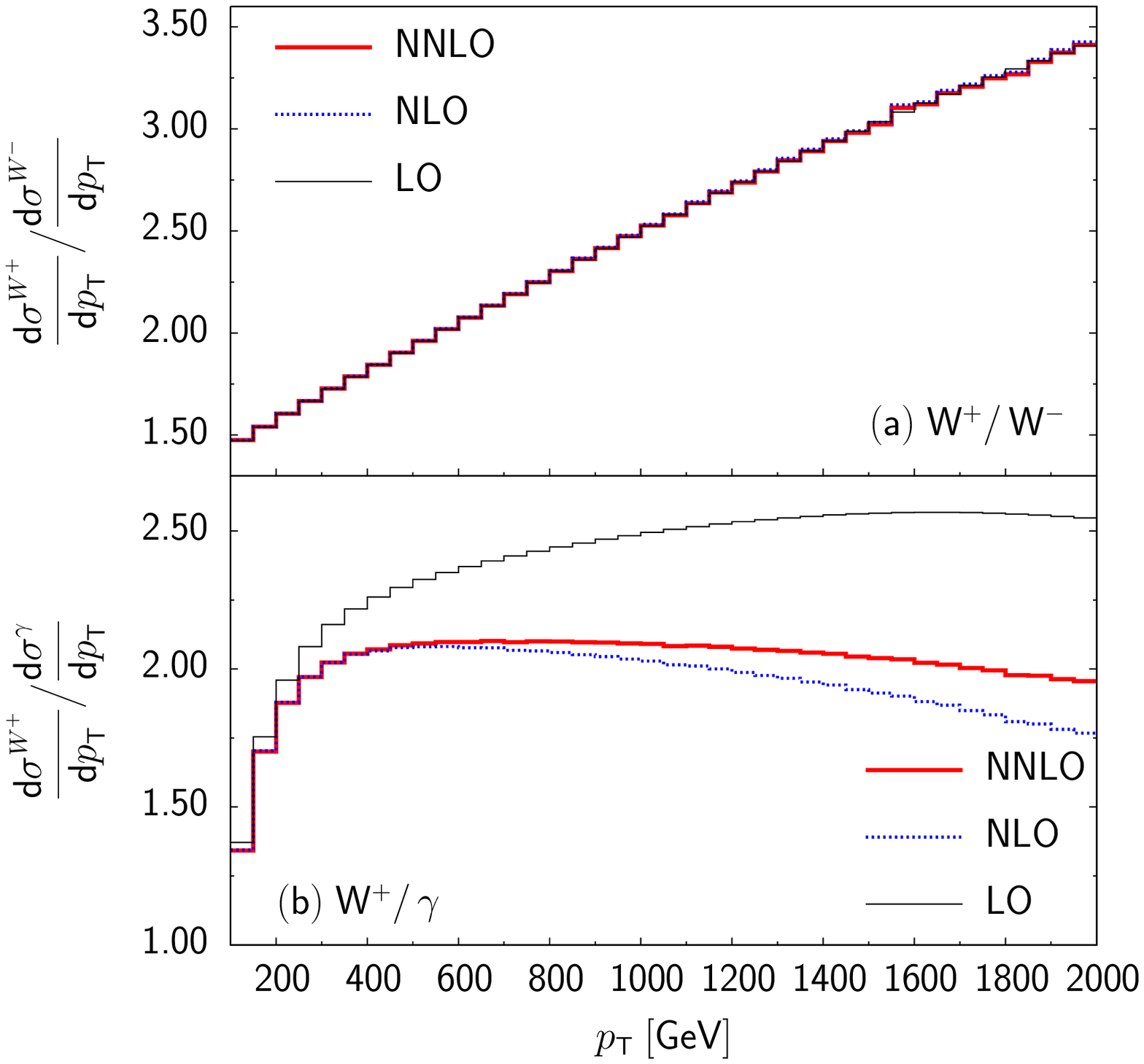, angle=0, width=12cm}
\end{center}
\vspace*{-2mm}
\caption{Ratio of the $\pT$-distributions
for the processes (a) $p p\rar W^+ j$ and  $p p\rar W^- j$ and
(b) $p p\rar W^+ j$ and  $p  p\rar \gamma j$ 
at $\sqrt{s}=14\TeV$:
LO (thin solid), NLO(dotted) and NNLO (thick solid) predictions.
}
\label{fig:ratiowplusminusgammalhc}
\end{figure}
Ratios of $\pT$-distributions for $W^+$, $W^-$, $Z$ bosons
\cite{Kuhn:2005az} and photons \cite{Kuhn:2005gv}, in contrast to the
distributions themselves are expected to be relatively insensitive to
QCD corrections and theoretical uncertainties associated with $\al_\rS$
and PDFs.  These ratios lead to important experimental tests of $W$
and $Z$ couplings in the high-energy region.
For $W^+$ and $W^-$ the ratio is presented
in~\reffi{fig:ratiowplusminusgammalhc}a.  The LO value increases from
1.5 at $\pT=100\GeV$ to 3.4 at $\pT=2\TeV$.  As already observed, the
(relative) electroweak corrections to the $W^+$- and $W^-$-boson
$\pT$-distributions are almost identical. In consequence, the LO, NLO
and NNLO curves in~\reffi{fig:ratiowplusminusgammalhc}a overlap.
In contrast, the impact of the electroweak corrections on the
$W^+/\gamma$ ratio (\reffi{fig:ratiowplusminusgammalhc}b) at the LHC
is clearly visible.  The LO prediction, ranging from 1.4 to 2.5,
receives a negative NLO correction that grows with $\pT$ and amounts to
$-0.5$ for $\pT=1\TeV$.  At $\pT=2\TeV$ the
difference between the NNLO and NLO curves is about $0.2$.

\begin{figure}[p]
\vspace*{2mm}
  \begin{center}
\epsfig{file=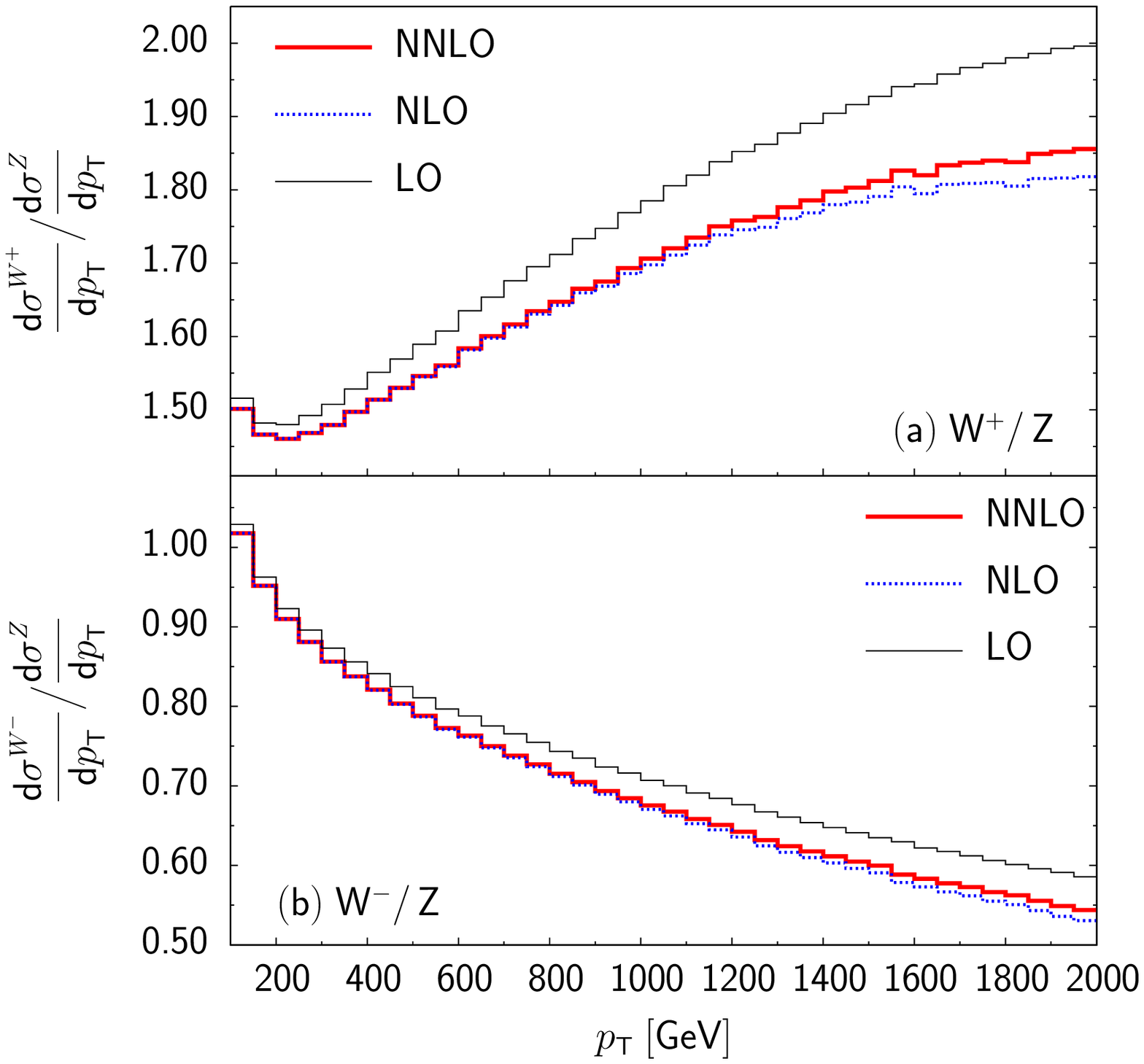, angle=0, width=12cm}
\end{center}
\vspace*{-2mm}
\caption{Ratio of the $\pT$-distributions
for the processes (a) $p p\rar W^+ j$ and  $p  p\rar Z j$ and
(b) $p p\rar W^- j$ and  $p  p\rar Z j$ 
at $\sqrt{s}=14\TeV$:
LO (thin solid), NLO(dotted) and NNLO (thick solid) predictions.
}
\label{fig:ratiowplusminuszlhc}
\end{figure}

The ratios of $\pT$-distributions for $W^+/Z$ and $W^-/Z$ are shown in
\reffi{fig:ratiowplusminuszlhc}a and \reffi{fig:ratiowplusminuszlhc}b,
respectively.  For the $W^+/Z$ ratio the LO prediction ranges from 1.5
to 2.  For $\pT \ge 1\TeV$ it is reduced by 
0.09 to 0.18
by the NLO
electroweak corrections.  The logarithmic two-loop corrections are
small.  
A qualitatively similar behaviour 
is observed for the $W^-/Z$
ratio.

\begin{figure}[p]
\vspace*{2mm}
  \begin{center}
\epsfig{file=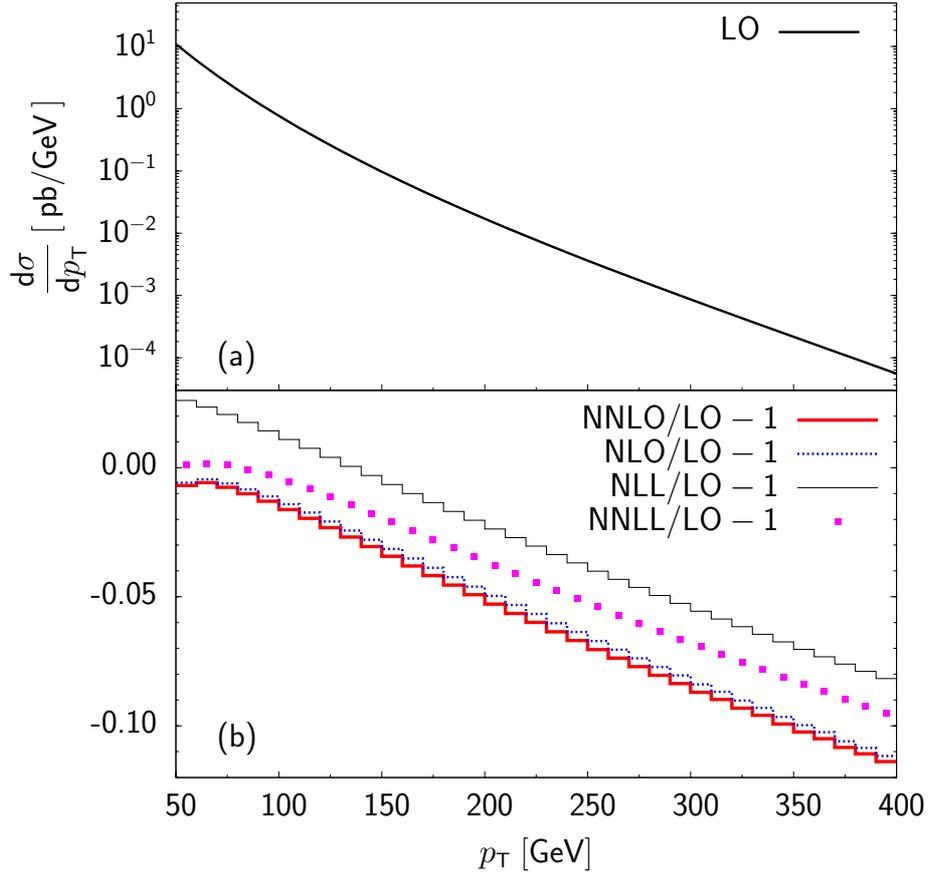, angle=0, width=12cm}
\end{center}
\vspace*{-2mm}
\caption{Transverse-momentum distribution for $W$-boson production at
the Tevatron.
(a) LO distribution for $p \bar p\rar W^{+ (-)} \ j$ (solid).
(b) Relative NLO (dotted), NLL (thin solid), NNLL (squares) and NNLO (thick solid)
electroweak correction \wrt the LO distribution for  $pp\rar W^{+(-)}
\ j$.
}
\label{fig:ptwtev}
\end{figure}

\begin{figure}[hp]
  \begin{center}
\epsfig{file=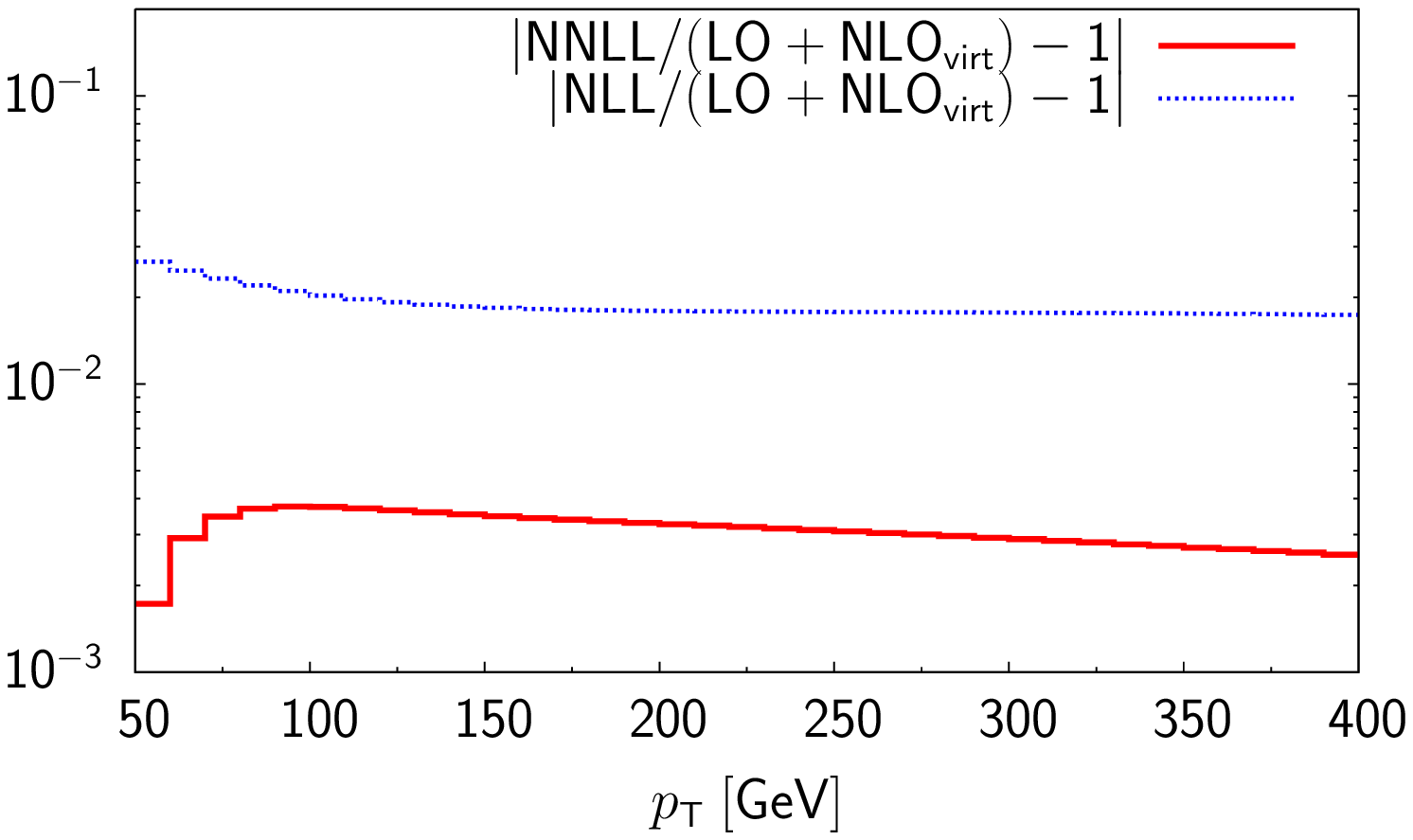, angle=0, width=12cm}
\end{center}
\vspace*{-2mm}
\caption{Relative precision of the high-energy approximations at one loop 
in the process $p \bar p\rar W^{+(-)} j$ at $\sqrt{s}= 2\TeV$
as a function of $\pT$: NNLL (solid) and NLL (dashed) \wrt the 
IR-finite part of the exact one-loop result (LO$+\mathrm{NLO}_{\mathrm{virt}}$).} 
\label{fig:ptqualNNLLtev}
\end{figure}

\begin{figure}[p]
\vspace*{2mm}
  \begin{center}
\epsfig{file=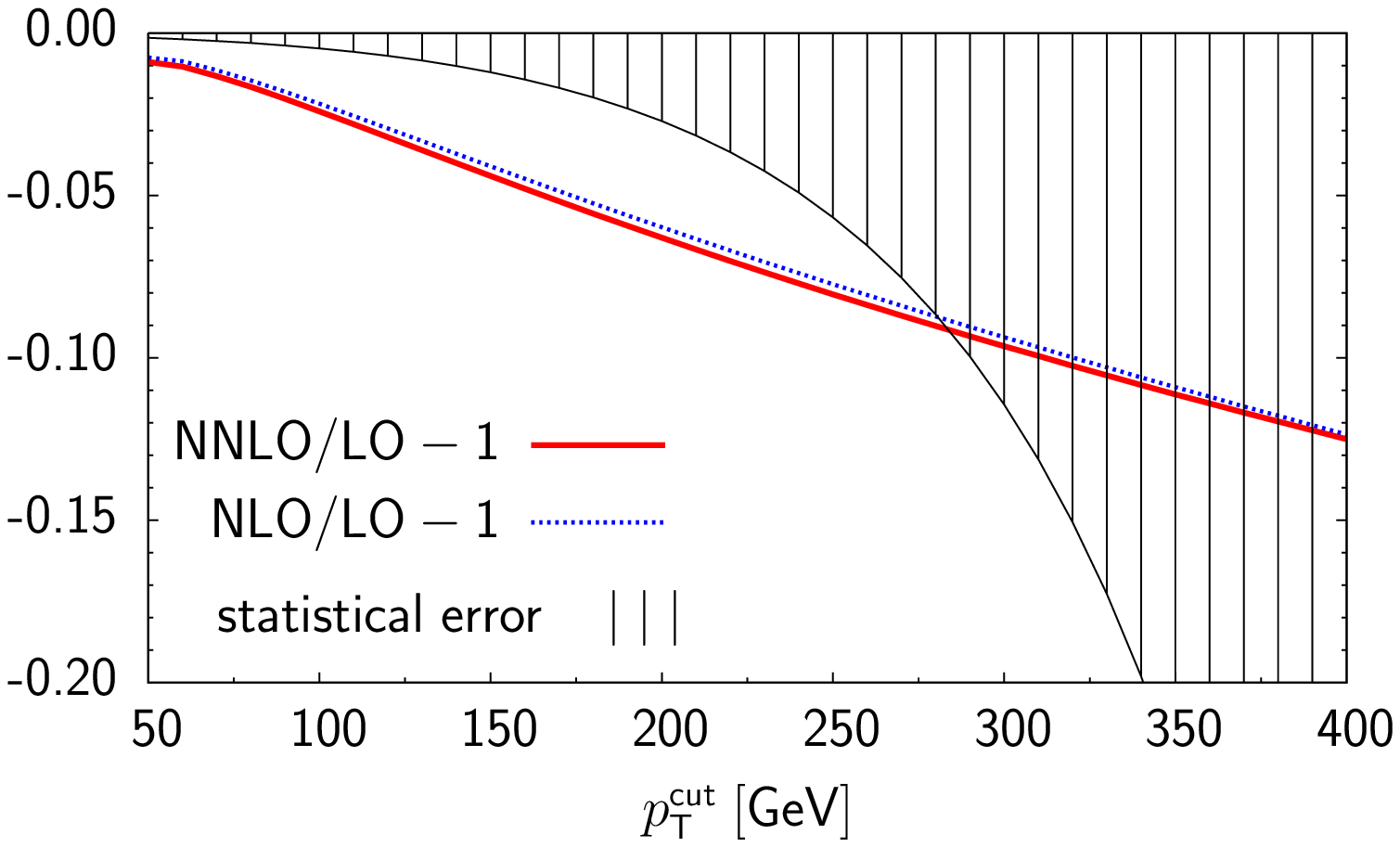, angle=0, width=12cm}
\end{center}
\vspace*{-2mm}
\caption{Relative NLO (dotted) and NNLO (solid) electroweak corrections \wrt the LO and statistical
  error (shaded area) for the unpolarized integrated cross section for 
$p \bar p\rar W^{+(-)}j$ at $\sqrt{s}= 2\TeV$ as a function of $\pTcut(W)$.}
\label{fig:cswtev}
\end{figure}

\begin{figure}[p]
\vspace*{2mm}
  \begin{center}
\epsfig{file=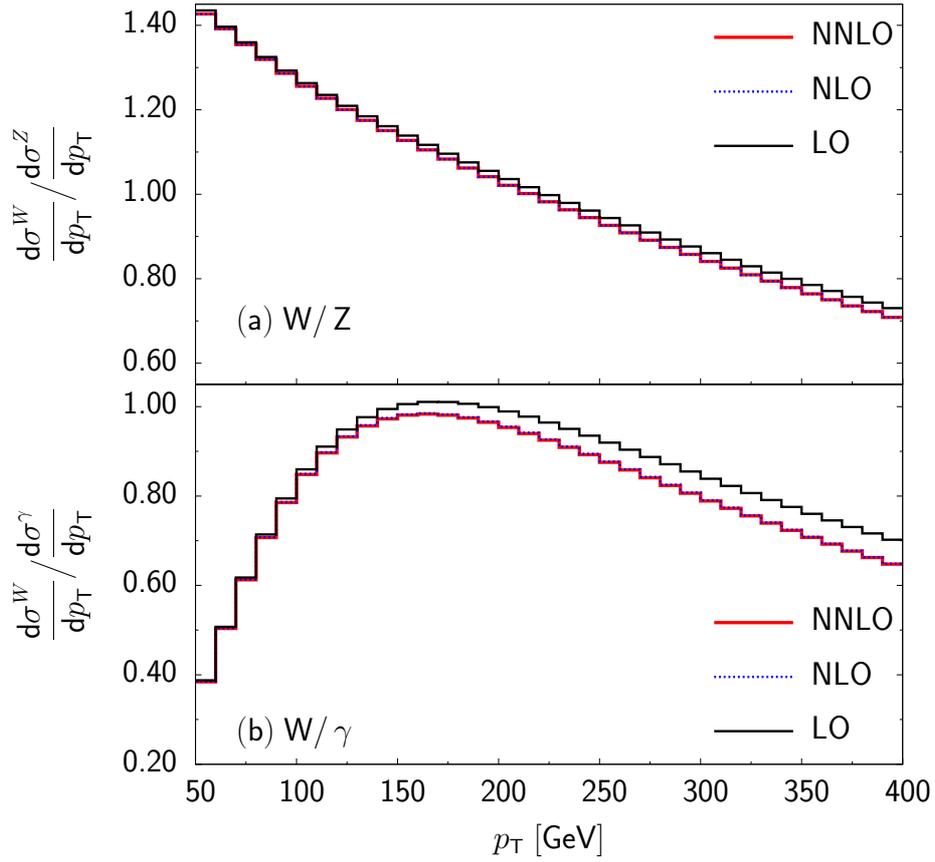, angle=0, width=12cm}
\end{center}
\vspace*{-2mm}
\caption{Ratio of the $\pT$-distributions
(a) for the processes $p \bar p\rar W^{+(-)} j$  and  $p \bar p\rar Z j$
and (b)  for the processes $p \bar p\rar W^{+(-)} j$ and $p \bar p\rar \gamma j$
at $\sqrt{s}=2\TeV$:
LO (thin solid), NLO(dotted) and NNLO (thick solid) predictions.
}
\label{fig:ratiowtev}
\end{figure}

The results of a similar analysis for $W^+$ production at the 
Tevatron ($\sqrt{s}=2\TeV$) are shown in \reffis{fig:ptwtev}--\ref{fig:ratiowtev} (the $\pT$-distributions for $W^+$ and $W^-$ production 
are obviously identical).
The LO $\pT$-distribution is shown in~\reffi{fig:ptwtev}a,
the relative NLO, NLL, NNLL and NNLO corrections 
in~\reffi{fig:ptwtev}b.
The NLO corrections grow with $\pT$ and reach $-11\%$ at $\pT=400\GeV$.
 The one-loop NNLL and NLL approximations describe the exact NLO results 
with $1\%$ and $3\%$ precision, respectively. 
The dominant two-loop effects have little impact on the size of the corrections.

The quality of the high-energy approximations \wrt the 
IR-finite part of the virtual corrections ($\mathrm{NLO}_{\mathrm{virt}}$)
is shown in \reffi{fig:ptqualNNLLtev}.
In the $\pT$-range under consideration both approximations
are less precise than at the LHC, nevertheless the NNLL approximation
in sufficient for all practical purposes.

In~\reffi{fig:cswtev} the relative NLO and NNLO corrections
to the integrated  cross section for $\pT \ge \pTcut$ 
are compared with the expected statistical error for an integrated luminosity 
$\mathcal{L}=7\mathrm{fb}^{-1}$ \cite{TEVlum}.
The size of the NLO electroweak corrections is above the
statistical error for a significant range of $\pT$-values.
Therefore they should be included in the analysis
when considering precision measurements.
In contrast, the impact of the dominant two-loop corrections is negligible.

The effect of the electroweak corrections on the ratios of $\pT$-distributions
for $W/Z$  and for $W/\gamma$ is shown 
in \reffi{fig:ratiowtev}a and \reffi{fig:ratiowtev}b, respectively.

\section{Summary}
\label{se:conc}
In this work the electroweak corrections to large transverse momentum 
production of
$W$ bosons at the hadron colliders Tevatron and LHC were evaluated. 
The contributions
from real and virtual photons cannot be separated in a gauge-invariant 
manner from purely weak corrections
and were thus included in our analysis. Soft and collinear
singularities were regulated by introducing a small quark mass and a
small photon mass and, alternatively, by using dimensional
regularization.
The real photon radiation was evaluated using the
dipole 
subtraction
formalism. The agreement between the results derived in the 
two regularization schemes has been an important cross check of the 
calculation. Numerous additional tests were performed to ensure the 
correctness of the result. 

At the Tevatron, $\pT$-values up to around $300\GeV$ can be reached 
with reasonable event rates. In this region the $\ord(\alpha)$
electroweak corrections 
reach up to $-10\%$ and are thus of relevance for precision 
measurements. Two-loop electroweak corrections are negligible at the Tevatron.
With $\pT$ below $400\GeV$ the relative rates for $W$, $Z$ and $\gamma$ production are 
hardly affected by electroweak corrections.

In contrast,
for transverse momenta in the TeV region accessible at the LHC, electroweak 
corrections play an important role.  
The  $\ord(\alpha)$  corrections lead to a reduction 
of the cross section by about 
$-15\%$ at transverse momenta of $500\GeV$  
and reach more than $-40\%$ at $2\TeV$. 
The logarithmically dominant terms were extracted from the 
exact expression of the virtual corrections 
and agreement was found with the predictions based on 
the process-independent analysis of electroweak Sudakov logarithms. 
If no cuts on real photons are applied, the contribution of the 
real photon emission is numerically small (about $1\%$) and almost
independent of $\pT$.
Numerically the NLL and NNLL approximations give a good 
description of the full $\ord(\alpha)$
result with an accuracy of about 
1--2\%.
Considering the large event rate at the LHC,
leading to a fairly good statistical precision
even at transverse momenta up to $2\TeV$,
we evaluated also the dominant (NLL) two-loop terms. 
In the high-$\pT$ region, these two-loop logarithmic effects increase the cross
section by 5--10\% and thus become of importance in precision studies.
We also studied the relative rates for $W^+$, $W^-$, $Z$ and
$\ga$ production, which are expected to be stable with respect to
QCD effects. 
The electroweak corrections cancel almost completely in the $W^+/W^-$
ratio. In contrast, their impact on the $W^+/Z$ and the $W^+/\gamma$ 
ratios is significant and  leads to a shift of $\ord(10\%)$
for $\pT \ge 1\TeV$. 

\section*{Acknowledgements}
We would like to thank S.~Dittmaier, B.~J\"ager and P.~Uwer for
helpful discussions.
This work was supported in part by
BMBF Grant No.\ 05HT4VKA/3, the Sonderforschungsbereich Transregio 9
and the DFG
Graduiertenkolleg ``Hochenergiephysik und Teilchenastrophysik''.

\newpage

\begin{appendix}

\section{Recombination and exclusive $Wj$ cross section}
\label{app:recomb}
As discussed in \refse{sse:kinematics}, the recombination prescription
that we use to regularize photon-quark final-state collinear singularities 
implies a different treatment of final-state quarks and gluons.
While for final-state gluons we apply a cut on $p_{\rT,\, g}$ within
the entire phase space, for final-state quarks
the recombination effectively removes the cut on
$p_{\rT,\, q}$  inside the collinear cone $R(q,\gamma)<R_\mathrm{sep}$.
As a consequence the recombined $g q' \to W^\si q \gamma$ cross section
\refeq{reccs} has a logarithmic dependence on the cut-off parameter $R_\mathrm{sep}$.
In order to quantify this  $R_\mathrm{sep}$-dependence, let us  
consider the contribution of real photon radiation 
inside the recombination cone. To this end, 
assuming that the cone is sufficiently small ($R_\mathrm{sep}\ll 1$), 
we adopt a collinear approximation
\beqar
\int_{R(q,\gamma)<R_\mathrm{sep}} \rd \hat{\si}^{g q' \to W^\si q \gamma}
&=&
\hat{\si}^{g q' \to W^\si q}\int_0^1 \rd z F_{q \gamma}(z),
\eeqar
where\footnote{Here we assume lowest-order kinematics, \ie
$p_{\rT,q}+p_{\rT,\gamma}=p_{\rT,W}$ in the collinear region.
} $z=
p_{\rT,\gamma}/(p_{\rT,q}+p_{\rT,\gamma})
=1-p_{\rT,q}/p_{\rT,W}$
is the photon momentum fraction  and
\cite{Catani:1996vz}
\beqar
F_{q \gamma}(z)
&=&
\frac{\alpha Q_q^2}{2\pi}P_{q\gamma }(z,\varepsilon)\frac{(4\pi\mu^2)^\varepsilon}{\Gamma(1-\varepsilon)}\int_0^{k^2_{\perp,\mathrm{max}}}\frac{\rd k^2_\perp}{(k^2_\perp)^{1+\varepsilon}}
\nl&=&
-\frac{\alpha Q_q^2}{2\pi}
P_{q\gamma }(z)
\frac{(4\pi)^\varepsilon}{\varepsilon\Gamma(1-\varepsilon)}
+\bar F_{q \gamma}(z,\mu^2)
\eeqar
with
\beqar
\bar F_{q \gamma}(z,\mu^2)
&=&
-\frac{\alpha Q_q^2}{2\pi}\left[
P_{q\gamma }(z)
\ln\left(\frac{\mu^2}{k^2_{\perp,\mathrm{max}}}\right)
-z\right].
\eeqar
Here $P_{q\gamma }(z,\varepsilon)=P_{q\gamma }(z)-\varepsilon z$ with $P_{q\gamma}(z)=[1+(1-z)^2]/z$ is the $q\to \gamma$ splitting function in $4-2\varepsilon$
dimensions,
$k_\perp$ is the photon tranverse momentum \wrt the photon-quark system,
and  $k_{\perp,\mathrm{max}}=z(1-z)R_\mathrm{sep}p_{\rT,W}$.
The $1/\varepsilon$ collinear singularity resulting from 
inclusive photon radiation, \ie integrating over the complete 
energy spectrum $0\le z \le 1$,
cancels against the virtual corrections.

The $R_\mathrm{sep}$-dependence of the recombined cross section 
\refeq{reccs}
is due to the fact that, inside the recombination cone 
quarks with  $p_{\rT,q}<\pTminj$ (or equivalently 
photons with $z> 1-\pTminj/p_{\rT,W}$) are not rejected. Thus the variation of
$\hat\sigma_{\mathrm{rec.}}$ induced by a rescaling $R_\mathrm{sep}\to \xi_\mathrm{sep} R_\mathrm{sep}$ amounts to 
\beqar
\frac{\Delta \hat\sigma^{g q' \to W^\si q\gamma}_{\mathrm{rec.}}}
{\hat\sigma^{g q' \to W^\si q}}
=\frac{\alpha Q_q^2}{2\pi}
\ln \xi^2_\mathrm{sep}
\int_{z_{\mathrm{min}}}^1 \rd z P_{q\gamma}(z)
\qquad\mbox{with}\quad z_{\mathrm{min}}=1-\pTminj/p_{\rT,W}.
\eeqar
For relatively small transverse momenta ($p_{\rT,W} \simeq 2 \pTminj$)
a rescaling of $R_\mathrm{sep}$ by a factor $\xi_\mathrm{sep}=10$
shifts the  $g q' \to W^\si q (\gamma)$ cross section by less than 
2 (0.5) permille for up- (down-) type quarks.
Moreover it is obvious that at high  $p_{\rT,W}$, where
$z_{\mathrm{min}}\to 1$, this effect tends to disappear.

Let us now compare the recombination procedure 
with a realistic definition of exclusive  $pp\to Wj$ production,
where final-state quarks  ($a=q$) and gluons ($a=g$)
are subject to the same cut $p_{\rT,a}> \pTminj$ 
within the entire phase space (including collinear quark-photon configurations).
Since the recombination procedure does not affect final-state gluons,
only channels involving final-state quarks need to be considered.
The difference between the recombined 
 $g q' \to W^\si q \gamma$ cross section \refeq{reccs} and the
exclusive cross section \refeq{exclcs}
corresponds to the contribution of
hard collinear photons with $R(q,\gamma)<R_\mathrm{sep}$ and
 $z_{\mathrm{min}}\le z\le 1$.
This collinear hard-photon radiation
can be described by means of quark fragmentation functions 
\cite{Glover:1993xc,Buskulic:1995au,Bourhis:1997yu,Fontannaz:2001ek,Klasen:2002xb,GehrmannDeRidder:2006vn} as
\beqar
\Delta \hat\sigma_{\mathrm{excl.}}
=
\hat\sigma^{g q' \to W^\si q \gamma}_{\mathrm{rec.}}
-
\hat\sigma^{g q' \to W^\si q \gamma}_{\mathrm{excl.}}
=
\hat{\si}^{g q' \to W^\si q}\int_{z_{\mathrm{min}}}^1 \rd z {\mathcal{D}_{q \gamma}}(z).
\eeqar
Here the effective quark fragmentation function 
$\mathcal{D}_{q \gamma}(z)=F_{q \gamma}(z)+{D}_{q \gamma}(z)$
consists of the perturbative contribution $F_{q \gamma}$ and the 
bare fragmentation function ${D}_{q \gamma}$. 
The collinear singularity resulting from the perturbative contribution
is factorized into the bare fragmentation function at the scale $\mu$, 
such that in the  $\MSBAR$ scheme \cite{Glover:1993xc}
\beqar
\mathcal{D}_{q \gamma}(z)&=&
{\bar F}_{q \gamma}(z,\mu^2)+{\bar D}_{q \gamma}(z,\mu^2),
\eeqar
and the renormalized fragmentation function ${\bar D}_{q \gamma}$ can be extracted from experimental measurements. Using the parametrization 
\cite{Buskulic:1995au,GehrmannDeRidder:2006vn}
\beqar
{\bar D}_{q \gamma}(z,\mu_0^2)=
\frac{\alpha Q_q^2}{2\pi}\left[-
P_{q\gamma}(z)
\ln(1-z)^2
-13.26\right],
\eeqar
obtained by the ALEPH collaboration at $\mu_0=0.14\,\GeV$, we arrive at
\beqar
{\mathcal D}_{q \gamma}(z)
&=&
\frac{\alpha Q_q^2}{2\pi}\left[
P_{q\gamma}(z)
\ln\left(\frac{z R_\mathrm{sep}p_{\rT,W}}{\mu_0}\right)^2
+z-13.26\right].
\eeqar
With this expression we derive 
a conservative upper bound for $\Delta \hat\sigma_{\mathrm{excl.}}$.
To this end we consider  $Q_q=2/3$, $R_\mathrm{sep}\simeq 1$, and a wide range of transverse momenta, $2 \pTminj\le p_{\rT,W} \le 2\TeV $.
With these parameters we obtain
\beqar
\frac{\Delta \hat\sigma_{\mathrm{excl.}}}{\hat\sigma}\lsim 2\times 10^{-3}.
\eeqar
We conclude that,
for $R_\mathrm{sep}\lsim\ord(1)$,
the recombined cross section
has a negligible dependence on the recombination parameter 
$R_\mathrm{sep}$
and provides a fairly precise description  of exclusive
$pp\to Wj$ production at high transverse momentum.

\section{Standard matrix elements}
\label{app:smel}
The algebraic expressions involving external momenta, Dirac matrices, spinors and gauge-boson polarization vectors have been reduced to a set of 10 standard matrix elements
\beqar
\smel_i &=& 
\bar{v}(p_{\qbar}) 
\smel_{i}^{\mu\nu}
\omega_\rL u(p_{q'})\, 
\varepsilon^*_\mu(p_W) \varepsilon^*_\nu(p_g),
\eeqar
with
\beqar
\smel^{\mu\nu}_{1} &=& \gamma^\mu(\ps_W-\ps_{\qbar})\gamma^\nu
,\nl
\smel^{\mu\nu}_{2} &=&(\ps_W-\ps_g)g^{\mu\nu}
,\nl
\smel^{\mu\nu}_{3} &=&  \gamma^\mu p_W^\nu
,\nl
\smel^{\mu\nu}_{4} &=&-  \gamma^\nu p_g^\mu 
,\nl
\smel^{\mu\nu}_{5} &=&  \gamma^\mu  p_{q'}^\nu
,\nl
\smel^{\mu\nu}_{6} &=&-   \gamma^\nu p_{\qbar}^\mu
,\nl
\smel^{\mu\nu}_{7} &=& (\ps_W-\ps_g) p_g^\mu p_W^\nu
,\nl
\smel^{\mu\nu}_{8} &=& (\ps_W-\ps_g) p_{\qbar}^\mu p_{q'}^\nu
,\nl
\smel^{\mu\nu}_{9} &=& (\ps_W-\ps_g) p_g^\mu p_{q'}^\nu
,\nl
\smel^{\mu\nu}_{10}&=&  (\ps_W-\ps_g) p_{\qbar}^\mu p_W^\nu
.
\eeqar
These algebraic expressions correspond to the massless subset of the standard matrix elements of \citere{Denner:1991kt}.

\section{Scalar loop integrals}
\label{app:loopint}
In this appendix we list the scalar loop integrals $\loops_{j}(M_V^2)$
that contribute to \refeq{algebraicred}.  The symbols 
$\loops_{j}$ are chosen in analogy with \citere{Kuhn:2005az}.
For convenience, to denote constant terms we define
\beq
\loops_{0}(M_V^2)=1.
\eeq
For the scalar integrals $A_0, B_0, C_0$ and $D_0$
we adopt the notation of {\sc FeynCalc} \cite{Mertig:1990an}.
However, we choose their normalization according to \citere{Denner:1991kt}, 
\ie we include the factor $(2\pi\mu)^{4-D}$ which is omitted in the 
conventions of {\sc FeynCalc}.

The UV-divergent one- and two-point functions
are denoted as
\beqar\label{loopinta}
\loops_{1a}(M_V^2) &=& B_0(m^2;M_V^2,m^2)
,\nl
\loops_{1b}(M_V^2) &=& B_0(m^2;M_W^2,m^2)=\loops_{1a}(M_W^2)
,\nl
\loops_{ 2}(M_V^2) &=& B_0(p_W^2;m^2,m^2)
,\nl
\loops_{ 3}(M_V^2) &=& B_0(p_W^2;M_W^2,M_V^2)
,\nl
\loops_{ 4}(M_V^2) &=& B_0(\shat;m^2,m^2)
,\nl
\loops_{5a}(M_V^2) &=& B_0(\uhat;M_V^2,m^2)
,\nl
\loops_{5b}(M_V^2) &=& B_0(\uhat;M_W^2,m^2)
=\loops_{5a}(M_W^2)
,\nl
\loops_{6a}(M_V^2) &=& B_0(\that;M_V^2,m^2)
,\nl
\loops_{6b}(M_V^2) &=& B_0(\that;M_W^2,m^2)
=\loops_{6a}(M_W^2)
.
\eeqar
The remaining loop integrals are free from 
UV singularities.
The following  three-point functions 
are finite if $M_V$ and the $W$-boson transverse momentum are non-vanishing:
\beqar\label{loopintb}
\loops_{7}(M_V^2) &=& C_0(\shat,m^2,m^2;m^2,m^2,M_V^2)
,\nl
\loops_{ 8}(M_V^2) &=& C_0(\uhat,p_W^2,m^2;M_V^2,m^2,m^2)
,\nl
\loops_{9a}(M_V^2) &=& C_0(\uhat,p_W^2,m^2;m^2,M_W^2,M_V^2)
,\nl
\loops_{9b}(M_V^2) &=& C_0(\uhat,p_W^2,m^2;m^2,M_V^2,M_W^2)
=\left.\loops_{9a}(M_V^2)\right|_{M^2_V\leftrightarrow M^2_W}
,\nl
\loops_{10}(M_V^2) &=& C_0(\that,p_W^2,m^2;M_V^2,m^2,m^2)
=\left.\loops_{8}(M_V^2)\right|_{\that\leftrightarrow \uhat}
,\nl
\loops_{11a}(M_V^2) &=& C_0(\that,p_W^2,m^2;m^2,M_W^2,M_V^2)
=\left.\loops_{9a}(M_V^2)\right|_{\that\leftrightarrow \uhat}
,\nl
\loops_{11b}(M_V^2) &=& C_0(\that,p_W^2,m^2;m^2,M_V^2,M_W^2)
=\left.\loops_{9b}(M_V^2)\right|_{\that\leftrightarrow \uhat}
,
\eeqar
In addition, the box diagrams b1--b3 in \reffi{fig:loopdiags}, provide 
the following combinations of
three- and four-point functions 
\newcommand{\laa}{m^2}
\newcommand{\lab}{m^2}
\newcommand{\lac}{m^2}
\beqar\label{loopintc}
\loops_{12}(M_V^2)
&=& 
D_0(m^2,0,p_W^2,m^2,\uhat,\shat;M_V^2,\lac,\lab,\laa)  
-\frac{1}
{\shat\uhat+(\that+\uhat)M_V^2}
\nl&&{}
\times\bigg[
{(\uhat-p_W^2) C_0(\uhat,p_W^2,m^2;M_V^2,\lab,m^2)}
+{\uhat C_0(\uhat,0,m^2;M_V^2,\lab,\lac)} 
\nl&&{}
+{(\shat-p_W^2) C_0(\shat,p_W^2,0;\lac,\laa,\lab)}
\bigg],
\nl
\loops_{13}(M_V^2) 
&=&
\left.\loops_{12}(M_V^2)\right|_{\that\leftrightarrow \uhat}
,\nl
\loops_{14a}(M_V^2) 
&=&
D_0(p_W^2,m^2,0,m^2,\that,\uhat;M_V^2,M_W^2,\lac,\lab)  
\nl&&{}
-\frac{\that C_0(\that,0,m^2;M_V^2,\lac,\lab)
+ \uhat C_0(\uhat,0,m^2;M_W^2,\lab,\lac)}{\that\uhat-\that M_W^2-\uhat M_V^2} 
,\nl
\loops_{14b}(M_V^2) 
&=&
\left.\loops_{14a}(M_V^2)\right|_{M_W^2\leftrightarrow M_V^2}
=
\left.\loops_{14a}(M_V^2)\right|_{\that\leftrightarrow \uhat}
.
\eeqar
For non-vanishing $M_V$ and $W$-boson transverse momentum,
the functions $\loops_{12}$--$\loops_{14b}$ are finite. 
The fact that the scalar four-point functions in \refeq{loopintc}
appear always in combination with three-point functions 
is due to the cancellation of the collinear singularities
that are associated with the $gq\bar q$ vertex \cite{Kuhn:2005az}.
Although these singularities  are present in 
individual $D_0$ and $C_0$ functions,
they always cancel in the complete result for box diagrams.

\section{Infrared singularities}
\label{app:IRsing}
The scalar integrals $\loops_{i}(M_V^2)$
in \refapp{app:loopint}
contain soft and collinear singularities
that appear when  $M_V=M_A\to 0$
and
$m\to 0$. 
As discussed in \refse{se:masssing}, these integrals 
are split into IR-singular (IR) and IR-finite (fin)
parts,
\beq
\loops_{i}(M_A^2)
= 
\loops_{i}^{\mathrm{IR}}
+
\loops_{i}^{\mathrm{\irfin}}
.
\eeq
The IR-singular parts depend on the scheme adopted to regularize soft and collinear singularities. The IR-finite parts are scheme independent
and free 
from  soft-collinear singularities, but can contain ultraviolet poles.

Let us start with the two-point functions \refeq{loopinta}. Here 
only $\loops_{1a}(M^2_A)$ gives rise to IR singularities. 
This integral is split into
\beqar
\loops_{1a,\mathrm{MR}}^{\mathrm{IR}}
&=&
-\ln\left(\frac{m^2}{M_W^2}\right)+1
,\nl 
\loops_{1a,\mathrm{DR}}^{\mathrm{IR}}
&=&
-\left(\frac{4\pi\mu^2}{M_W^2}\right)^\varepsilon
\frac{\Gamma(1+\varepsilon)}{\varepsilon}-1
,\nl 
\loops_{1a}^{\mathrm{\irfin}}
&=&
\left(\frac{4\pi\mu^2}{M_W^2}\right)^\varepsilon
\frac{\Gamma(1+\varepsilon)}{\varepsilon}+1
.
\eeqar
We note that within dimensional regularization
the UV and IR singularities 
cancel each other 
and the massless two-point function vanishes, 
$\loops_{1a,\mathrm{DR}}^{\mathrm{IR}}+\loops_{1a}^{\mathrm{\irfin}}=0$.
The three-point functions 
$\loops_{9b}(M_A^2)$ and
$\loops_{11b}(M_A^2)$ are free 
from IR singularities and the singularities 
originating from 
$\loops_{8}(M_A^2)$ and
$\loops_{10}(M_A^2)$ do not need to be considered since the 
coefficients associated with these scalar integrals are of order $M_A^2$
(see \refapp{app:coeff}).
The remaining three-point functions in \refeq{loopintb}
contain soft and collinear singularities. For them we find
\newcommand{\Li}{\mathrm{Li}}
\beqar
\loops_{7,\mathrm{MR}}^{\mathrm{IR}}
&=&
\frac{1}{\shat}
\left[
-\frac{1}{2}\ln^2\left(\frac{M_W^2}{m^2}\right)
+\ln\left(\frac{M_W^2}{\la^2}\right)
\ln\left(\frac{-\shat}{m^2}\right)
\right]
,\nl
\loops_{7,\mathrm{DR}}^{\mathrm{IR}}
&=&
\left(\frac{4\pi\mu^2}{M_W^2}\right)^\varepsilon\frac{\Gamma(1+\varepsilon)}{\shat}
\left[
\frac{1}{\varepsilon^2}
-\frac{1}{\varepsilon}
\ln\left(\frac{-\shat}{M_W^2}\right)
\right]
,\nl
\loops_{7}^{\mathrm{\irfin}}
&=&
\frac{1}{\shat}
\left[
\frac{1}{2}\ln^2\left(\frac{-\shat}{M_W^2}\right)
-\frac{\pi^2}{6}
\right]
,
\eeqar
and
\beqar
\loops_{9a,\mathrm{MR}}^{\mathrm{IR}}
&=&
\frac{1}{\uhat-M_W^2}\left\{
\frac{1}{2}\left[
\ln\left(\frac{M^2_W}{\la^2}\right)
-\frac{1}{2}\ln\left(\frac{M^2_W}{m^2}\right)
\right]
\ln\left(\frac{M^2_W}{m^2}\right)
\right.\nl&&{}\left.
+
\ln\left(\frac{M^2_W}{\la^2}\right)
\ln\left(1-\frac{\uhat}{M^2_W}\right)
\right\}
,\nl
\loops_{9a,\mathrm{DR}}^{\mathrm{IR}}
&=&
\left(\frac{4\pi\mu^2}{M^2_W}\right)^\varepsilon\frac{\Gamma(1+\varepsilon)}{\uhat-M_W^2}
\left[
\frac{1}{2\varepsilon^2}
-\frac{1}{\varepsilon}
\ln\left(1-\frac{\uhat}{M^2_W}\right)
\right]
,\nl
\loops_{9a}^{\mathrm{\irfin}}
&=&
\frac{1}{\uhat-M_W^2}
\left[
\ln^2\left(1-\frac{\uhat}{M^2_W}\right)
+
\Li_2\left(\frac{\uhat}{M^2_W}\right)
\right]
,
\eeqar
where
$\Li_2(x)=-\int_0^x\rd t \ln(1-t)/t$.
The finite and singular parts for
$\loops_{11a}(M_A^2)
=\left.\loops_{9a}(M_A^2)
\right|_{\uhat\to\that}$
are constructed in the same way.

The singular parts of the subtracted four-point functions  \refeq{loopintc}
can be 
related to the ones of the three-point functions,
\beqar
\loops_{12}^{\mathrm{IR}}
&=&
\left.\loops_{13}^{\mathrm{IR}}\right|_{\that\leftrightarrow \uhat}
=\frac{1}{\uhat}\loops_{7}^{\mathrm{IR}}
,\nl
\loops_{14a}^{\mathrm{IR}}
&=&
\left.\loops_{14b}^{\mathrm{IR}}\right|_{\that\leftrightarrow \uhat}
=
\frac{1}{\that}\loops_{9a}^{\mathrm{IR}}
,
\eeqar
in both regularization schemes.
This implicitly defines the remainders as 
\beqar
\loops_{12}^{\mathrm{\irfin}}
&=&
\left.\loops_{13}^{\mathrm{\irfin}}\right|_{\that\leftrightarrow \uhat}
=
\loops_{12}(M_A^2)
-
\frac{1}{\uhat}\loops_{7}^{\mathrm{IR}}
,\nl
\loops_{14a}^{\mathrm{\irfin}}
&=&
\left.\loops_{14b}^{\mathrm{\irfin}}\right|_{\that\leftrightarrow \uhat}
=
\loops_{14a}(M_A^2)
-\frac{1}{\that}\loops_{9a}^{\mathrm{IR}}.
\eeqar
Using the explicit analytic expressions for the infrared singular four-point and three-point functions \cite{Beenakker:1988jr,Dittmaier:2003bc}
we obtain
\beqar
\loops_{12}^{\mathrm{\irfin}}
&=&
\frac{1}{\shat\uhat}
\biggl[
\frac{1}{2}\ln^2\left(\frac{-\shat}{M^2_W}\right)
-\ln^2\left(\frac{\shat}{\uhat}\right)
-2\Li_2\left(1-\frac{M^2_W}{\shat}\right)
-2\Li_2\left(1-\frac{M^2_W}{\uhat}\right)
-\frac{\pi^2}{2}
\biggr]
,\nl
\loops_{14a}^{\mathrm{\irfin}}
&=&
\frac{1}{\that(\uhat-M_W^2)}
\biggl[
2\ln\left(1-\frac{\uhat}{M^2_W}\right)
\ln\left(\frac{-\that}{M^2_W}\right)
-\frac{1}{2}\ln^2\left(\frac{-\that}{M^2_W}\right)
\nl&&{}
+\Li_2\left(\frac{\uhat}{M^2_W}\right)-\frac{\pi^2}{2}
\biggr]
.
\eeqar

\section{Explicit result for the virtual corrections} 
\label{app:coeff}
\newcommand{\different}[1]{#1}
In this appendix we present explicit analytic expression
for the functions $H_1^{\mathrm{I}}(M_V^2) $ defined in \refeq{unpolfunct}.
These functions describe the contribution of the 
unrenormalized Feynman diagrams of \reffi{fig:loopdiags}
to the unpolarized cross section.
They consist of linear combinations of
the scalar integrals defined in \refapp{app:loopint},
\beqar
   H_1^{\mathrm{I}}(M_V^2) &=& 
 \sum_{j} 
K_j^{\mathrm{I}}(M_V^2)\,
\mathrm{Re}\left[\loops_j(M_V^2)\right]
\qquad\mbox{for}\quad 
\mathrm{I}=\mathrm{A,N,X,Y}.
\eeqar
The coefficients of the function 
$H_1^{\mathrm{A}}(M_V^2)$ read
\beqar\label{nloabcoeff}
K^{\mathrm{A}}_{0 }(M_V^2) 
&=& 
-\frac{4 \shat^2+3(\that^2+\uhat^2)}{\that \uhat}
+\shat\Bigg(\frac{1}{\shat+\that}+\frac{1}{\shat+\uhat}-\frac{5}{\uhat}-\frac{5}{\that}+\frac{4}{\that+\uhat}\Bigg)
,\nl
K^{\mathrm{A}}_{1a}(M_V^2) 
&=&  
M_V^2
\Bigg\{-\Bigg[\frac{3 \shat}{{{(\shat+\that)}^2}}+\frac{3 \shat}{{{(\shat+\uhat)}^2}}\Bigg]
+
\Bigg(\frac{1}{\shat+\that}+\frac{1}{\shat+\uhat}\Bigg)
\nl&&{}
-2\Bigg(\frac{ \shat+\uhat}{{\that^2}}+\frac{ \shat+\that}{{\uhat^2}}\Bigg)
+\frac{ 2 {\shat^2} (2\shat+\that+\uhat)}{ \that \uhat (\shat+\that)  (\shat+\uhat)}
\Bigg\}
+4\frac{ (\shat+\that)^2+(\shat+\uhat)^2}{ \that \uhat}
,\nl
K^{\mathrm{A}}_{1b}(M_V^2) 
&=&0 
,\nl
K^{\mathrm{A}}_{2 }(M_V^2) 
&=& 
p_W^2 \Bigg[ \frac{6 \shat M_V^2}{{{(\shat+\that)}^3}}+\frac{6 \shat M_V^2}{{{(\shat+\uhat)}^3}}+\frac{2 \shat M_V^2}{{{(\shat+\that)}^2} \uhat}+\frac{2 \shat M_V^2}{
{{(\shat+\uhat)}^2\that}}+\frac{4 (\shat+\that+\uhat)}{{{(\that+\uhat)}^2}}
-\frac{3}{\that}
\nl&&{}
-\frac{3}{\uhat}
+\frac{2 \shat+\that-2 M_V^2}{{{(\shat+\that)}^2}}
+\frac{2 \shat+\uhat-2 M_V^2}{{{(\shat+\uhat)}^2}}-\frac{\shat
(2\shat+\that+\uhat) (2 M_V^2+3 \shat)}{\that \uhat (\shat+\that) (\shat+\uhat)}\Bigg]
,\nl
K^{\mathrm{A}}_{3 }(M_V^2) 
&=&  
0
,\nl
K^{\mathrm{A}}_{4 }(M_V^2) 
&=& 
-\frac{4 \shat (\shat+2\that+2\uhat)}{{{(\that+\uhat)}^2}}
,\nl
K^{\mathrm{A}}_{5a }(M_V^2) 
&=& 
-\frac{6 M_V^2 \shat \uhat }{{{(\shat+\that)}^3}}
+\frac{M_V^2 (2 \uhat-5 \shat)-\shat
\uhat}{{{(\shat+\that)}^2}}
+\frac{2 M_V^2 (\shat+\that+\uhat) }{{\uhat^2}}-\frac{M_V^2+4 \shat+\uhat}{\shat+\that}
,\nl
K^{\mathrm{A}}_{5b }(M_V^2) 
&=&0
,\nl 
K^{\mathrm{A}}_{6a }(M_V^2) 
&=& 
K^{\mathrm{A}}_{5a }(M_V^2)\Big|_{\that\leftrightarrow\uhat} 
,\nl
K^{\mathrm{A}}_{6b }(M_V^2) 
&=&0
,\nl
K^{\mathrm{A}}_{7 }(M_V^2) 
&=&
-\frac{\shat }{\that \uhat} \Bigg[2 (\shat+M_V^2)(\that+\uhat) +{\that^2}+{\uhat^2}\Bigg]
,\nl
K^{\mathrm{A}}_{8 }(M_V^2) 
&=&
\frac{p_W^2 M_V^2}{\uhat(\uhat-p_W^2)^3}
\Bigg[
2\that M_V^2 (\uhat-\shat-\that)
-4 p_W^2\shat (\shat+\that+M_V^2)
\Bigg]
,\nl
K^{\mathrm{A}}_{9a }(M_V^2) 
&=&  
K^{\mathrm{A}}_{9b }(M_V^2) 
=0 
,\nl
K^{\mathrm{A}}_{10}(M_V^2) 
&=& K^{\mathrm{A}}_{8}(M_V^2)\Big|_{\that\leftrightarrow\uhat} 
,\nl
K^{\mathrm{A}}_{11a}(M_V^2) 
&=&  
K^{\mathrm{A}}_{11b}(M_V^2) 
=0
,\nl
K^{\mathrm{A}}_{12 }(M_V^2) 
&=& 
-\frac{M_V^2(\that+\uhat)+\shat\uhat
}{\that \uhat}
\Bigg[2{{(\shat+M_V^2)}}(\shat+M_V^2+\that)+{\that^2}\Bigg] 
,\nl
K^{\mathrm{A}}_{13 }(M_V^2) 
&=& K^{\mathrm{A}}_{12}(M_V^2) \Big|_{\that\leftrightarrow\uhat} 
,\nl
K^{\mathrm{A}}_{14a}(M_V^2) 
&=& K^{\mathrm{A}}_{14b}(M_V^2) =
0
.
\eeqar
The only difference between $H_1^{\mathrm{A}}(M_V^2)$ 
and the equally named function in \citere{Kuhn:2005az}
is due to the fact that  
$H_1^{\mathrm{A}}(M_V^2)$ in \citere{Kuhn:2005az} 
includes the contribution of the fermionic wave-function 
renormalization constants,
which modify the coefficients
$K^{\mathrm{A}}_{0}$ and $K^{\mathrm{A}}_{1a}$
(see eqs.~(54) and (55) in \citere{Kuhn:2005az}).

For the coefficients of the function 
$H_1^{\mathrm{N}}(M_V^2)$ we obtain
\beqar
K^{\mathrm{N}}_{0 }(M_V^2) 
&=& 
\frac{4 {\shat}}{\that \uhat}(\shat+\that+\uhat)
-2 \shat \Bigg(\frac{1}{\shat+\uhat}+\frac{1}{\shat+\that}+\frac{2}{\that+\uhat}\Bigg)+2
\Bigg(\frac{\that}{\uhat}+\frac{\uhat}{\that}\Bigg)
,\nl
\different{K^{\mathrm{N}}_{1a}(M_V^2)} 
&=&  
-\frac{M_V^2}{2}\Biggl\{
\frac{4\shat}{\that\uhat}
+(\that\uhat-2\shat(\shat+\that+\uhat))
\biggl[
\frac{1}{\that(\shat+\uhat)^2}
+\frac{1}{\uhat(\shat+\that)^2}
\biggr]
\Biggr\}
\nl&&{}
-\frac{\shat(4\shat+3\that+3\uhat)+\that^2+\uhat^2}{\that\uhat}
,\nl
\different{K^{\mathrm{N}}_{1b}(M_V^2)} 
&=&  
-\left. K^{\mathrm{N}}_{1a}(M_V^2) \right|_{M_V^2\leftrightarrow M_W^2}
,\nl
K^{\mathrm{N}}_{2}(M_V^2) 
&=& 
-K^{\mathrm{A}}_{2}(M_V^2) 
,\nl
\different{K^{\mathrm{N}}_{3}(M_V^2)} 
&=& {(M_W^2+M_V^2)} \Bigg[\frac{3 \shat \that}{{{(\shat+\uhat)}^3}}+\frac{3 \shat \uhat}{{{(\shat+\that)}^3}}\Bigg]
-\frac{1}{\that\uhat}
\Bigg[ \frac{1}{(\shat+\that)^2}+ \frac{1}{(\shat+\uhat)^2}\Bigg]
\nl&&{}\times
\Bigg\{
{\shat^4}
-2 {\that^2} {\uhat^2}
+\shat^2(\that+\uhat)( 2\shat+\that+\uhat )
+(M_W^2+M_V^2) 
\nl&&{}\times
\Bigg[{\shat^2}(\shat+\that+\uhat) 
- \that\uhat( 2 \shat-\that -\uhat )\Bigg]
\Bigg\}
,\nl
K^{\mathrm{N}}_{4}(M_V^2) 
&=& 
-K^{\mathrm{A}}_{4}(M_V^2) 
,\nl
\different{K^{\mathrm{N}}_{5a}(M_V^2)} 
&=& 
\frac{2 \shat (\shat+\that)-2 \that \uhat}{{{2(\shat+\that)}^2}}
-\Biggl[
\frac{\shat+\that}{\uhat}
-\frac{2\shat}{\shat+\that}
-\frac{\uhat(2\shat+\that)}{2(\shat+\that)^2}
\Biggr]
-M_V^2\Biggl[
\frac{1}{\uhat}-\frac{2\shat+\that}{2(\shat+\that)^2}
\Biggr]
\nl&&{}+(2M_V^2-M_W^2)\Bigg[
\frac{2\shat}{(\shat+\that)^2}
+\frac{\uhat(2\shat-\that)}{(\shat+\that)^3}
\Bigg]
,\nl
\different{K^{\mathrm{N}}_{5b}(M_V^2)} 
&=& 
\frac{2 \shat (\shat+\that)-2 \that \uhat}{{{2(\shat+\that)}^2}}
+\Biggl[
\frac{\shat+\that}{\uhat}
-\frac{2\shat}{\shat+\that}
-\frac{\uhat(2\shat+\that)}{2(\shat+\that)^2}
\Biggr]
+M_W^2\Biggl[
\frac{1}{\uhat}-\frac{2\shat+\that}{2(\shat+\that)^2}
\Biggr]
\nl&&{}-M_V^2\Bigg[
\frac{2\shat}{(\shat+\that)^2}
+\frac{\uhat(2\shat-\that)}{(\shat+\that)^3}
\Bigg]
,\nl
\different{K^{\mathrm{N}}_{6a}(M_V^2)} 
&=& K^{\mathrm{N}}_{5a }(M_V^2)\Big|_{\that\leftrightarrow\uhat} 
,\nl
\different{K^{\mathrm{N}}_{6b}(M_V^2)} 
&=& K^{\mathrm{N}}_{5b }(M_V^2)\Big|_{\that\leftrightarrow\uhat} 
,\nl
K^{\mathrm{N}}_{7}(M_V^2) 
&= &
-K^{\mathrm{A}}_{7}(M_V^2) 
,\nl
K^{\mathrm{N}}_{8}(M_V^2) 
&=& 
-K^{\mathrm{A}}_{8}(M_V^2)
,\nl
\different{K^{\mathrm{N}}_{9a}(M_V^2)} 
&=& 
\uhat-\frac{\shat^2-\shat\that}{2\that}+\frac{{\shat^2}+ \shat\that}{2\uhat}
+{(M_W^2+M_V^2)} \Bigg[\frac{2 {\shat^2}+\that \shat+{\that^2}}{2\that \uhat}
-\frac{\that-\shat}{2\that}\Bigg] 
\nl&&{}
-\frac{2 M_V^2\that \uhat}{{{(\shat+\that)}^2}}
+\frac{{M_V^2} }{\shat+\that}
\Bigg[\frac{{M_W^2 \shat^2}}{\that \uhat}
-\frac{M_V^2\uhat(2\shat-\that)}{{(\shat+\that)}^2}
-\frac{2(M_W^2+M_V^2) \shat}{\shat+\that}
\Bigg]
,\nl
\different{K^{\mathrm{N}}_{9b}(M_V^2)} 
&=&  
\left. K^{\mathrm{N}}_{9a}(M_V^2) \right|_{M_V^2\leftrightarrow M_W^2}
,\nl
K^{\mathrm{N}}_{10}(M_V^2) 
&=&  K^{\mathrm{N}}_{8 }(M_V^2)\Big|_{\that\leftrightarrow\uhat} 
=-K^{\mathrm{A}}_{10}(M_V^2) 
,\nl
\different{K^{\mathrm{N}}_{11a}(M_V^2)} 
&=& 
 K^{\mathrm{N}}_{9a }(M_V^2)\Big|_{\that\leftrightarrow\uhat}
,\nl
\different{K^{\mathrm{N}}_{11b}(M_V^2)} 
&=& 
 K^{\mathrm{N}}_{9b }(M_V^2)\Big|_{\that\leftrightarrow\uhat}
,\nl
K^{\mathrm{N}}_{12}(M_V^2) 
&=& 
-K^{\mathrm{A}}_{12}(M_V^2) 
,\nl
K^{\mathrm{N}}_{13}(M_V^2) 
&=& K^{\mathrm{N}}_{12}(M_V^2)\Big|_{\that\leftrightarrow\uhat} 
=-K^{\mathrm{A}}_{13}(M_V^2) 
,\nl
\different{K^{\mathrm{N}}_{14a}(M_V^2)} 
&=& 
\frac{{M_W^2}\that+M_V^2\uhat-\that \uhat}{2\that \uhat} 
\Bigg[2 {M_W^2}{M_V^2}+(2 \shat+\that+\uhat) ({M_W^2}+{M_V^2})
-2\that \uhat
\nl&&{}
-\shat (\that+\uhat)\Bigg]
,\nl
\different{K^{\mathrm{N}}_{14b}(M_V^2)} 
&=&  
\left. K^{\mathrm{N}}_{14a}(M_V^2) \right|_{M_V^2\leftrightarrow M_W^2}
.
\eeqar
For $M_V=M_W$  the function $H_1^{\mathrm{N}}(M_V^2)$ 
is identical to the equally named function in \citere{Kuhn:2005az}.

The only non-vanishing coefficients of the function
$H_1^{\mathrm{X}}(M_V^2)$ read
\beqar\label{xcoeff}
K^{\mathrm{X}}_{0 }(M_V^2) 
&=& 
-\frac{\that^2+\uhat^2+\shat(\that+\uhat)}{\that\uhat}
,\nl
K^{\mathrm{X}}_{1a }(M_V^2) 
&=& 
-\left[K^{\mathrm{X}}_{5a }(M_V^2) +K^{\mathrm{X}}_{6a}(M_V^2)\right]
,\nl
K^{\mathrm{X}}_{5a }(M_V^2) 
&=& 
\frac{2(M_V^2-\uhat)(\shat+\that)}{\uhat^2}
,\nl
K^{\mathrm{X}}_{6a}(M_V^2) 
&=& K^{\mathrm{X}}_{5a }(M_V^2)\Big|_{\that\leftrightarrow\uhat} 
.
\eeqar

Finally, for $H_1^{\mathrm{Y}}(M_V^2)$ we obtain
\beqar
\different{K^{\mathrm{Y}}_{0 }(M_V^2)} 
&=& 
\frac{(\uhat-\that)(\shat+\that+\uhat)}{\that \uhat}
,\nl
\different{{K^{\mathrm{Y}}_{1a}(M_V^2)}} 
&=&  
M_V^2\Biggl[
\frac{2}{\that}
+\frac{3\shat}{(\shat+\that)^2}
-\frac{1}{\shat+\that}
-\frac{2}{\uhat}
-\frac{3\shat}{(\shat+\uhat)^2}
+\frac{1}{\shat+\uhat}
\Biggr]
\nl&&{}
+M_V^2\Biggl[
\frac{2(\shat+\uhat)}{\that^2}
-\frac{2(\shat+\that)}{\uhat^2}
\Biggr]
,\nl
\different{{K^{\mathrm{Y}}_{1b}(M_V^2)}} 
&=&  
-M_W^2\Biggl[
\frac{2}{\that}
+\frac{3\shat}{(\shat+\that)^2}
-\frac{1}{\shat+\that}
-\frac{2}{\uhat}
-\frac{3\shat}{(\shat+\uhat)^2}
+\frac{1}{\shat+\uhat}
\Biggr]
\nl&&{}
-\Biggl[
\frac{2(\shat+\uhat)}{\that}
-\frac{2(\shat+\that)}{\uhat}
\Biggr]
,\nl
\different{K^{\mathrm{Y}}_{2}(M_V^2)} 
&=& 0
,\nl
\different{{K^{\mathrm{Y}}_{3}(M_V^2)}} 
&=& 
\frac{2(M_W^2-M_V^2)(\that-\uhat)(\shat+\that+\uhat)}{(\shat+\that)^3(\shat+\uhat)^3}
\biggl[-7\shat^3+(\that\uhat-6\shat^2)(\that+\uhat)
\nl&&{}
-\shat(2\that^2-\that\uhat+2\uhat^2)
\biggr]
,\nl
\different{K^{\mathrm{Y}}_{4}(M_V^2)} 
&=& 0
,\nl
\different{{K^{\mathrm{Y}}_{5a}(M_V^2)}} 
&=& 
-\frac{2\shat^2+3\that\uhat+2\shat(\that+\uhat)}{(\shat+\that)^2}
-M_W^2\biggl[
\frac{4\shat}{(\shat+\that)^2}
+\frac{2\uhat(2\shat-\that)}{(\shat+\that)^3}
\biggr]
\nl&&{}
-M_V^2\biggl[
\frac{2\shat+\that}{(\shat+\that)^2}
-\frac{2}{\uhat}
-\frac{2(\shat+\that)}{\uhat^2}
\biggr]
,\nl
\different{{K^{\mathrm{Y}}_{5b}(M_V^2)}} 
&=& 
\frac{2\shat^2+3\that\uhat+2\shat(\that+\uhat)}{(\shat+\that)^2}
+M_V^2\biggl[
\frac{4\shat}{(\shat+\that)^2}
+\frac{2\uhat(2\shat-\that)}{(\shat+\that)^3}
\biggr]
\nl&&{}
+M_W^2\biggl[
\frac{2\shat+\that}{(\shat+\that)^2}
-\frac{2}{\uhat}
\biggr]
-\frac{2(\shat+\that)}{\uhat}
,\nl
\different{{K^{\mathrm{Y}}_{6a}(M_V^2)}} 
&=& - K^{\mathrm{Y}}_{5a }(M_V^2)\Big|_{\that\leftrightarrow\uhat} 
,\nl
\different{{K^{\mathrm{Y}}_{6b}(M_V^2)}} 
&=& -K^{\mathrm{Y}}_{5b }(M_V^2)\Big|_{\that\leftrightarrow\uhat} 
,\nl
\different{K^{\mathrm{Y}}_{7}(M_V^2)} 
&= &0
,\nl
\different{K^{\mathrm{Y}}_{8}(M_V^2)} 
&=& 0
,\nl
\different{K^{\mathrm{Y}}_{9a}(M_V^2)} 
&=& 
\frac{1}{(\shat+\that)^3 \that\uhat}
\Biggl\{
2M_V^4\that\uhat\biggl[2\shat^2-\that\uhat+2\shat(\that+\uhat)\biggr]
\nl&&{}
-M_V^2(\shat+\that)\biggl[
(\shat+\that)^2(2\shat^2+\shat\that+\that^2)
+(\shat-\that)(\shat+\that)^2\uhat
-4\that^2\uhat^2
\nl&&{}
+2M_W^2\shat\bigg(
\shat(\shat+\that)-2\that\uhat
\biggr)
\biggr]
-(\shat+\that)^3
\biggl[
\shat^2(\that-\uhat)+2\that\uhat^2
\nl&&{}
+\shat\that(\that+\uhat)
+M_W^2\biggl(
2\shat^2+\that(\that-\uhat)+\shat(\that+\uhat)
\biggr)
\biggr]
\Biggr\}
,\nl
\different{{K^{\mathrm{Y}}_{9b}(M_V^2)}} 
&=&  
-\left. K^{\mathrm{Y}}_{9a}(M_V^2) \right|_{M_V^2\leftrightarrow M_W^2}
,\nl
\different{K^{\mathrm{Y}}_{10}(M_V^2)} 
&=&  0
,\nl
\different{K^{\mathrm{Y}}_{11a}(M_V^2)} 
&=& 
- K^{\mathrm{Y}}_{9a }(M_V^2)\Big|_{\that\leftrightarrow\uhat}
,\nl
\different{K^{\mathrm{Y}}_{11b}(M_V^2)} 
&=& 
- K^{\mathrm{Y}}_{9b }(M_V^2)\Big|_{\that\leftrightarrow\uhat}
,\nl
\different{K^{\mathrm{Y}}_{12}(M_V^2)} 
&=& 0
,\nl
\different{K^{\mathrm{Y}}_{13}(M_V^2)} 
&=&0
,\nl
\different{{K^{\mathrm{Y}}_{14a}(M_V^2)}} 
&=& 
-2{K^{\mathrm{N}}_{14a}(M_V^2)} 
,\nl
\different{{K^{\mathrm{Y}}_{14b}(M_V^2)}} 
&=&  
2{K^{\mathrm{N}}_{14b}(M_V^2)}= -\left. K^{\mathrm{Y}}_{14a}(M_V^2) \right|_{M_V^2\leftrightarrow M_W^2}
.
\eeqar

\section{Real corrections}
\label{app:diptables}
In Table \ref{table:massreg} and Table \ref{table:dimreg} we list the dipoles that were used 
to calculate the subtraction terms in \refeq{MDip} for the massive regularization and 
in \refeq{MDip:dimreg} for the dimensional regularization, respectively.
We give references to the explicit formulae for the dipole terms and the phase-space mappings 
in the original paper \cite{Dittmaier:1999mb} and \cite{Catani:1996vz,Catani:2002hc}.

\btab[ph]
\bce
\begin{tabular}{|c|c|c|c|}
\hline
Dipole & Type (emitter, spectator) & eq.~no. & $\tilde{\Phi}_{2,nm}$ \\
\hline
\hline
$\gsub{}{ab,\tau}$ 
& massless IS, massless IS  & (3.22)  & (3.25)--(3.27)          \\
\hline
$\gsub{}{aW,\tau}$
& massless IS, massive  FS    & ({\rm A}.1)  & (3.12)          \\
\hline
$\gsub{}{Wa,\tau}$
& massive FS, massless IS & ({\rm A}.1)   & (3.12)         \\
\hline
$\gsub{}{ak,\tau}$ 
& massless IS, massless FS    & (3.9) & (3.12)           \\
\hline
$\gsub{}{ka,\tau}$
& massless FS, massless IS& (3.9)  & (3.12)          \\
\hline
$\gsub{}{kW,\tau}$
& massless FS, massive FS & (4.4)   &  (4.5) \\
\hline
$\gsub{}{Wk,\tau}$
& massive FS, massless FS  & (4.4) &  (4.5)\\
\hline
\end{tabular}
\ece
\caption{\label{table:massreg}Dipole subtraction terms from~\citere{Dittmaier:1999mb} used to calculate
  $\mathsf{M}^{a b}_{\rm sub}$ in \refeq{MDip} for the massive regularization 
(IS = initial-state, FS = final-state).}
\etab

\btab[ph]
\bce
\begin{tabular}{|c|c|c|c|}
\hline
Dipole & Type (emitter, spectator) & eq. nos. & $\TPHI{2,nm}$\\
\hline 
\hline
 $\Cdip{a \ga, b}{}$ 
& massless IS, massless IS& 
(5.136), (5.145) & (5.137), (5.139),   \\
&  & in Ref.~\cite{Catani:1996vz}& (5.140) in Ref.~\cite{Catani:1996vz} \\
\hline
$\Cdip{a \ga}{W,}$ &
massless IS, massive FS & 
(5.71), (5.81) & (5.73), (5.74) \\
&  &  in Ref.~\cite{Catani:2002hc} &  in Ref.~\cite{Catani:2002hc}\\
\hline
$\Cdip{a}{\ga W,}$ & 
massive FS, massless IS  &
(5.40), (5.50) & (5.42), (5.43) \\  
& &  in Ref.~\cite{Catani:2002hc}&  in Ref.~\cite{Catani:2002hc}\\
\hline
 $\Cdip{a \ga}{k,}$ & 
massless IS, massless FS & 
(5.61), (5.65) & (5.62)-(5.64)  \\
&   & in Ref.~\cite{Catani:1996vz}& in Ref.~\cite{Catani:1996vz}\\ 
\hline
 $\Cdip{a}{\ga k,}$ &
massless FS, massless IS &  
(5.36), (5.39) & (5.37), (5.38)  \\
&   &  in Ref.~\cite{Catani:1996vz}&  in Ref.~\cite{Catani:1996vz}\\ 
\hline
$\Cdip{}{\ga k, W,}$  &
massless FS, massive FS &
(5.2), (5.16)& (5.3), (5.7), (5.9)  \\
& &  in Ref.~\cite{Catani:2002hc} &  in Ref.~\cite{Catani:2002hc}\\ 
\hline
$\Cdip{}{\ga W, k,}$ &
massive FS, massless FS & 
(5.2), (5.16)& (5.3), (5.7), (5.9)    \\
&  &  in Ref.~\cite{Catani:2002hc} &  in Ref.~\cite{Catani:2002hc}  \\
\hline
\end{tabular}
\ece
\caption{\label{table:dimreg}Dipole expressions from~\citeres{Catani:1996vz,Catani:2002hc} 
used to calculate $\mathsf{M}_{\rm sub}^{a b}$ in~(\ref{MDip:dimreg}) 
for the dimensional regularization.}
\etab

\end{appendix}


\begin{thebibliography}{99}
\bibitem{NLOrealcorrexp}
The expressions for the functions $H^{i,\bar q q'}_\rr$
in~\refeq{MReal} are available from the
authors upon request.
\end{thebibliography}
\end{document}